%% file: main.tex
\documentclass[11pt,letterpaper,reqno]{amsart}

\usepackage[foot]{amsaddr}

\usepackage[textwidth = 2cm]{todonotes}

\usepackage{aliascnt}
\usepackage{xparse} 

\makeatletter
\let\AC@oldnewtheorem\newtheorem
\RenewDocumentCommand{\newtheorem}{s m o m o}{%
  \IfBooleanTF{#1}{%
    \AC@oldnewtheorem*{#2}{#4}%
  }{%
    \IfNoValueTF{#3}{%
      \IfNoValueTF{#5}{%
        \AC@oldnewtheorem{#2}{#4}%
      }{%
        \AC@oldnewtheorem{#2}{#4}[#5]%
      }%
    }{%
      \newaliascnt{#2}{#3}%
      \AC@oldnewtheorem{#2}[#2]{#4}%
      \aliascntresetthe{#2}%
    }%
  }%
}
\makeatother

\usepackage{amsmath}
\usepackage{amsfonts}
\usepackage{amssymb}
\usepackage{amsthm}
\usepackage{mathtools}
\usepackage{verbatim}
\usepackage{dsfont}
\usepackage[letterpaper,margin=1in]{geometry}
\usepackage{microtype}
\usepackage[libertine]{newtxmath}
\usepackage{newtxtext}
\allowdisplaybreaks

\usepackage[tt=false]{libertine} 
\usepackage{caption}
\usepackage{bm}
\usepackage{bbm}
\usepackage{hyperref, color}
\hypersetup{colorlinks=true,citecolor=blue, linkcolor=blue, urlcolor=blue}
\usepackage[linesnumbered,boxed,ruled,vlined]{algorithm2e}
\usepackage{algorithmicx}
\usepackage[numbers]{natbib}
\usepackage{xcolor}
\usepackage{enumerate} 
\usepackage[shortlabels]{enumitem}
\usepackage{tabularx}
\usepackage{array}
\usepackage{cleveref}
\usepackage{pifont}
\usepackage{xifthen}
\usepackage{xstring}

\usepackage{stackengine}
\usepackage{scalerel}
\usepackage{graphicx}
\usepackage{float}
\usepackage{tikz}
\usepackage{adjustbox}
\usepackage{fancyhdr, comment, environ}

\usetikzlibrary{positioning}
\usetikzlibrary{tikzmark, fit, backgrounds}
\usetikzlibrary{matrix,fit,backgrounds}  

\DeclareMathAlphabet{\mathcal}{OMS}{cmsy}{m}{n}

\newlist{loopblock}{itemize}{1}
\setlist[loopblock]{
  label={},
  leftmargin=2.5em,
  itemindent=0pt,
  itemsep=0.35em,
  topsep=0.35em
}

\newcolumntype{L}[1]{>{\raggedright\arraybackslash}p{#1}}
\newcolumntype{C}[1]{>{\centering\arraybackslash}m{#1}}
\newcolumntype{R}[1]{>{\raggedleft\arraybackslash}p{#1}}

\usepackage{makecell}
\usepackage{footnote}
\makesavenoteenv{tabular}

\newtheorem{theorem}{Theorem}[section]

\newtheorem*{claim*}{Claim}

\newtheorem{lemma}[theorem]{Lemma}
\newtheorem{proposition}[theorem]{Proposition}
\newtheorem{corollary}[theorem]{Corollary}
\theoremstyle{definition}

\newtheorem{definition}[theorem]{Definition}

\newtheorem*{remark*}{Remark}

\newcommand{\R}{\mathbb{R}}
\newcommand{\one}{\mathbf{1}}

\def\^#1{\mathbb{#1}} 
\def\*#1{\mathbf{#1}} 
\def\+#1{\mathcal{#1}} 
\def\-#1{\mathrm{#1}} 
\def\=#1{\boldsymbol{#1}} 
\def\@#1{\mathscr{#1}}

\newcommand{\abs}[1]{\left | #1 \right |}

\renewcommand{\Pr}[2][]{ \ifthenelse{\isempty{#1}}
  {\mathop{\mathbf{Pr}}\left[#2\right]} {\mathop{\mathbf{Pr}}_{#1} \left[#2\right]} }
\newcommand{\E}[2][]{ \ifthenelse{\isempty{#1}}
  {\mathbf{E}\left[#2\right]}
  {\mathbf{E}_{#1}\left[#2\right]} }

\newcommand{\diag}{\mathsf{diag}}

\newcommand{\norm}[1]{\left\Vert#1\right\Vert}

\makeatletter
\renewcommand{\paragraph}{%
  \@startsection{paragraph}{4}{\z@}%
    {1ex \@plus .5ex \@minus .2ex}%
    {-1em}%
    {\normalfont\normalsize\bfseries}%
}
\makeatother

\title[Global Sinkhorn Convergence and Spectral Stability of Matrix Scaling]
{Global Sinkhorn Convergence and Spectral Stability \\of Matrix Scaling}

\author{Kun He}
\address[Kun He]{Renmin University of China. \textnormal{E-mail: \url{hekun2023@ruc.edu.cn}}}

\begin{document}

\begingroup
\let\Small\normalsize
\begin{abstract}
We give a global spectral bound for the Sinkhorn--Knopp (SK) algorithm.
Let $A$ be a row-stochastic nonnegative $n\times n$ matrix admitting
a finite doubly stochastic scaling $B$. Define
\[
 g=1-s_2(B)^2>0,
 \qquad
 D=\min\left\{\|(u,v)\|_\infty:
 B=\operatorname{diag}(e^u)A\operatorname{diag}(e^{-v})\right\}.
\]
From the standard initialization, SK achieves total marginal
$\ell_1$ error $\varepsilon$ in
\[
 O\!\left(\frac{D+\log n+\log(1/\varepsilon)}{g}\right)
\]
iterations. A scale-invariant potential contracts
globally according to the spectrum of $B$.
The linear dependence on $D$ is necessary in the fixed-gap regime:
we construct strictly positive inputs with constant input and endpoint
spectral gaps, $D=\Theta(n)$, and an $\Omega(n)$ iteration lower bound
at error $n^{-2}$. Our upper bound gives $O(n)$ iterations on this family.

We next give spectral certificates for scalability and short scaling
distance. For a row-stochastic input $A$ with positive column sums
$c_j$, set
\[
 K=\frac{\max_jc_j}{\min_jc_j},\qquad
 \sigma=s_2\!\left(A\operatorname{diag}(c)^{-1/2}\right).
\]
The input-only condition $\sqrt K\,\sigma<1$ certifies exact scalability
and polynomial scaling distortion, with $D=O(\log n)$ for fixed $K$
and a fixed positive spectral margin.
Given a finite scaling $B$, the weaker condition
$\sqrt K\,\sigma s_2(B)<1$ gives the same structural bounds.
With a constant endpoint gap, these bounds imply logarithmic SK
iteration complexity. Boundary constructions show that the strict
spectral inequalities cannot uniformly be replaced by nonstrict ones.

For the structural bounds, we propagate scaling factors using
two-sided buffered robust vertex expansion.
For dense inputs satisfying the expansion conditions, macroscopic
seeds yield constant distortion.
This recovers logarithmic SK convergence above density $1/2$ and,
under an additional input spectral condition, guarantees it at any
fixed positive density.
\end{abstract}
\endgroup

\maketitle

\setcounter{tocdepth}{1}

\clearpage
\tableofcontents
\pagebreak

\input{Introduction.tex}
\input{Preliminary}
\input{Global_convergence}
\input{buffered_expansion}
\input{Spectral_Condition}

\input{Lowerbounds}
\input{Dense_Matrix}


\clearpage
\appendix
\input{MissingProof}

\clearpage


\section*{AI methodology}
We used ChatGPT (GPT-5.4, GPT-5.5, GPT-5.6, and GPT-6 Pro) to assist in constructing counterexamples for the lower bounds
and developing the arguments connecting spectral gaps to buffered
robust vertex expansion, and structural stability to the convergence
rate of the Sinkhorn--Knopp algorithm.
These tools materially affected Sections \ref{sec-spectralgap-brvx}, \ref{sec-structural stability-convergence}, \ref{sec-lowerbounds}, and \ref{sec-dense-lowerbound}. The author independently verified the correctness and originality of all content, including references.

\bibliographystyle{alpha}
\bibliography{refs}

\end{document}

%% file: Introduction.tex
\section{Introduction}\label{sec-intro}
\medskip
\noindent 

Given a nonnegative matrix $A$ and prescribed row and column sums,
the matrix scaling problem asks for positive diagonal matrices
$X$ and $Y$ such that $XAY$ matches these marginals.
This problem arises throughout statistics, optimization, and
numerical computation. Classical applications include adjusting
contingency tables to specified marginal totals
\cite{deming1940least}.
A prominent modern application is entropically regularized
optimal transport, where marginal constraints are enforced
through diagonal scaling
\cite{cuturi2013sinkhorn,altschuler2017near}.
This connection has made matrix scaling widely useful for
comparing distributions in machine learning and computer vision.
For further background and applications, see
\cite{idel2016review}.

We focus on scaling a nonnegative $n\times n$ matrix to a doubly
stochastic matrix, whose row and column sums all equal one.
For a nonnegative $n\times n$ matrix $P$, write
$r_i(P)=\sum_j P_{ij}$ and $c_j(P)=\sum_i P_{ij}$, and let
$r(P)$ and $c(P)$ be the corresponding marginal vectors.
A matrix is row-stochastic if $r(P)=\mathbf{1}$,
column-stochastic if $c(P)=\mathbf{1}$, and doubly stochastic
if both conditions hold. We focus on scaling to doubly
stochastic matrices.
We measure approximation quality by the total marginal
$\ell_1$ error and seek an output satisfying
\[
 \|r(P)-\mathbf{1}\|_1+\|c(P)-\mathbf{1}\|_1
 \leq\varepsilon.
\]
Throughout the discussion below, $0<\varepsilon\leq1$.

The classical Sinkhorn--Knopp (SK) algorithm
\cite{sinkhorn1967diagonal,sinkhorn1967concerning}, also known as
RAS \cite{bacharach1965estimating} or iterative proportional
fitting \cite{ruschendorf1995convergence}, alternates row and
column normalizations. Each update is simple, preserves the
support of the input, and can be implemented in time linear in the
number of nonzero entries. 
We count a single row or column normalization as one iteration.
Our indexing starts from the row-normalized input $A^{(0)}$,
followed by alternating column and row normalizations.
Thus, $A^{(2t)}$ denotes the matrix after $t$ complete
column--row cycles.
The low cost of these updates,
together with their parallel implementation, makes SK a standard
method for matrix scaling. Greedy coordinate variants, including
Greenkhorn \cite{altschuler2017near}, provide another approach to
enforcing the marginal constraints.

Convex optimization gives further algorithms for matrix scaling.
\cite{cohen2017matrix} develop a
box-constrained Newton method and an interior-point method,
while \cite{allen2017much} give several faster
algorithms for general scaling problems and important special
cases. Writing $\ell=\operatorname{nnz}(A)$, when an exact scaling
exists with diagonal factors bounded above and below by polynomial
factors in $n$, the box-constrained Newton method yields a running time of
$\widetilde{O}(\ell\log^2(n/\varepsilon))$ in the doubly stochastic
setting, and \cite{allen2017much} obtain
$\widetilde{O}(\ell+n^{4/3})$.
\cite{cohen2017matrix} also give an interior-point alternative with running
time $\widetilde{O}(\ell^{3/2}\log(n/\varepsilon))$.
These are total running times, with additional logarithmic
dependence on the dimension and relevant input parameters
suppressed; the bound in \cite{allen2017much} also suppresses
polylogarithmic dependence on $1/\varepsilon$.
Such results make control of the scaling factors important beyond
the analysis of alternating normalization.

Spectral analysis offers a way to relate algorithmic behavior to
the structure of a particular scaling instance. The associated
spectral quantities describe how mass mixes across rows and
columns and how strongly the marginal constraints couple
different parts of the matrix. They also enter the curvature of
convex formulations of scaling. This perspective can explain why
some instances converge much faster than worst-case accuracy
bounds suggest, while also providing information about the
conditioning of the scaling solution.

Classical convergence theory already distinguishes global
contraction from the behavior near the limiting scaling.
For positive matrices, Franklin and Lorenz
\cite{franklin1989scaling} establish global geometric convergence
in Hilbert's projective metric, with contraction bounds determined
by cross-ratios of the input entries. A complementary analysis
concerns the spectrum of the limiting matrix. For fully
indecomposable square matrices, Knight
\cite{knight2008sinkhorn} identifies $s_2(B)^2$ as the governing
asymptotic spectral factor per complete row--column normalization
cycle. Here, $B$ is the doubly stochastic scaling and $s_2(B)$
denotes its second-largest singular value, with singular values
counted with multiplicity. This identifies the local contraction
mechanism, but does not quantify how long the algorithm takes to
enter the asymptotic regime.

Spectral assumptions can also yield structural bounds on the
scaling solution itself. Kwok, Lau, and Ramachandran
\cite{kwok2019spectral} relate an input spectral gap to linear
convergence of a scaling gradient flow and its discretization,
conditioning of the scaling matrices, and capacity bounds.
Their analysis assumes that the input is already nearly balanced,
with the permitted imbalance constrained by the spectral gap.
In the square setting, let $\varepsilon_0$ denote the relative
initial marginal imbalance and $\lambda$ their spectral-gap
parameter. Their guarantees require
$\lambda^2\geq C\varepsilon_0\log n$
for a sufficiently large absolute constant $C$.
For a fixed spectral gap, this requires
$\varepsilon_0=O(1/\log n)$, so the permitted initial imbalance
must decrease with the dimension.
Under this assumption, the scaling matrices have condition
numbers bounded by $1+O(\varepsilon_0\log n/\lambda)$.
Thus, these guarantees combine spectral mixing with an initial
near-balance condition that becomes more restrictive as the
dimension grows.
Franks and Moitra \cite{franks2020rigorous} connect quantum
expansion with geodesic strong convexity and use this connection
to obtain stability and linear convergence guarantees for
operator scaling under suitable quantitative conditions.
Their analysis also gives statistical and computational
guarantees for Tyler's M-estimator.

Recent work also gives global and nonasymptotic spectral convergence bounds.
\cite{qu2026sinkhorn} establish global linear convergence of
Sinkhorn's algorithm for nonnegative matrices admitting finite
scalings and having connected support graphs.
Their bounds involve the algebraic connectivity of an associated
weighted bipartite graph and scaling-dependent condition measures;
the refined estimates retain polynomial dependence on these measures.
\cite{gao2026tight} give a nonasymptotic analysis of the local linear
regime and an explicit bound on the preceding global phase.
Their bounds depend on the normalized spectrum of the optimal
scaling, the size of suitable logarithmic scaling vectors, and
the smallest target marginal.
These results quantify both spectral contraction and the cost of
reaching a regime where that contraction is effective.

The preceding results motivate a global question: how do the distance
to the scaling solution and its spectral mixing jointly control the
entire SK trajectory?
For a row-stochastic matrix $A$ with a finite doubly stochastic
scaling $B$, define its optimal logarithmic scaling radius by
\[
 D(A;B)=
 \min_{\substack{u,v\in\mathbb R^n\\
 B=\operatorname{diag}(e^u)A\operatorname{diag}(e^{-v})}}
 \max\{\|u\|_\infty,\|v\|_\infty\}.
\]
This radius measures the initial adjustment required by the scaling
factors, after minimizing over all logarithmic representations of the scaling.
The endpoint gap $g=1-s_2(B)^2$ measures the mixing of the exact scaling.
We seek a global iteration bound that displays both parameters.

The structural problem is then to certify that the radius is small.
Bounds on all diagonal scaling products,
\[
 T^{-1}\le x_i y_j\le T\qquad\forall i,j\in[n],
\]
give $D(A;B)\le\log T$ by a common centering of
$\log x_i$ and $-\log y_j$.
We call such product bounds structural stability under diagonal
scaling. They also imply $T^{-1}A_{ij}\le B_{ij}\le TA_{ij}$.
Thus polynomial distortion gives logarithmic scaling radius,
and constant distortion gives constant radius.
Our spectral criteria seek this control while allowing an initial
marginal imbalance that remains fixed as the dimension grows.

Combinatorial expansion also provides structural control of
scaling. For symmetric graph adjacency matrices, Rudelson,
Samorodnitsky, and Zeitouni \cite{rudelson2016hafnians} show that
linear minimum degree and an appropriate strong vertex-expansion
condition imply polynomial lower bounds on the diagonal entries
of a symmetric doubly stochastic scaling. Their proof already
propagates scaling-factor bounds through growing vertex sets.
Our weighted, nonsymmetric formulation and its auxiliary-matrix
reduction are described in the technique overview below.

Recent results for dense inputs reveal a sharp dependence of SK
convergence on combinatorial structure.
For fixed $\gamma,\rho\in(0,1]$, a matrix
$D\in[0,1]^{n\times n}$ is $(\gamma,\rho)$-dense if every row and
column contains at least $\lceil\gamma n\rceil$ entries of value
at least $\rho$. 
He \cite{he2026phase} establishes a sharp
distinction between the regimes above and below density $1/2$.
For fixed $\gamma>1/2$ and $\rho>0$, SK reaches total marginal
$\ell_1$ error at most $\varepsilon$ in
$O_{\gamma,\rho}(\log n+\log(1/\varepsilon))$ iterations.
Conversely, for every fixed $0<\gamma<1/2$ and sufficiently small
$\varepsilon$, there are exactly scalable dense inputs requiring
$\Omega(n/\varepsilon)$ iterations.
These lower bounds show that density alone cannot guarantee
rapid convergence below the threshold.
A finer understanding therefore calls for conditions that capture
how matrix mass connects subsets of rows and columns.
The resulting structural question is which quantitative expansion
properties explain rapid convergence in the dense regime and
continue to guarantee it for inputs below density $1/2$.

These considerations lead to three questions:
\begin{enumerate}
\item Can the endpoint spectral gap control the whole SK trajectory,
with an explicit dependence on the logarithmic scaling radius?
\item Is dependence on that radius necessary even when both the
normalized input and the exact scaling have constant spectral gaps?
\item Which input-only or mixed spectral conditions certify a small
radius, and how do these certificates explain fast convergence for
dense inputs and further instances below density $1/2$?
\end{enumerate}

\subsection{Main Results}
Our results first control the SK dynamics in terms of the scaling
radius and endpoint gap, then provide structural certificates for
the radius. Throughout, the input is row-stochastic; a general input
with positive row sums is first row-normalized.
We write $D=D(A;B)$ and $g=1-s_2(B)^2$ when the instance is clear.

\paragraph{Global convergence from scaling radius and endpoint gap.}
The first theorem applies to every nonnegative square input with a
finite doubly stochastic scaling and $g>0$.
It requires neither a density condition nor an expansion assumption.

\begin{theorem}[Global convergence in terms of scaling radius and gap]
\label{thm:global-sk-logradius-gap}
Let $n\ge2$ and let $A\in\mathbb R_{\ge0}^{n\times n}$ be
row-stochastic. Suppose that $B$ is a doubly stochastic positive
diagonal scaling of $A$, and write
\[
 D=D(A;B),\qquad g=1-s_2(B)^2>0.
\]
Starting from $A^{(0)}=A$, alternately normalize columns and rows.
For every $0<\varepsilon\le1$ and every integer $t$ satisfying
\begin{equation}\label{eq:global-sk-logradius-gap-rounds}
 t\ge\left\lceil\frac{2}{g}
 \left(8D+\log\frac{16n^3}{\varepsilon^2}\right)\right\rceil,
\end{equation}
the row-stochastic matrix $A^{(2t)}$ satisfies
\[
 \|r(A^{(2t)})-\mathbf1\|_1
 +\|c(A^{(2t)})-\mathbf1\|_1\le\varepsilon.
\]
Consequently, the number of required iterations is
\[
 O\!\left(\frac{D+\log n+\log(1/\varepsilon)}g\right).
\]
\end{theorem}

We compare the global bound in
\Cref{thm:global-sk-logradius-gap} with
\cite[Theorem~3.2]{gao2026tight} using the same row-stochastic
input $A$, unit target marginals $p=q=\mathbf1$, and finite
scaling $B$.
Their $\varepsilon$-approximation criterion is the maximum of
the two marginal $\ell_2$ errors. Taking their tolerance to be
$\varepsilon/(2\sqrt n)$ ensures total marginal $\ell_1$ error
at most $\varepsilon$, at the cost of an additional $O(\log n)$
term in their bound.
With $D=D(A;B)$ and $g=1-s_2(B)^2$, their theorem then gives
\[
 O\!\left(
 \frac{nD^2}{g^4}
 +\frac{\log n+\log(1+D)+\log(1/\varepsilon)}{g}
 \right)
\]
complete cycles, up to one normalization step from the indexing
convention. Our bound is
\[
 O\!\left(\frac{D+\log n+\log(1/\varepsilon)}{g}\right).
\]
In particular, a fixed endpoint gap and $D=O(\log n)$ give a
logarithmic iteration bound here, whereas substitution into
their general bound gives an $O(n\log^2 n)$ global term.
The dependence on the endpoint gap is consistent with the local
spectral factor $s_2(B)^2$ identified by
\cite{knight2008sinkhorn}; the new global estimate also controls
the preceding displacement through $D$.

\paragraph{A matching obstruction at constant spectral gaps.}
The scaling-radius term is necessary even when the input and endpoint
mix rapidly. The following positive family will also witness the
boundary obstruction for the mixed spectral certificate below.

\begin{proposition}
\label{prop:product-threshold-counterexample}
For every positive integer $n$ divisible by $12$, there exists a strictly
positive row-stochastic matrix $A\in\mathbb{R}_{>0}^{n\times n}$ with
\[
 \min_j c_j(A)=\frac12,\qquad
 \max_j c_j(A)=2,\qquad
 K\triangleq\frac{\max_j c_j(A)}{\min_j c_j(A)}=4.
\]
Let $C=XAY$ be its doubly stochastic scaling, and define
\[
 Q=\operatorname{diag}(c_1(A),\ldots,c_n(A)),
 \qquad B=AQ^{-1/2}.
\]
Then
\begin{equation}\label{eq:ptc-spectral-equality}
 s_2(B)=s_2(C)=\frac1{\sqrt2},
 \qquad \sqrt K\,s_2(B)s_2(C)=1,
\end{equation}
but
\begin{equation}\label{eq:ptc-exact-distortion}
 \max_{i,j}\frac{C_{ij}}{A_{ij}}
 =\max_{i,j}\frac{A_{ij}}{C_{ij}}
 =1+\frac{2^n}{3}.
\end{equation}
Consider the iteration of the SK algorithm and let $A^{(0)}=A$. Define
\[
 T(\varepsilon)=\min\left\{t\in\mathbb{Z}_{\ge0}:\sum_{j\in [n]}\left|A^{(t)}([n],j)-1\right| + \sum_{i\in [n]}\left|A^{(t)}(i,[n])-1\right|\le\varepsilon\right\}.
\]
Then $T(n^{-2})=\Omega(n)$.
\end{proposition}

This family also shows that the linear dependence on the scaling
radius in \Cref{thm:global-sk-logradius-gap} is necessary in the
worst case.
Its exact distortion formula implies that
$D_n=D(A_n;C_n)=\Theta(n)$, while the endpoint gap is fixed at $1/2$.
At accuracy $\varepsilon=n^{-2}$,
\Cref{thm:global-sk-logradius-gap} therefore gives an $O(n)$
upper bound on the number of normalization steps, matching the
$\Omega(n)$ lower bound in
\Cref{prop:product-threshold-counterexample}.
The following corollary makes this matching bound explicit.

\begin{corollary}
\label{cor:fixed-gap-logradius-tight}
Let $A_n$ and $C_n$ be the input and its doubly stochastic scaling
from \Cref{prop:product-threshold-counterexample}, where
$n\in12\mathbb Z_{>0}$. Write
\[
 D_n=D(A_n;C_n),\qquad L_n=\log\left(1+\frac{2^n}{3}\right).
\]
Then
\begin{equation}\label{eq:counterexample-logradius-bounds}
 \frac{L_n}{2}\le D_n\le L_n,
 \qquad 1-s_2(C_n)^2=\frac12.
\end{equation}
In particular, $D_n=\Theta(n)$.

Moreover, let $T(\varepsilon)$ be the stopping time defined in
\Cref{prop:product-threshold-counterexample} for the input $A_n$,
counting each row or column normalization as one iteration.
Then, as $n\to\infty$ through multiples of $12$,
\begin{equation}\label{eq:fixed-gap-logradius-matching-time}
 T(n^{-2})=\Theta(n).
\end{equation}
Thus the linear dependence on $D$ in
\Cref{thm:global-sk-logradius-gap} is necessary in the worst case,
even when the endpoint gap is fixed and
$\log(1/\varepsilon)=O(\log n)$.
\end{corollary}

\paragraph{Spectral certificates for a small scaling radius.}
The global convergence theorem reduces the remaining structural
question to controlling $D(A;B)$.
We give two sufficient spectral criteria for $D(A;B)=O(\log n)$ at
fixed marginal imbalance and fixed positive spectral margin.
Throughout this discussion, $A$ is row-stochastic; for a general
input with positive row sums, first normalize its rows and apply
the criteria to the resulting matrix.

\begin{theorem}\label{cor-k-sigmaa}
Let $n\ge2$ and $K\ge1$. Let $A\in\mathbb{R}_{\ge0}^{n\times n}$
be row-stochastic, with positive column sums, and
\begin{align}
\frac{\max_j A([n],j)}{\min_j A([n],j)} = K.
\end{align}
Define
$M=A\left(\operatorname{diag}\left(A([n],1),\ldots,A([n],n)\right))\right)^{-1/2}$ and $\sigma=s_2(M)$.
If $1-\sqrt K\,\sigma>0$,
then there exists a doubly-stochastic matrix $B=XAY$, where $X=\diag(x_1,\ldots,x_n)$ and $Y=\diag(y_1,\ldots,y_n)$ are positive diagonal
matrices.
Moreover, there are constants $C_0\ge1$ and $\beta>0$, depending only on
$K$ and $1-\sqrt K\,\sigma$, such that with $T=C_0n^\beta$,
\begin{align*}
 T^{-1}\le x_i y_j\le T,
 \qquad
 T^{-1}A_{ij}\le B_{ij}\le T A_{ij}
 \qquad \forall i,j\in [n].
\end{align*}
\end{theorem}

\Cref{cor-k-sigmaa} is an input-only certificate.
For fixed $K$ and fixed
$\delta=1-\sqrt K\,\sigma>0$, it establishes
finite exact scalability and
\[
 D(A;B)\le\log C_0+\beta\log n
 =O_{K,\delta}(\log n).
\]
Both quantities in its hypothesis are determined by $A$; the
criterion does not require computing $B$ or its endpoint gap.
\Cref{thm:global-sk-logradius-gap} consequently gives
\[
 O_{K,\delta}\!\left(
 \frac{\log n+\log(1/\varepsilon)}{g}
 \right)
\]
iterations whenever $g>0$.
In particular, if $g$ is bounded below by a positive constant
independent of $n$, the SK algorithm achieves total marginal
$\ell_1$ error at most $\varepsilon$ within
$O(\log n+\log(1/\varepsilon))$ iterations.

Given a finite doubly stochastic scaling $B$, information about
its spectrum permits a weaker radius certificate.
The following theorem replaces the input-only hypothesis
$\sqrt K\,\sigma<1$ by
\[
 \sqrt K\,\sigma\sigma_\perp<1,
 \qquad \sigma_\perp=s_2(B).
\]
For fixed $K$ and a fixed positive margin, it again guarantees
$D(A;B)=O(\log n)$.
This mixed criterion uses endpoint information and is therefore
an analytical guarantee, whereas \Cref{cor-k-sigmaa} can be
verified from the input alone.

\begin{theorem}\label{thm:polynomial-scaling-gap-seeded}
Let $n\ge2$ and $K\ge1$. Let $A\in\mathbb{R}_{\ge0}^{n\times n}$
be row-stochastic, with positive column sums, and
\begin{align}\label{def-k-ratio}
\frac{\max_j A([n],j)}{\min_j A([n],j)} = K.
\end{align}
Let $B=XAY$ be doubly-stochastic, where $X=\diag(x_1,\ldots,x_n)$ and $Y=\diag(y_1,\ldots,y_n)$ are positive diagonal
matrices. 
Define
\begin{align}\label{eq-defmsigma}
M=A\left(\operatorname{diag}\left(A([n],1),\ldots,A([n],n)\right))\right)^{-1/2},\qquad \sigma=s_2(M),\qquad \sigma_{\perp}=s_2(B).
\end{align}
and assume
$\Delta\triangleq1-\sqrt K\,\sigma\sigma_{\perp}>0$.
Then there are constants $C_0\ge1$ and $\beta>0$, depending only on
$K$ and $\Delta$, such that with $T=C_0n^\beta$,
\begin{align}\label{eq-distrotion-t}
 T^{-1}\le x_i y_j\le T,
 \qquad
 T^{-1}A_{ij}\le B_{ij}\le T A_{ij}
 \qquad \forall i,j\in [n].
\end{align}
Moreover, consider the SK algorithm with input $A$.
For every $0<\varepsilon\le1$, there exists some integer
\begin{equation*}
 t = O_{K,\Delta}\left(
 \frac{\log n+\log(1/\varepsilon)}{1-\sigma_{\perp}^4}
 \right)
\end{equation*}
such that 
\begin{align}\label{eq-errorbound-maintheorem}
\norm{r\left(A^{(2t)}\right)- \boldsymbol{1}}_{1} + \norm{c\left(A^{(2t)}\right)- \boldsymbol{1}}_{1} \leq \varepsilon.
\end{align}
\end{theorem}

For fixed $K$ and
$\Delta=1-\sqrt K\,\sigma\sigma_\perp>0$,
\Cref{thm:polynomial-scaling-gap-seeded} gives
\[
 D(A;B)\le\log C_0+\beta\log n
 =O_{K,\Delta}(\log n)
\]
and guarantees total marginal $\ell_1$ error at most
$\varepsilon$ after
\[
 O_{K,\Delta}\!\left(
 \frac{\log n+\log(1/\varepsilon)}
      {1-\sigma_\perp^4}
 \right)
\]
full column--row cycles.
In particular, if $1-\sigma_\perp^4\ge\kappa$
for a fixed constant $\kappa>0$, then
$O_{K,\Delta,\kappa}(\log n+\log(1/\varepsilon))$
cycles suffice.

The strict inequality in
\Cref{thm:polynomial-scaling-gap-seeded} cannot be replaced by a
nonstrict one in a uniform guarantee.
Indeed, \Cref{prop:product-threshold-counterexample} gives a boundary
family with $K=4$ and $\sigma=\sigma_\perp=1/\sqrt2$ for which
$\sqrt K\,\sigma\sigma_\perp=1$, the scaling distortion is
exponential in $n$, and SK requires $\Omega(n)$ iterations to
achieve total marginal $\ell_1$ error at most $n^{-2}$.
This establishes worst-case boundary sharpness.

\vspace{0.3cm}
\paragraph{Boundary obstruction for the input-only certificate.} 
The input-only criterion in \Cref{cor-k-sigmaa} has an analogous
worst-case boundary obstruction. For every fixed $K>1$, at
$\sigma=1/\sqrt K$ there are positive matrices whose scaling
distortion is exponential in $n$ and for which SK requires
$\Omega_K(n^3)$ iterations to achieve total marginal $\ell_1$
error at most $n^{-2}$.

\begin{proposition}
\label{prop:critical-spectral-counterexample}
Fix $K>1$. For every positive integer $n$ divisible by $4$, there exists
a row-stochastic matrix $A\in\mathbb{R}_{>0}^{n\times n}$ such that the following conditions hold:
\begin{enumerate}
\item \label{first-prop-critical-spectral-counterexample}
The ratio of the largest to the smallest column sum of $A$ is exactly $K$: 
\begin{equation}\label{eq-k-exactratio}
\frac{\max_{j\in[n]} A([n],j)}{\min_{j\in[n]} A([n],j)} = K.
\end{equation}
\item  \label{second-prop-critical-spectral-counterexample}
Define
$B=A(\operatorname{diag}(A([n],1),\ldots,A([n],n)))^{-1/2}$.
Then 
\begin{equation}\label{eq-stwobk-exactratio}
 s_2(B)=\frac{1}{\sqrt{K}}.
\end{equation}
\item  \label{third-prop-critical-spectral-counterexample}
Let $C$ be the doubly stochastic matrix satisfying $C=XAY$ where $X,Y$ are positive diagonal
matrices. Then
\begin{equation}\label{eq-cvsa-exactratio}
 \max_{i,j}\frac{C_{ij}}{A_{ij}}=\Theta_K(2^n),
 \qquad
 \max_{i,j}\frac{A_{ij}}{C_{ij}}=\Theta_K(2^n).
\end{equation}
\item  \label{fourth-prop-critical-spectral-counterexample}
Consider the iteration of the SK algorithm and let $A^{(0)}=A$. Define
\[
 T(\varepsilon)=\min\left\{t\in\mathbb{Z}_{\ge0}:\sum_{j\in [n]}\left|A^{(t)}([n],j)-1\right| + \sum_{i\in [n]}\left|A^{(t)}(i,[n])-1\right|\le\varepsilon\right\}.
\]
Then $T\left(n^{-2}\right)=\Omega\left(n^3\right)$.
\end{enumerate}
\end{proposition}

\paragraph{Application to dense inputs.}
We next apply the framework for scaling radius and endpoint gap to dense matrices.
In this setting, the structural argument produces macroscopic
seeds, yielding both a constant scaling radius and a constant
endpoint gap under the hypotheses below.

\begin{theorem}
\label{cor:dense-seed}
Fix $\gamma,\rho\in(0,1]$.
For all sufficiently large $n$, let $D\in[0,1]^{n\times n}$ be $(\gamma,\rho)$-dense.
Define
\begin{align*}
 r&\triangleq \left(D(1,[n]),\cdots,D(n,[n])\right),\qquad c\triangleq \left(D([n],1),\cdots,D([n],n)\right).\\
 E&\triangleq\operatorname{diag}(r)^{-1/2}
 D\operatorname{diag}(c)^{-1/2},
 \qquad  \;
 \sigma\triangleq s_2(E).
\end{align*}
Then we have the following results:
\begin{enumerate}[(a)]
\item If $\sigma < \gamma\rho - \tau$ for some positive constant $\tau$, then the SK algorithm achieves the error $\varepsilon$ in $O_{\rho,\gamma,\tau}(\log n + \log 1/\varepsilon)$ iterations. \label{item-sigmaleqgammarhominustau}
\item If $\gamma> 1/2$, then the SK algorithm achieves the error $\varepsilon$ in $O_{\rho,\gamma}(\log n + \log 1/\varepsilon)$ iterations. \label{item-gammalargeroneovertwo}
\end{enumerate}
\end{theorem}

For $A=\operatorname{diag}(r)^{-1}D$ and its doubly stochastic
scaling $B$, the proof supplies $D(A;B)=O(1)$ and
$g=1-s_2(B)^2=\Omega(1)$, with constants depending on the fixed
density parameters and, in part~(a), the spectral margin.
The global convergence theorem then gives the stated
$O(\log n+\log(1/\varepsilon))$ bound.
Part~(b) recovers the density-above-$1/2$ guarantee of
\cite{he2026phase}; part~(a) gives an additional spectral route
to the same convergence bound at any fixed positive density.
Both follow from buffered robust vertex expansion, with the
large seed size strengthening the general bound on the radius
to a constant bound.

The next theorem gives a boundary obstruction at the specific
parameters $\gamma=1/2$ and $\rho=1$.
At $\sigma=\gamma\rho=1/2$, a positive dense family requires
$\Omega(n^2)$ iterations to achieve total marginal $\ell_1$
error at most $1/(400n)$.
Thus the strict spectral condition in \Cref{cor:dense-seed}
cannot uniformly be relaxed to equality.

\begin{theorem}\label{thm-tight-dense}
For every
sufficiently large even integer $n$, there exist a $(1/2,1)$-dense matrix
$D_n\in(0,1]^{n\times n}$ satisfying the following properties:
\begin{enumerate}
\item
Define
\[
\begin{aligned}
 R_n&=\operatorname{diag}\bigl(r_1(D_n),\ldots,r_n(D_n)\bigr),\\
 C_n&=\operatorname{diag}\bigl(c_1(D_n),\ldots,c_n(D_n)\bigr),\\
 E_n&=R_n^{-1/2}D_nC_n^{-1/2}.
\end{aligned}
\]
Then $\sigma_n\triangleq s_2(E_n)=1/2$.
\item
Consider the iteration of the SK algorithm and let $A_n^{(0)}=R_n^{-1}D_n$.
If
\[
 \sum_{i=1}^{n}\left|r_i(A_n^{(t)})-1\right|
 +
 \sum_{j=1}^{n}\left|c_j(A_n^{(t)})-1\right|
 \le \frac{1}{400n},
\]
then $t =  \Omega\left(n^2\right)$.
\end{enumerate}
\end{theorem}

\paragraph{Applications beyond SK}
The radius certificates also supply the structural promise used
by optimization-based scaling algorithms.
Let $\ell=\operatorname{nnz}(A)$. A logarithmic scaling radius gives
running times
$\widetilde O(\ell\log^2(n/\varepsilon))$ via
\cite[Theorem~4.5]{cohen2017matrix} and
$\widetilde O(\ell+n^{4/3})$ via \cite{allen2017much}, with
polylogarithmic dependence on $1/\varepsilon$ suppressed in the
latter bound.
These are consequences for existing algorithms of controlling
the scaling radius.

Our large-radius family gives a complementary application to
iteration lower bounds.
\Cref{prop:bounded-log-iterations} applies to sequences that
start from $M^{(0)}=A$, remain in the positive diagonal scaling
family $M^{(t)}=X_tAY_t$, and change the logarithm of each entry
by at most an absolute constant per update.
On the family in \Cref{prop:product-threshold-counterexample},
reaching total marginal $\ell_1$ error $n^{-2}$ requires
$\Omega(n)$ such updates.
For these algorithms, we establish lower bounds on the number of
accepted outer updates, starting from the initialization prescribed
by each algorithm.

\Cref{prop:bounded-log-iterations} states the quantitative bound;
\Cref{cor:three-scaling-algorithms} gives the consequences for the
specified implementations and initializations.
The formal bounded-update proposition is stated with its proof in
\Cref{sec-lowerbounds}.

\subsection{Overview of the Techniques}\label{subsec:technique-overview}

The analysis has a dynamical part and a structural part.
The dynamical part gives a global convergence bound in terms of the
endpoint gap $g=1-s_2(B)^2$ and the logarithmic scaling distance $D$.
A matching family shows why dependence on $D$ is necessary even when
the endpoint mixes rapidly. The structural part then gives spectral
certificates for $D=O(\log n)$ and, for dense inputs satisfying the
required expansion conditions, $D=O(1)$.
This order separates the general convergence theorem from the
conditions used to bound its initial-distance parameter.

\paragraph{A global potential for the SK dynamics.}
Let $B=XAY$ be a finite doubly stochastic scaling of the row-stochastic
input $A$, with $s_2(B)<1$. For $w>0$, define
\[
 \mathcal A(w)=\operatorname{diag}((Bw)^{-1})
 B\operatorname{diag}(w),
 \qquad
 \mathcal F(w)=[B^\top(Bw)^{-1}]^{-1}.
\]
One complete column--row normalization cycle sends
$\mathcal A(w)$ to $\mathcal A(\mathcal F(w))$.
We track both $w$ and its coordinatewise reciprocal through
the scale-invariant potential
\[
 \Phi(w)=\langle w^2\rangle\langle w^{-2}\rangle-1,
 \qquad \langle z\rangle=\frac1n\sum_jz_j.
\]
For the two-step walk $P_B=B^\top B$, arithmetic--harmonic mean
inequalities give the coordinatewise bounds
\[
 \mathcal F(w)\le P_Bw,
 \qquad
 \mathcal F(w)^{-1}\le P_B(w^{-1}).
\]
The endpoint spectrum contracts centered second moments under $P_B$.
The scalar inequality
\[
 \langle w\rangle^2\langle w^{-1}\rangle^2
 \le\frac{\langle w^2\rangle\langle w^{-2}\rangle+1}{2}
\]
couples the remaining mean terms and yields
\[
 \Phi(\mathcal F(w))
 \le\frac{1+s_2(B)^4}{2}\Phi(w)
 \qquad\text{for every }w>0.
\]
This contraction applies from the standard initialization
$w_0=Y^{-1}\mathbf1$, for which $\mathcal A(w_0)=A$.
Choosing logarithmic scaling vectors of norm $D$ gives
$\max_jw_{0,j}/\min_jw_{0,j}\le e^{2D}$ and hence
$\Phi(w_0)\le e^{4D}$.
Converting the potential to total marginal error gives
\[
 O\!\left(\frac{D+\log n+\log(1/\varepsilon)}{g}\right)
 \qquad\text{complete cycles},\qquad g=1-s_2(B)^2,
\]
since $1-s_2(B)^4=g(1+s_2(B)^2)$.
The theorem applies to every row-stochastic nonnegative square input
with a finite doubly stochastic scaling and a positive endpoint gap.
Structural information enters by controlling $D$.

\paragraph{A matching family with constant spectral gaps.}
The family in \Cref{prop:product-threshold-counterexample} separates
rapid mixing at the endpoint from the adjustment needed to reach it.
Positive block matrices reduce both the scaling equations and the
actual SK updates to low-dimensional recurrences.
Unequal block sizes force an exponentially small block mass to become
macroscopic, while a doubly stochastic tensor factor fixes both
relevant second singular values at $1/\sqrt2$.
The largest logarithmic entrywise distortion is $\Theta(n)$;
the corresponding logarithmic scaling distance satisfies $D=\Theta(n)$.
Along the SK trajectory, the small block's odds can grow by at most
a constant factor per cycle, so attaining error $n^{-2}$ requires
$\Omega(n)$ iterations.
The global convergence theorem gives $O(n)$ on this same family.
Thus the dependence on $D$ is necessary when the endpoint gap is
fixed. This family does not establish joint optimality in
$D$, $g$, the dimension, and the accuracy.

\paragraph{Buffered expansion and propagation of scaling factors.}
The central structural tool is two-sided buffered robust vertex
expansion. For an active cutoff $m\geq1$, its forward condition requires
\[
 A(I,[n]\setminus J)\geq\phi|I|
 \qquad\text{whenever}\qquad
 m\leq|I|\leq qn,\quad |J|\leq(1+\eta)|I|,
\]
and the reverse condition exchanges rows
and columns. Here $q>1/2$, $\eta,\phi>0$, and $q(1+\eta)<1$.
The condition controls the mass escaping a buffer
larger than the starting set. This extra room allows sets to grow
even when the thresholds defining them deteriorate at each step.

To see the mechanism, consider a scaling $B=XAY$ and normalize the
diagonal factors so that $\min_i x_i=\min_j y_j=\kappa$.
The seed assumption supplies at least $m$ rows with
$x_i\geq H_x/\kappa$ and at least $m$ columns with
$y_j\geq H_y/\kappa$, for fixed positive $H_x,H_y$.
Suppose a seed set $R$ of at least $m$ rows satisfies $x_i\geq h$.
Buffered expansion and an upper bound on the corresponding
$B$-mass force a larger set of columns on which $y_j=O(1/h)$.
Column normalization then forces these columns to receive enough
mass from a still larger set of rows where $x_i\geq h/L$, for a
constant $L$. Thus a two-step exploration increases the set size
by a constant factor while losing only a constant factor in the
scaling threshold. A reverse exploration starts from a large-$y$
seed and produces small-$x$ rows.

After $O(\log(n/m))$ steps, the two explorations yield large row
and column sets on which both scaling factors are small.
Their sizes and the marginal constraints force $\Omega(n)$
$B$-mass between them, while $B_{ij}=x_iA_{ij}y_j$ bounds this
mass by $\kappa^2n$ times the accumulated threshold losses.
This forces a lower bound on $\kappa$ and subsequently bounds
every product $x_i y_j$.
In our applications, the resulting distortion is polynomial in
$n/m$: singleton seeds give polynomial distortion, whereas seeds
of linear size give constant distortion.
The threshold-propagation mechanism is related to that used for
symmetric graph adjacency matrices in \cite{rudelson2016hafnians}.
Our structural theorem, \Cref{thm:seeded-multiscale}, treats weighted,
nonsymmetric matrices through two-sided mass-escape inequalities,
with an explicit active cardinality cutoff $m$ and a bound that
tracks the seed size.
In particular, macroscopic seeds give constant distortion, while
singleton seeds give polynomial distortion.
The theorem only requires exact column normalization and suitable
bounds on row masses. This permits its application to the
reversible auxiliary matrix built from $B^\top A$ below, whose
scaling controls ratios of the original diagonal factors without
requiring the original input to be nearly balanced.

\paragraph{Verifying expansion from spectral information.}
We obtain the required escape bounds by applying a centered
singular-value estimate to weighted indicator vectors.
Centering retains the variance factors associated with both
sets, allowing us to handle nonuniform marginals.
For a row-stochastic matrix with column-sum ratio at most $K$,
we combine this estimate with bounds on the mass of a column
subset that use the actual ratio $K$.
The resulting expansion criterion has threshold
\[
 \theta_K(q,\alpha)
 =\sqrt{\frac{1-\alpha q}{K\alpha(1-q)}},
 \qquad \alpha=1+\eta,
\]
which approaches $1/\sqrt K$ as $\alpha\downarrow1$.
Thus a strict margin below $1/\sqrt K$ supplies buffered
expansion with quantitative parameters, without requiring the
column sums to approach equality.
The weighted version of the argument also applies before row
normalization, which will be useful for dense inputs.

\paragraph{Combining the input and endpoint spectra.}
The mixed spectral condition is verified through an auxiliary
matrix. Suppose $B=XAY$ is doubly stochastic. Write
$Q=\operatorname{diag}(c_1(A),\ldots,c_n(A))$,
$M=AQ^{-1/2}$, $\sigma=s_2(M)$, and
$\sigma_\perp=s_2(B)$.
Set $P=B^\top A$. This matrix is row-stochastic and has exactly
the same column sums as $A$.
Writing $e=n^{-1/2}\mathbf1$ and
$f=n^{-1/2}Q^{1/2}\mathbf1$, the trivial singular directions align:
\[
 PQ^{-1/2}-ef^\top
 =(B^\top-ee^\top)(M-ef^\top).
\]
Consequently,
$s_2(PQ^{-1/2})\leq\sigma\sigma_\perp$.
This explains the product of the two nontrivial singular values
in our criterion: $\sqrt K\,\sigma\sigma_\perp<1$ certifies
buffered expansion of $P$.

The scaling relation supplies the additional identity
\[
 Y^{-1}P=A^\top XA,
 \qquad Y^{-1}PY=P^\top.
\]
In particular, $P$ is reversible with weights proportional to
$1/y_j$. We exploit this identity by lazifying the auxiliary
matrix: $\widehat A=(1-\tau)I+\tau P$, for a suitable constant
$\tau>0$. This produces nearly uniform auxiliary margins while
retaining quantitative expansion, and
$Y^{-1}\widehat A Y=\widehat A^\top$.
The reciprocal diagonal factors $Y^{-1}$ and $Y$ provide
singleton seeds after normalization.
The multiscale structural theorem therefore bounds every ratio
$y_j/y_k$ by a polynomial.
These bounds return to the original scaling through the identity
\[
 x_i y_j=\sum_k B_{ik}\frac{y_j}{y_k},
\]
whose right-hand side is a convex combination.
The near-uniform margins needed in this reduction are created
on the auxiliary matrix; the original input can retain constant
marginal imbalance.

For the input-only condition $\sqrt K\,\sigma<1$, the spectral
verifier applies directly to $A$.
Two-sided buffered expansion implies strict Hall inequalities
and hence the existence of an exact doubly stochastic scaling.
We can then apply the mixed criterion using $s_2(B)\leq1$.
Thus the input-only guarantee does not presuppose scalability.

\paragraph{Dense inputs: macroscopic seeds and spectral transfer.}
For dense inputs satisfying our expansion criteria, the structural
argument starts from much larger seeds.
Writing $d=\gamma\rho$ and row-normalizing $D$ to obtain $A$,
density gives $A_{ij}\leq1/(dn)$ and column sums in
$[d,d^{-1}]$.
Together with the scaling marginals, these bounds produce
linearly many seeds for both diagonal factors.
Only constantly many propagation steps are then needed, giving
$T=O(1)$.

The expansion hypothesis follows directly by counting large
entries when $\gamma>1/2$; at arbitrary fixed positive density,
it follows when $s_2(E)$ lies below $\gamma\rho$ by a fixed
positive margin, where $E$ is the normalization of $D$ by both
marginals.
To complete the convergence argument, we compare Dirichlet
forms of the associated two-step walks. The comparison gives
\[
 1-s_2(B)^2
 \geq\frac{(\gamma\rho)^5}{T^2}
 \bigl(1-s_2(E)^2\bigr).
\]
The input gap is constant in each case, through the spectral
promise or overlap of large-entry neighborhoods.
Constant distortion transfers this gap to the endpoint, and the
global potential argument yields logarithmic SK convergence.

\paragraph{Further boundary examples and bounded updates.}
The other boundary examples exhibit different dynamics.
At the input-only threshold, an invariant cross-ratio and a reciprocal
drift estimate keep the marginal error above $\Omega_K(n/(t+1))$
over the relevant time range, yielding the cubic lower bound at
accuracy $n^{-2}$.
For the critical dense family, a fractional-linear change of variable
gives an exact scalar contraction with factor $1-\Theta(n^{-2})$,
yielding the quadratic lower bound.
These arguments follow the actual normalization trajectory,
including its initialization.

To obtain the lower bounds for specified bounded-update methods,
we prove a global error bound within a positive diagonal scaling family.
At the optimum, the Hessian of the exponential scaling objective
is the weighted bipartite Laplacian associated with the exact scaling.
Comparing the Hessian along rays with this Laplacian and integrating
the resulting gradient bound shows that sufficiently small marginal
error forces small logarithmic entrywise distance from the exact scaling.
The constant-gap family begins at logarithmic distance $\Theta(n)$.
Consequently, sequences that remain in this diagonal scaling family
and change every entry by at most a constant multiplicative factor
per accepted update need $\Omega(n)$ such updates.
\Cref{prop:bounded-log-iterations,cor:three-scaling-algorithms}
state the assumptions and the specific implementations to which
this argument applies.

ChatGPT assisted with counterexample construction and with developing
the spectral-to-expansion and stability-to-convergence arguments;
the author independently verified their correctness and originality,
as detailed in the AI methodology statement.

\subsection{Organization of the Paper}

\Cref{sec-preliminary} presents the preliminaries.
\Cref{sec-structural stability-convergence} proves the global spectral
convergence theorem for SK.
\Cref{sec-brve} develops the buffered expansion framework and its
structural stability guarantees, together with spectral criteria
for verifying the required expansion.
\Cref{sec-spectralconditions} derives the input-only and mixed
spectral certificates for general matrices.
\Cref{sec-lowerbounds} proves the matching constant-gap family,
the boundary constructions, and the lower bounds for specified
bounded-update implementations.
\Cref{sec-upperbound} applies the framework to dense matrices.
\Cref{sec-bve-stricthall,sec-spectralgapcomparison} provide the proofs
of the strict Hall implication and the spectral-gap comparison.

\newpage

%% file: Preliminary.tex
\section{Preliminary}\label{sec-preliminary}

Throughout this paper, 
we use $\mathbb{Z}_{>0}$ to denote the set of strictly positive integers. Let $\mathbb{R}_{>0}$ and $\mathbb{R}_{\ge 0}$ denote the sets of strictly positive and nonnegative real numbers, respectively. 
For any integers $m, n \in \mathbb{Z}_{>0}$, we use $\mathbb{R}_{>0}^m$ to represent the set of $m$-dimensional vectors with strictly positive entries. Similarly, $\mathbb{R}_{\ge 0}^{m \times n}$ denotes the set of $m \times n$ matrices with nonnegative entries.

All logarithms are natural.
We write $\one$ for the all-ones vector of the appropriate dimension.
Vector powers and reciprocals are understood componentwise.
Vectors are denoted in boldface and are assumed to be column vectors unless otherwise specified. 
For any vector $\={v} \in \mathbb{R}^n$, let $v_i$ denote its $i$-th component. 
We write 
$\langle \={v}\rangle \triangleq \frac{1}{n}\sum_{j=1}^n v_j$.

For any matrix $A \in \mathbb{R}^{n \times n}$ for which the eigenvalues are real
, let $\lambda_1, \dots, \lambda_n \in \mathbb{R}$ be the eigenvalues of $A$. We sort them such that they are in non-increasing order. 
We use $\lambda_2(A)$ to denote the second largest eigenvalue of $A$.

For a nonnegative matrix $A\in\R_{\geq0}^{n\times n}$ and sets
$I,J\subseteq[n]$, write
\[
A(I,J)\triangleq\sum_{i\in I,\,j\in J}A_{ij},
 \qquad
 N_A(I)\triangleq\{j\in[n]:A_{ij}>0\text{ for some }i\in I\}.
\]
We identify a nonnegative matrix $A$ with its weighted bipartite
graph on the row and column vertex sets, where row vertex $i$
and column vertex $j$ are joined by an edge of weight $A_{ij}$
whenever $A_{ij}>0$.

Throughout this paper, unless explicitly specified otherwise,
we normalize the positive diagonal factors
$X=\diag(x_1,\ldots,x_n)$ and $Y=\diag(y_1,\ldots,y_n)$
in a scaling $B=XAY$ so that $\min_i x_i=\min_j y_j$.
This entails no loss of generality: replacing $(X,Y)$ by
$(tX,t^{-1}Y)$ with $t=\sqrt{\min_j y_j/\min_i x_i}$
leaves $B$ unchanged and makes the two minima equal.

For simplicity, we use $\ell_1$ error in this paper. 
When we say a matrix $A \in \mathbb{R}_{\ge 0}^{n \times n}$ achieve the $\varepsilon$ error, we mean
\begin{align}
\sum_{i\in [n]}\abs{r_i(A) - 1} + \sum_{j\in [n]}\abs{c_j(A) - 1} \leq \varepsilon.
\end{align}
Our result can be extended to $\ell_2$ error easily.

\begin{definition}[Optimal logarithmic scaling radius]
\label{def:optimal-logscaling-radius}
Let $A\in\mathbb R_{\ge0}^{n\times n}$ be row-stochastic, and let
$B$ be a doubly stochastic matrix that is a positive diagonal scaling
of $A$. Define
\begin{equation}\label{eq:def-optimal-logscaling-radius}
 D(A;B)\triangleq
 \min_{\substack{u,v\in\mathbb R^n\\
 B=\operatorname{diag}(e^u)A\operatorname{diag}(e^{-v})}}
 \max\{\|u\|_\infty,\|v\|_\infty\}.
\end{equation}
Here exponentials are taken componentwise. The minimum is over all
logarithmic representations; no normalization of the diagonal factors
is imposed in this minimization.
\end{definition}

The minimum is attained. Indeed, $A$ and $B$ have the same support,
and the feasible set is the nonempty closed affine set specified by
\[
 u_i-v_j=\log\frac{B_{ij}}{A_{ij}}
 \qquad\text{whenever }A_{ij}>0.
\]
Intersecting this set with a closed norm ball containing one feasible
pair gives a nonempty compact set on which the norm attains its minimum.
The definition allows all feasible pairs, including the gauge change
$(u,v)\mapsto(u+a\mathbf1,v+a\mathbf1)$, solely for evaluating
$D(A;B)$. Diagonal factors used subsequently may be returned to the
usual convention $\min_i x_i=\min_j y_j$ by reciprocal rescaling,
which leaves all products $x_i y_j$ unchanged.

If $B=XAY$ and $T^{-1}\le x_i y_j\le T$ for all $i,j$, then
\begin{equation}\label{eq:radius-from-product-bounds}
 D(A;B)\le\log T.
\end{equation}
Indeed, put $u_i=\log x_i$ and $v_j=-\log y_j$.
The cross differences satisfy $|u_i-v_j|\le\log T$, so the
diameter of all coordinates of $(u,v)$ is at most $2\log T$.
A common shift centering these coordinates therefore gives a feasible
pair of norm at most $\log T$.
The following lemma is fundamental tool in spectral graph theory, providing the classical variational characterization of the spectral gap for symmetrizable stochastic matrices, which serves as the cornerstone for our subsequent analysis.

\begin{lemma}\label{lem-variational_characterization}
Let $P \in \mathbb{R}_{\geq 0}^{n \times n}$ be a row-stochastic matrix, i.e., $r_i(P) = 1$ for each $i\in [n]$.
Assume that there exists a strictly positive probability vector $\pi \in \mathbb{R}^n_{>0}$ with $\| \pi \|_1 = 1$ such that $P$ is reversible with respect to $\pi$, i.e.,
$ \pi_i P_{ij} = \pi_j P_{ji}$ for all $i, j \in [n]$.
For any vector $x \in \mathbb{R}^n$, define
\begin{equation}\label{eq-def-epxx-varpix}
\begin{aligned}
    \mathcal{E}_P(x,x) &\triangleq \frac{1}{2} \sum_{i,j=1}^n \pi_i P_{i,j} (x_i - x_j)^2, \\
    \mathrm{Var}_\pi(x) &\triangleq \frac{1}{2} \sum_{i,j=1}^n \pi_i \pi_j (x_i - x_j)^2 = \min_{c \in \mathbb{R}} \sum_{i=1}^n \pi_i (x_i - c)^2.
\end{aligned}
\end{equation}
Then
\begin{equation*}
    1 - \lambda_2(P) = \inf_{\substack{x \in \mathbb{R}^n \\ \mathrm{Var}_\pi(x) > 0}} \frac{\mathcal{E}_P(x,x)}{\mathrm{Var}_\pi(x)} = \min_{\substack{x \neq 0 \\ \langle x, \mathbf{1} \rangle_\pi = 0}} \frac{\mathcal{E}_P(x,x)}{\|x\|_\pi^2},
\end{equation*}
where $\langle x, \mathbf{1} \rangle_\pi = \sum_{i=1}^n \pi_i x_i = 0$ denotes that $x$ is orthogonal to the all-ones vector $\mathbf{1}$ under the $\pi$-weighted inner product, and $\|x\|_\pi^2 = \sum_{i=1}^n \pi_i x_i^2$.
\end{lemma}

The following definition is the key concept used in our proof.

\begin{definition}[Two-sided buffered robust vertex expansion]
\label{def:buffered-expansion}
Let $n\ge2$, $q\in(1/2,1)$, and $\eta,\phi>0$. 
Assume
$q(1+\eta)<1$.  Let $A\in\R_{\geq0}^{n\times n}$ be row-stochastic.  We
say that $A$ satisfies $(q,\eta,\phi)$-two-sided buffered robust vertex
expansion if the following two conditions hold.

For every nonempty $I\subseteq[n]$ with $|I|\le qn$ and every
$J\subseteq[n]$ with $|J|\le(1+\eta)|I|$,
\begin{equation}
 A(I,[n]\setminus J)\ge\phi|I|. \label{eq:buffered-forward}
\end{equation}
For every nonempty $J\subseteq[n]$ with $|J|\le qn$ and every
$I\subseteq[n]$ with $|I|\le(1+\eta)|J|$,
\begin{equation}
 A([n]\setminus I,J)\ge\phi|J|. \label{eq:buffered-reverse}
\end{equation}
\end{definition}

Here, \eqref{eq:buffered-forward} requires every row set $I$ with
$0<|I|\le qn$ to send at least $\phi|I|$ mass outside every
column buffer containing at most $(1+\eta)|I|$ columns.
Since $A$ is row-stochastic, $A(I,[n])=|I|$, so this is
at least a fraction $\phi$ of the total mass of $I$.
Similarly, \eqref{eq:buffered-reverse} requires every column set $J$ with
$0<|J|\le qn$ to receive at least $\phi|J|$ mass from outside
every row buffer containing at most $(1+\eta)|J|$ rows.
Here, the lower bound is normalized by the number of columns
$|J|$.

These conditions strengthen support expansion by requiring
a quantitative lower bound on the mass escaping each buffer.
In particular, the first inequality implies
\[
|N_A(I)|>(1+\eta)|I|
\qquad\text{whenever }0<|I|\le qn.
\]
Indeed, otherwise the admissible choice $J=N_A(I)$ would give
$A(I,[n]\setminus J)=0$, contradicting
\eqref{eq:buffered-forward} because $\phi|I|>0$.
The second inequality yields the analogous expansion property
for column neighborhoods.

\newpage

%% file: Global_convergence.tex
\section{Global convergence of Sinkhorn--Knopp}
\label{sec-structural stability-convergence}

This section proves \Cref{thm:global-sk-logradius-gap}.
We first establish a formulation using bounds on all diagonal
scaling products. This part applies independently of the expansion
conditions developed in the next section.
At the end of the section, we obtain the main theorem by choosing
$T=e^{2D(A;B)}$.

\begin{theorem}
\label{thm:endpoint-gap-global-sinkhorn}
Let $n\ge2$ and let $A\in\mathbb R_{\ge0}^{n\times n}$ be row-stochastic.
Suppose that $S=XAY$ is doubly stochastic, where
$X=\operatorname{diag}(x_1,\ldots,x_n)$ and
$Y=\operatorname{diag}(y_1,\ldots,y_n)$ are positive diagonal matrices.
Assume that $s_2(S)<1$, $T\ge1$ and
$T^{-1}\le x_i y_j\le T$ for each $i,j\in[n]$.
Starting from $A^{(0)}=A$, alternately normalize columns and rows,
so that $A^{(2t)}$ is the matrix after $t$ full column--row cycles.
For every $0<\varepsilon\le1$ and every integer $t$ satisfying
\begin{equation}\label{eq:simple-endpoint-rounds}
 t\ge\left\lceil\frac{2}{1-s_2(S)^4}
 \left(4\log T+\log\frac{16n^3}{\varepsilon^2}\right)\right\rceil,
\end{equation}
the matrix $A^{(2t)}$ is row-stochastic and satisfies
\begin{equation}\label{eq-maxcolumnerroratwotlevarepsilonovern}
 \max_{j\in[n]}
 \left|\sum_iA^{(2t)}_{ij}-1\right|
 \le\frac{\varepsilon}{n}.
\end{equation}
In particular, the required number of full column--row cycles is
\[
 O\!\left(\frac{\log n+\log T+\log(1/\varepsilon)}
 {1-s_2(S)^4}\right).
\]
\end{theorem}

For $w>0$, define
\begin{align}\label{eq-def-mwfw}
 M(w)\triangleq\operatorname{diag}((Sw)^{-1})S\operatorname{diag}(w),
 \qquad
 \mathcal F(w)\triangleq[S^\top(Sw)^{-1}]^{-1}.
\end{align}
By $w>0$ and that $S$ is nonnegative and doubly stochastic,
one can verify $(Sw)_i>0$ for each $i\in [n]$.
Hence, $ M(w)$ and $\mathcal F(w)$ are well-defined.
Furthermore, one can also prove $(\mathcal F(w))_i >0$ for each $i\in [n]$.

Define the scale-invariant potential
\begin{equation}\label{eq:simple-moment-potential}
 \Phi(w)\triangleq
 \langle w^2\rangle\langle w^{-2}\rangle-1.
\end{equation}

\begin{lemma}
\label{lem:simple-moment-contraction}
Let $S$ be nonnegative and doubly stochastic, with $s_2(S)<1$. Then, for every $w>0$,
\begin{equation}\label{eq:simple-global-contraction}
 \Phi(\mathcal F(w))
 \le\left(1-\frac{1-(s_2(S))^4}{2}\right)\Phi(w).
\end{equation}
\end{lemma}

\begin{proof}
Set $P=S^\top S$ and $\theta=(s_2(S))^4$.
Since $S$ is doubly stochastic,
\[
P\mathbf1=S^\top S\mathbf1=S^\top\mathbf1=\mathbf1.
\]
Moreover, $P$ is symmetric positive semidefinite, with eigenvalues
\[
 1=\left(s_1(S)\right)^2>\left(s_2(S)\right)^2
 \ge\cdots\ge\left(s_n(S)\right)^2\ge0.
\]
Thus the spectral theorem implies that $\mathbf1^\perp$ is
$P$-invariant and that
\[
\left\|P\big|_{\mathbf1^\perp}\right\|_2=s_2(S)^2.
\]
Consequently,
\begin{align}\label{eq-pznorm-contraction}
\|Pz\|_2^2\le\theta\|z\|_2^2
\qquad\text{whenever }z\perp\mathbf1.
\end{align}
For an arbitrary $z\in\mathbb R^n$, write
\[
\mu=\langle z\rangle,\qquad z_0=z-\mu\mathbf1,
\]
Hence, $z_0\perp\mathbf1$.
Since $P$ is doubly stochastic,
\[
Pz=\mu\mathbf1+Pz_0,
\qquad Pz_0\perp\mathbf1.
\]
Hence, \eqref{eq-pznorm-contraction} gives
\begin{equation}\label{eq:simple-centered-moment}
\begin{aligned}
 \langle(Pz)^2\rangle
 &=\mu^2+\frac1n\|Pz_0\|_2^2\\
 &\le\mu^2+\theta\bigl(\langle z^2\rangle-\mu^2\bigr)\\
 &=\theta\langle z^2\rangle+(1-\theta)\langle z\rangle^2.
\end{aligned}
\end{equation}

Recall that
\[
P=S^\top S,
\qquad
\mathcal F(w)=\bigl[S^\top(Sw)^{-1}\bigr]^{-1},
\]
where all reciprocals are taken componentwise.
Fix $w\in\mathbb R_{>0}^n$ and write $h=Sw$.
For every row $i$, row-stochasticity gives an index $k$ with
$S_{ik}>0$. Since $w_k>0$,
\[
 h_i=\sum_k S_{ik}w_k>0
 \qquad\forall i\in[n].
\]
Similarly, we also have
$\sum_i S_{ij}/h_i>0$ for every $j$.
Thus, all the
reciprocals below are well-defined.

We first prove $\mathcal F(w)\le Pw$.
Fix $j\in[n]$. The entries in the $j$-th column of $S$
are nonnegative and sum to $1$. Hence Cauchy--Schwarz gives
\begin{align*}
1=\left(\sum_i S_{ij}\right)^2=\left(
  \sum_i \sqrt{S_{ij}h_i}
         \sqrt{\frac{S_{ij}}{h_i}}
  \right)^2\le
  \left(\sum_i S_{ij}h_i\right)
  \left(\sum_i\frac{S_{ij}}{h_i}\right).
\end{align*}
Dividing both sides by the positive quantity $\sum_i S_{ij}/h_i$,
we have
\begin{align}\label{eq-sumsijoverhiinverseleqsumsijhi}
\left(\sum_i\frac{S_{ij}}{h_i}\right)^{-1}
\le \sum_i S_{ij}h_i.
\end{align}
Recall $h=Sw$. We have
\[
\mathcal F(w)
=\bigl[S^\top(Sw)^{-1}\bigr]^{-1}
=\bigl[S^\top h^{-1}\bigr]^{-1},
\]
Fix $j\in[n]$. Then
\[
(S^\top h^{-1})_j
=\sum_{i=1}^n (S^\top)_{ji}(h^{-1})_i
=\sum_{i=1}^n \frac{S_{ij}}{h_i}.
\]
Taking its reciprocal therefore gives
\begin{align}\label{eq-fjw-sumsijoverhiinverse}
\mathcal F_j(w)
=\frac1{(S^\top h^{-1})_j}
=\frac1{\displaystyle\sum_{i=1}^n S_{ij}/h_i}
=\left(\sum_{i=1}^n\frac{S_{ij}}{h_i}\right)^{-1}.
\end{align}
By \eqref{eq-sumsijoverhiinverseleqsumsijhi} and the definitions of $h$ and $P$, we have
\[
\mathcal F_j(w)
=\left(\sum_i\frac{S_{ij}}{h_i}\right)^{-1}
\le \sum_i S_{ij}h_i
=(S^\top h)_j
=(S^\top Sw)_j
=(Pw)_j.
\]
Since $j$ was arbitrary, we conclude that
$\mathcal F(w)\le Pw$
componentwise.

We next prove $(\mathcal F(w))^{-1}\le P(w^{-1})$.
For each fixed $i\in[n]$, the entries in the $i$-th row
of $S$ are nonnegative and sum to $1$.
The function $t\mapsto t^{-1}$ is convex on $(0,\infty)$,
so Jensen's inequality gives
\[
\frac1{h_i}
=\left(\sum_k S_{ik}w_k\right)^{-1}
\le \sum_k S_{ik}\frac1{w_k}.
\]
Multiplying this inequality by $S_{ij}\ge0$ and summing over
$i$, we obtain, for every $j\in[n]$,
\begin{align*}
 \sum_i\frac{S_{ij}}{h_i}
 &\le\sum_iS_{ij}\sum_kS_{ik}\frac1{w_k}\\
 &=\sum_k\left(\sum_iS_{ij}S_{ik}\right)\frac1{w_k}
 =\bigl[P(w^{-1})\bigr]_j.
\end{align*}
Combined with \eqref{eq-fjw-sumsijoverhiinverse},
we have
\begin{align*}
\frac1{\mathcal F_j(w)}=\sum_i \frac{S_{ij}}{h_i}\leq \bigl[P(w^{-1})\bigr]_j.
\end{align*}
Therefore,
$(\mathcal F(w))^{-1}
\le P(w^{-1})$ componentwise.

Write
\begin{equation}\label{eq-def-abpw}
 a=\langle w\rangle,\qquad b=\langle w^{-1}\rangle,
 \qquad p=\langle w^2\rangle,\qquad q=\langle w^{-2}\rangle.
\end{equation}
Hence, we have
\[
ab=\left(\frac1n\sum_iw_i\right)\left(\frac1n\sum_i\frac1{w_i}\right)=\frac1{n^2}\sum_{i,j}\frac{w_i}{w_j}.
\]
Thus,
\[\frac1{n^2}\sum_{i,j}\frac12
\left(\frac{w_i}{w_j}+\frac{w_j}{w_i}\right)
=\frac12(ab+ab)=ab.\]
Therefore,
\begin{align*}
 a^2b^2
 &=\left[
 \frac1{n^2}\sum_{i,j}\frac12
 \left(\frac{w_i}{w_j}+\frac{w_j}{w_i}\right)
 \right]^2.
\end{align*}
Since $t\mapsto t^2$ is convex, Jensen's inequality with
uniform weights $1/n^2$ gives
\[
\left(\frac1{n^2}\sum_{i,j}\left(\frac{w_i}{w_j}+\frac{w_j}{w_i}\right)\right)^2
\le \frac1{n^2}\sum_{i,j}\left(\frac{w_i}{w_j}+\frac{w_j}{w_i}\right)^2.
\]
Hence,
\begin{align}\label{eq-asquarebsquare-upperbound}
 a^2b^2
 &\le\frac1{n^2}\sum_{i,j}\frac14
 \left(\frac{w_i}{w_j}+\frac{w_j}{w_i}\right)^2.
\end{align}
By \eqref{eq-def-abpw},
\[
p=\frac1n\sum_{i=1}^n w_i^2,
\qquad
q=\frac1n\sum_{i=1}^n w_i^{-2}.
\]
We have
\[
\frac1{n^2}\sum_{i,j}\frac{w_j^2}{w_i^2}
=\left(\frac1n\sum_j w_j^2\right)
 \left(\frac1n\sum_i w_i^{-2}\right)
=pq.
\]
Consequently,
\begin{align*}
 &\frac1{4n^2}\sum_{i,j}
 \left(\frac{w_i}{w_j}+\frac{w_j}{w_i}\right)^2\\
 &\qquad=\frac1{4n^2}\sum_{i,j}
 \left(\frac{w_i^2}{w_j^2}+\frac{w_j^2}{w_i^2}+2\right)
 =\frac{pq+1}{2}.
\end{align*}
Combined with \eqref{eq-asquarebsquare-upperbound}, we have
\begin{align}\label{eq:simple-moment-coupling}
 a^2b^2
 &\le\frac1{n^2}\sum_{i,j}\frac14
 \left(\frac{w_i}{w_j}+\frac{w_j}{w_i}\right)^2
 =\frac{pq+1}{2}.
\end{align}

Recall 
$0<\mathcal F(w)\le Pw$.
Applying \eqref{eq:simple-centered-moment} with $z=w$,
we therefore obtain
\begin{align*}
\langle(\mathcal F(w))^2\rangle\le\langle(Pw)^2\rangle\le\theta\langle w^2\rangle
 +(1-\theta)\langle w\rangle^2=\theta p+(1-\theta)a^2.
\end{align*}
Similarly, we also have
\begin{align}
\langle(\mathcal F(w))^{-2}\rangle\le\left\langle\bigl(P(w^{-1})\bigr)^2\right\rangle\le\theta\langle w^{-2}\rangle
 +(1-\theta)\langle w^{-1}\rangle^2=\theta q+(1-\theta)b^2.
\end{align}
Set
\[
 U\triangleq\langle(\mathcal F(w))^2\rangle
                 \langle(\mathcal F(w))^{-2}\rangle.
\]
The preceding bounds give
\[
 U\le(\theta p+(1-\theta)a^2)
       (\theta q+(1-\theta)b^2).
\]
Expanding this product yields
\begin{align*}
 \theta pq+(1-\theta)a^2b^2-U
 &\ge\theta(1-\theta)
 \bigl(pq+a^2b^2-pb^2-a^2q\bigr)\\
 &=\theta(1-\theta)(p-a^2)(q-b^2).
\end{align*}
Moreover, by \eqref{eq-def-abpw} and Jensen's inequality,
we have
\[
p =\frac1n\sum_iw_i^2 \geq \left(\frac1n\sum_iw_i\right)^2 = a^2
\]
Similarly, $q\ge b^2$.
Together with $0\le\theta<1$, we obtain
\[
\theta(1-\theta)(p-a^2)(q-b^2)\ge0.
\]
Therefore,
\[
\langle(\mathcal F(w))^2\rangle\langle(\mathcal F(w))^{-2}\rangle
\le \theta pq+(1-\theta)a^2b^2.
\]
Combining this inequality with
\eqref{eq:simple-moment-potential} and
\eqref{eq:simple-moment-coupling}, we obtain
\begin{align*}
 \Phi(\mathcal F(w))+1
 &\le\theta pq+(1-\theta)a^2b^2\\
 &\le\theta pq+\frac{(1-\theta)(pq+1)}2\\
 &=\frac{1+\theta}{2}pq+\frac{1-\theta}{2}.
\end{align*}
Moreover, by \eqref{eq:simple-moment-potential} and \eqref{eq-def-abpw},
\begin{align*}
\Phi(w)=
 \langle w^2\rangle\langle w^{-2}\rangle-1 = pq - 1.
\end{align*}
Thus, 
\begin{align*}
 \Phi(\mathcal F(w))
 &\le\frac{1+\theta}{2}(pq-1)\\
 &=\frac{1+\theta}{2}\Phi(w).
\end{align*}
The lemma is proved.
\end{proof}

\begin{proof}[Proof of Theorem~\ref{thm:endpoint-gap-global-sinkhorn}]
For every $w>0$, \eqref{eq-def-mwfw} gives
\begin{align}\label{eq-mwij-sijwjswi}
\left(M(w)\right)_{ij}=\frac{S_{ij}w_j}{(Sw)_i}.
\end{align}
Consequently,
\[
\sum_j\left(M(w)\right)_{ij}
=\frac{\sum_jS_{ij}w_j}{(Sw)_i}
=1.
\]
Thus $M(w)$ is row-stochastic.
Moreover, 
by \eqref{eq-def-mwfw} we also have
\begin{align*}
\frac1{\left(\mathcal F(w)\right)_j}
&=\bigl[S^\top(Sw)^{-1}\bigr]_j=\sum_i(S^\top)_{ji}\frac1{(Sw)_i}=\sum_i\frac{S_{ij}}{(Sw)_i}.
\end{align*}
Consequently, the $j$-th column sum of $M(w)$ is
\begin{align*}
\sum_i\bigl(M(w)\bigr)_{ij}
&=\sum_i\frac{S_{ij}w_j}{(Sw)_i}=w_j\sum_i\frac{S_{ij}}{(Sw)_i}=\frac{w_j}{\left(\mathcal F(w)\right)_j}>0.
\end{align*}
The positivity follows from $w_j>0$ and $\left(\mathcal F(w)\right)_j>0$.

Let $c_j(w)$ denote $\sum_i\left(M(w)\right)_{ij}$.
Let $\widetilde M(w)$ be the matrix obtained by normalizing
the columns of $M(w)$. 
Formally, 
\begin{align*}
\left(\widetilde M(w)\right)_{ij}=\frac{\left(M(w)\right)_{ij}}{c_j(w)}=\frac{S_{ij}w_j}{(Sw)_i}
  \frac{\left(\mathcal F(w)\right)_j}{w_j}=\frac{S_{ij}\left(\mathcal F(w)\right)_j}{(Sw)_i}.
\end{align*}
Equivalently,
\[
\widetilde M(w)
=\operatorname{diag}((Sw)^{-1})
 S\operatorname{diag}(\mathcal F(w)).
\]
Hence,
\[
\sum_j\left(\widetilde M(w)\right)_{ij}
=\frac{\sum_jS_{ij}\left(\mathcal F(w)\right)_j}{(Sw)_i}
=\frac{(S\mathcal F(w))_i}{(Sw)_i}>0.
\]
The positivity follows from $(Sw)_i>0$, $\left(\mathcal F(w)\right)_j>0$, $S_{i,j}\geq 0$, and $\sum_j S_{i,j} = 1$.
Let $r_i(w)\triangleq\sum_j\left(\widetilde M(w)\right)_{ij}$.
Thus,
\begin{align*}
\frac{\left(\widetilde M(w)\right)_{ij}}{r_i(w)}
&=\frac{S_{ij}\left(\mathcal F(w)\right)_j}{(Sw)_i}
  \frac{(Sw)_i}{(S\mathcal F(w))_i}=\frac{S_{ij}\left(\mathcal F(w)\right)_j}{(S\mathcal F(w))_i}=M(\mathcal F(w))_{ij}.
\end{align*}
Hence one full column--row normalization cycle sends
$M(w)$ to $M(\mathcal F(w))$.

Now set
\[
w_0=Y^{-1}\mathbf1,
\qquad
w_{t+1}=\mathcal F(w_t)\qquad \forall t\ge 0.
\]
Since $A\mathbf1=\mathbf1$ and $S=XAY$,
\[
 Sw_0=XAYY^{-1}\mathbf1=X\mathbf1.
\]
Hence, by \eqref{eq-def-mwfw},
\begin{align*}
 M(w_0)
 &=\operatorname{diag}((Sw_0)^{-1})
   S\operatorname{diag}(w_0)\\
 &=X^{-1}(XAY)Y^{-1}=A=A^{(0)}.
\end{align*}
Recall that one full column--row normalization cycle sends
$M(w)$ to $M(\mathcal F(w))$.
Starting from $A^{(0)}=M(w_0)$, induction yields
\begin{equation}\label{eq:simple-actual-iterates}
A^{(2t)}=M(w_t),\qquad \forall t\ge 0.
\end{equation}

\vspace{0.5cm}

In the following, we bound $
\left|\sum_{i=1}^n\left(M(w)\right)_{ij}-1\right|$.
Recall $T^{-1}\le x_i y_j\le T$ for each $i,j\in[n]$.
For any fixed $i$, 
we have
\[
\frac{y_j}{y_k}=\frac{x_i y_j}{x_i y_k}\le T^2 ,\qquad \forall\ j,k\in [n].
\]
Hence, $w_0=Y^{-1}\mathbf1$ gives 
\begin{align}\label{eq-ratiomaxminwletsquare}
 \frac{\max_j w_{0,j}}{\min_j w_{0,j}}\le T^2.
\end{align}
Note that
\[
\langle w_0^2\rangle
=\frac1n\sum_j w_{0,j}^2\le (\max_j w_{0,j})^2,
\qquad
\langle w_0^{-2}\rangle
=\frac1n\sum_j w_{0,j}^{-2}\le (\min_j w_{0,j})^{-2}.
\]
Hence, by \eqref{eq:simple-moment-potential} and \eqref{eq-ratiomaxminwletsquare}, 
\begin{align*}
 \Phi(w_0)
 &\le\left(\frac{\max_jw_{0,j}}{\min_jw_{0,j}}\right)^2-1\\
 &\le T^4-1<T^4.
\end{align*}
By Lemma~\ref{lem:simple-moment-contraction},
\begin{align*}
 \Phi(w_t)
 &\le\left(\frac{1+s_2(S)^4}{2}\right)^t\Phi(w_0)\\
 &\le T^4\exp\!\left(-\frac{1-s_2(S)^4}{2}t\right).
\end{align*}
Consequently, \eqref{eq:simple-endpoint-rounds} ensures
\begin{equation}\label{eq:simple-small-potential}
 \Phi(w_t)\le\frac{\varepsilon^2}{16n^3}.
\end{equation}

Set $w=w_t$ and
rescale it so that $\langle w\rangle=1$.
By \eqref{eq-mwij-sijwjswi} and the definition of $\Phi$,
this rescaling leaves both $M(w)$ and $\Phi(w)$ unchanged.
Since $\langle w\rangle=1$,
Jensen's inequality gives 
\[
\langle w^{-2}\rangle
=\frac1n\sum_i w_i^{-2}
\ge\left(\frac1n\sum_i w_i\right)^{-2}
=\langle w\rangle^{-2}
=1.
\]
Hence,
\[
 \langle(w-\mathbf1)^2\rangle = \langle w^2\rangle - 2 \langle w\rangle +1
 =\langle w^2\rangle-1 \leq \langle w^2\rangle\langle w^{-2}\rangle -1
 \le\Phi(w).
\]
Hence, \eqref{eq:simple-small-potential} gives
\[
|w_j-1|^2
\le\sum_{i=1}^n(w_i-1)^2
=n\langle(w-\mathbf1)^2\rangle
\le n\Phi(w)\le\frac{\varepsilon^2}{16n^2},\qquad \forall j\in [n].
\]
Thus,
\[
|w_j-1|
\le\frac{\varepsilon}{4n},
\qquad \forall j\in[n].
\]
We obtain $w_j \in [1-\varepsilon/(4n),1+\varepsilon/(4n)]$.
Thus,
\[
\left(1-\frac{\varepsilon}{4n}\right)\sum_{j=1}^n S_{ij}
\le \sum_{j=1}^n S_{ij}w_j
\le
\left(1+\frac{\varepsilon}{4n}\right)\sum_{j=1}^n S_{ij},\qquad \forall\ i\in [n].
\]
Using $\sum_{j=1}^n S_{ij}=1$ and
$(Sw)_i=\sum_{j=1}^n S_{ij}w_j$, this becomes
\[
1-\frac{\varepsilon}{4n}
\le (Sw)_i
\le 1+\frac{\varepsilon}{4n},
\qquad \forall\ i\in[n].
\]
Thus, for $S_{ij}>0$, \eqref{eq-mwij-sijwjswi} gives
\begin{align}\label{eq-absmwijoversijminusonelevarepsilon}
 \left|\frac{\left(M(w)\right)_{ij}}{S_{ij}}-1\right|
 =\left|\frac{w_j}{(Sw)_i}-1\right|
 \le\frac{\varepsilon/(2n)}{1-\varepsilon/(4n)}
 \le\frac{\varepsilon}{n}.
\end{align}

In addition, by \eqref{eq-mwij-sijwjswi}, we have $M(w)_{ij}=0$ if $S_{ij}=0$.
Combined with
$\sum_{i=1}^n S_{ij}=1$, we have
\begin{align*}
 \sum_iM(w)_{ij}-1
 &=\sum_i\bigl(M(w)_{ij}-S_{ij}\bigr)\\
 &=\sum_{i:S_{ij}>0}
 S_{ij}\left(\frac{M(w)_{ij}}{S_{ij}}-1\right).
\end{align*}
Applying the triangle inequality and using $S_{ij}\ge0$, we obtain
\[
\left|\sum_{i=1}^n\left(M(w)\right)_{ij}-1\right|
\le
\sum_{i:S_{ij}>0}
S_{ij}\left|\frac{M(w)_{ij}}{S_{ij}}-1\right|.
\]
Consequently, by \eqref{eq-absmwijoversijminusonelevarepsilon} and $\sum_{i=1}^n S_{ij}=1$, we obtain
\[
\left|\sum_{i=1}^n\left(M(w)\right)_{ij}-1\right|
\le
\frac{\varepsilon}{n}\sum_{i:S_{ij}>0}S_{ij}
=\frac{\varepsilon}{n}.
\]
Together with \eqref{eq:simple-actual-iterates}, this proves \eqref{eq-maxcolumnerroratwotlevarepsilonovern}.
The theorem is immediate.
\end{proof}

\begin{proof}[Proof of \Cref{thm:global-sk-logradius-gap}]
Choose a minimizing pair $(u,v)$ in the definition of $D(A;B)$.
Temporarily set
\[
 \widetilde X=\operatorname{diag}(e^u),\qquad
 \widetilde Y=\operatorname{diag}(e^{-v}),\qquad T=e^{2D}.
\]
Return these factors to the usual normalization by setting
\[
 a=\sqrt{\frac{\min_j\widetilde y_j}
                 {\min_i\widetilde x_i}},
 \qquad X=a\widetilde X,\quad Y=a^{-1}\widetilde Y.
\]
Then $B=XAY$, $\min_i x_i=\min_j y_j$, and, for every $i,j$,
\[
 e^{-2D}\le x_i y_j
 =\widetilde x_i\widetilde y_j=e^{u_i-v_j}\le e^{2D}.
\]
These product bounds hold for all index pairs, including those outside
the support of $A$. Thus \Cref{thm:endpoint-gap-global-sinkhorn}
applies. Since
\[
 1-s_2(B)^4=g(2-g)\ge g,\qquad 4\log T=8D,
\]
condition \eqref{eq:global-sk-logradius-gap-rounds} is sufficient
for the threshold in that theorem. The resulting columnwise error is
at most $\varepsilon/n$, while all row errors vanish, which gives
the stated total marginal error bound. The bound on the number of
cycles follows because $n\ge2$ and $0<\varepsilon\le1$.
\end{proof}

\newpage

%% file: buffered_expansion.tex
\section{Buffered robust vertex expansion}\label{sec-brve}

\subsection{Structural stability from buffered expansion}

In this subsection, we show that buffered robust vertex expansion implies structural stability.

\begin{theorem}
\label{thm:seeded-multiscale}
Let $n\ge2$, $\kappa>0$, $q\in(1/2,1)$, $\eta,\phi\in(0,1)$, and
$0<c\le1$.  Assume
\begin{equation}
 q(1+\eta)<1,
 \qquad
 c(1+\eta)>1,
 \qquad
 (1+c)(1+\eta)\lfloor qn\rfloor>n. \label{eq:seeded-parameter-assumptions}
\end{equation}
Let $A\in\R_{\geq0}^{n\times n}$ be row-stochastic.  Let $B=XAY$, where $X=\diag(x_1,\ldots,x_n)$ and
$Y=\diag(y_1,\ldots,y_n)$ are positive diagonal matrices with $\min_{i\in[n]}x_i=\min_{j\in[n]}y_j = \kappa$. 
Suppose that, for
some $0<s_{-}\le1\le s_{+}<\infty$ and some $b_+\ge1$,
\begin{align}
 &s_{-}\le A([n],j)\le s_{+}
 \qquad  \qquad \quad \, \ \  \forall j\in[n]. \label{eq:seeded-column-bounds}\\
& B([n],j)=1, B(j,[n])\le b_+
 \qquad \forall j\in[n]. \label{eq:seeded-B-margins}
\end{align}
Let $1\le m\le\lfloor qn\rfloor$ be an integer, and suppose that the
following conditions hold:
\begin{enumerate}[label={\rm (\roman*)}]
\item For every $I\subseteq[n]$ with $|I|\ge m$,
\begin{equation}
 c|I|\le B(I,[n])\le c^{-1}|I|. \label{eq:seeded-row-mass}
\end{equation}

\item The following buffered vertex-expansion bounds hold on the active
scales:
\begin{align}
 A(I,[n]\setminus J)&\ge\phi|I|
 &&\forall I,J\subseteq[n]\text{ with }m\le|I|\le qn
 \text{ and }|J|\le(1+\eta)|I|, \label{eq:seeded-forward-bve}\\
 A([n]\setminus I,J)&\ge\phi|J|
 &&\forall I,J\subseteq[n]\text{ with }m\le|J|\le qn
 \text{ and }|I|\le(1+\eta)|J|. \label{eq:seeded-reverse-bve}
\end{align}

\item There are sets $R_0,C_0\subseteq[n]$ and constants $H_x,H_y>0$ such
that $|R_0|,|C_0|\ge m$ and \label{item-percolation-iii}
\begin{equation}
 x_i\ge \frac{H_x}{\kappa}\quad\forall i\in R_0,
 \qquad
 y_j\ge \frac{H_y}{\kappa}\quad\forall j\in C_0. \label{eq:seeded-seed-assumption}
\end{equation}
\end{enumerate}
Then there are constants $C\ge1$ and $\beta>0$, depending only on
$c,\eta,\phi,s_{-},s_{+},b_+,H_x$, and $H_y$, and a number $T\ge1$ such that
\begin{equation}
 T\le\frac{Cn}{(1+c)(1+\eta)\lfloor qn\rfloor-n}
 \left(\frac{qn}{m}\right)^\beta. \label{eq:seeded-distortion-rate}
\end{equation}
\begin{equation}
 \frac1T\le x_i y_j\le T,
 \qquad
 \frac{A_{ij}}{T}\le B_{ij}\le T A_{ij}
 \qquad \forall i,j\in[n]. \label{eq:seeded-structure-stability}
\end{equation}
\end{theorem}

\vspace{0.3cm}

The following quantities are used in the proof of \Cref{thm:seeded-multiscale}.
Set
\begin{equation}
 \alpha\triangleq1+\eta,
 \qquad
 M\triangleq\lfloor qn\rfloor,
 \qquad
 \omega\triangleq(1+c)\frac{\alpha M}{n}-1. \label{eq:seeded-omega}
\end{equation}
The range of $m$ in the theorem gives $1\le m\le M$.  Assumption
\eqref{eq:seeded-parameter-assumptions} gives
\begin{align}\label{eq-def-qalphaomega}
q\alpha<1,\quad  c\alpha>1, \quad \omega>0.
\end{align}
Define
\begin{align}
 \delta&\triangleq\frac{c\alpha-1}{2\alpha},
 \qquad  \qquad \qquad
 g\triangleq\alpha(c-\delta)=\frac{1+c\alpha}{2}. \label{eq:seeded-delta-g}\\
 K&\triangleq
 \left\lceil\frac{\log(M/m)}{\log g}\right\rceil,\qquad  \qquad
 L_x\triangleq\frac{2s_{+}}{\delta\phi},
 \qquad  \qquad
 L_y\triangleq\frac{2}{\delta\phi}.  \label{eq:seeded-K-loss}
\end{align}
Hence, by \eqref{eq-def-qalphaomega}, \eqref{eq:seeded-delta-g} and $0<c\le1$, we have
$0<\delta<(c/2)\le1/2$, and
$g>1$. 
By $\phi\in(0,1)$, $s_{+} \geq 1$, $0<\delta \le1/2$ and \eqref{eq:seeded-K-loss}, we have
$L_x,L_y>1$.  
In addition, since
$M/m\ge1$ and $g> 1$, the quantity $K$ in \eqref{eq:seeded-K-loss} is therefore a
well-defined nonnegative integer.

\begin{lemma}\label{lem-rtojrplus}
Suppose that the conditions of~\Cref{thm:seeded-multiscale} are satisfied.
 Let
$R\subseteq[n]$ and $h>0$ satisfy $|R|\geq m$ and 
$h\leq x_i$ for each $i\in R$.
Define
\begin{equation}\label{eq-def-j}
\begin{aligned}
 J&\triangleq\left\{j\in[n]:y_j\le\frac{2}{c\phi h}\right\},\\
 R_+&\triangleq\left\{i\in[n]:x_i\ge\frac{h}{L_x}\right\}.
\end{aligned}
\end{equation}
Then we have $|J|>\alpha\min\left\{|R|,M\right\}$ and $|R_+|>g\min\left\{|R|,M\right\}$.
\end{lemma}
\begin{proof}
To prove this lemma,
it suffices to assume 
$|R|\le M$ and prove $|J|>\alpha|R|,|R_+|>g|R|$. Indeed, suppose that $|R|>M$.  Choose a subset
$R'\subseteq R$ with $|R'|=M$.  Since $R'\subseteq R$, we have
$x_i\ge h$ for every $i\in R'$.  
If $|J|>\alpha|R'|=\alpha M$ and $|R_+|>g|R'|=gM$,
we immediately have $|J|>\alpha\min\left\{|R|,M\right\},|R_+|>g\min\left\{|R|,M\right\}$, since $|R|>M$ implies $\min\{|R|,M\}=M$.  
Hence, we assume
$m\le |R|\le M$ for the remainder of the proof.

We first prove $|J|>\alpha|R|$.  Suppose instead that
$|J|\le\alpha|R|$.  Since $m\le|R|\le M\le qn$,
$|J|\le\alpha|R|=(1+\eta)|R|$ and
\eqref{eq:seeded-forward-bve} gives
$A(R,[n]\setminus J)\ge\phi|R|$.  
For each $i\in R$ and
$j\notin J$, we have
$x_i\ge h$ and $y_j>2/(c\phi h)$.  Using these inequalities, the
nonnegativity of $A$, and \eqref{eq:seeded-forward-bve}, we obtain
\begin{align*}
 B(R,[n])
 &\ge B(R,[n]\setminus J)
 =\sum_{i\in R,\,j\notin J}x_iA_{ij}y_j\\
 &>\frac{2}{c\phi}A(R,[n]\setminus J)
 \ge\frac{2}{c}|R|.
\end{align*}
This contradicts the upper bound $B(R,[n])\le c^{-1}|R|$ in
\eqref{eq:seeded-row-mass}.  Thus $|J|>\alpha|R|$.

In the next, we prove $|R_+|>g|R|$.
For each $i\notin R_+$ and $j\in J$,
\eqref{eq:seeded-K-loss} and \eqref{eq-def-j} give
\[
 x_i y_j < \frac{h}{L_x}\cdot \frac{2}{c\phi h} 
 = \frac{\delta\phi h}{2s_{+}}\frac{2}{c\phi h}
 =\frac{\delta}{cs_{+}}.
\]
Because $A([n],j)\le s_{+}$ for every $j$ by
\eqref{eq:seeded-column-bounds}, it follows that
\[
 B([n]\setminus R_+,J) =\sum_{i\not\in R_{+},j\in  J}x_iA_{i,j}y_{j} 
 \le\frac{\delta}{cs_{+}}A([n],J)
 \le\frac{\delta}{c}|J|.
\]
Furthermore, \eqref{eq:seeded-B-margins} gives
$B([n],J)=|J|$.  Therefore
\begin{equation}
 B(R_+,J) = B([n],J) - B([n]\setminus R_+,J)\ge\frac{c-\delta}{c}|J|. \label{eq:seeded-forward-overlap}
\end{equation}

We next verify that $|R_+|\ge m$.  If $|R_+|<m$, choose a set
$Q\subseteq[n]$ of cardinality $m$ containing $R_+$.  Such a set
exists because $m\le\lfloor qn\rfloor\le n$.  Since $|J|>\alpha|R|$ and
$|R|\ge m$, \eqref{eq:seeded-forward-overlap} and \eqref{eq:seeded-delta-g} give
\[
 B(Q,[n])\ge B(R_+,J)
 >\frac{c-\delta}{c}\alpha m
 =\frac{g}{c}m
 >\frac1c m.
\]
This contradicts the upper bound in \eqref{eq:seeded-row-mass} applied to
$Q$.  Hence $|R_+|\ge m$.
By \eqref{eq:seeded-row-mass}, 
$c^{-1}|R_+|
 \ge B(R_+,[n])$.
Combining this with \eqref{eq:seeded-forward-overlap} and
$|J|>\alpha|R|$ yields
\[
 \frac{|R_+|}{c}
 \ge B(R_+,[n])
 \ge B(R_+,J)
 >\frac{c-\delta}{c}\alpha|R|
 =\frac{g}{c}|R|.
\]
Consequently, $|R_+|>g|R|$. 
The lemma is proved.
\end{proof}

\begin{lemma}\label{lem-jtorjplus}
Suppose that the conditions of~\Cref{thm:seeded-multiscale} are satisfied.
Let
$J\subseteq[n]$ and $h>0$ satisfy
$|J|\ge m$ and 
$y_j\ge h$ for each $j\in J$.
Define 
\begin{equation}\label{eq-def-r}
\begin{aligned}
 J_+&\triangleq\left\{j\in[n]:y_j\ge\frac{h}{L_y}\right\},\\
 R&\triangleq\left\{i\in[n]:x_i\le\frac{2}{\phi h}\right\}.
\end{aligned}
\end{equation}
Then we have $|R|>\alpha\min\left\{|J|,M\right\}$ and $|J_+|>g\min\left\{|J|,M\right\}$.
\end{lemma}
\begin{proof}
To prove this lemma,
it suffices to assume 
$|J|\le M$ and prove $|R|>\alpha|J|,|J_+|>g|J|$. Indeed, suppose that $|J|>M$.  Choose a subset
$J'\subseteq J$ with $|J'|=M$.  Since $J'\subseteq J$, we have
$y_j\ge h$ for every $j\in J'$.  
If $|R|>\alpha|J'|=\alpha M$ and $|J_+|>g|J'|=gM$,
we immediately have $|R|>\alpha\min\left\{|J|,M\right\},|J_+|>g\min\left\{|J|,M\right\}$, since $|J|>M$ implies $\min\{|J|,M\}=M$.  
Hence, we assume
$m\le |J|\le M$ for the remainder of the proof.

At first, we prove $|R|>\alpha|J|$.  Suppose instead that
$|R|\le\alpha|J|$.  Since $m\le|J|\le M\le qn$,
$|R|\le\alpha|J|=(1+\eta)|J|$ and
\eqref{eq:seeded-reverse-bve} gives
$A([n]\setminus R,J)\ge\phi|J|$.
For each $i\notin R$ and $j\in J$, we have
$y_j\ge h$ and $x_i>2/(\phi h)$. Using these inequalities, the
nonnegativity of $A$, and \eqref{eq:seeded-reverse-bve}, we obtain
\begin{align*}
 B([n],J)
 &\ge B([n]\setminus R,J)
 =\sum_{i\notin R,\,j\in J}x_iA_{ij}y_j\\
 &>\frac{2}{\phi}A([n]\setminus R,J)
 \ge2|J|.
\end{align*}
Combined with $0<c\leq 1$,
we have $
B([n],J) \ge 2|J|$.
This contradicts $B([n],J)=|J|$, which follows from
\eqref{eq:seeded-B-margins}.  Thus
$|R|>\alpha|J|$.

In the next, we prove $|J_+|>g|J|$.
For each $i\in R$ and $j\notin J_+$,
\eqref{eq:seeded-K-loss} and \eqref{eq-def-r} give \begin{align}
x_i y_j< \frac{2}{\phi h}\cdot\frac{h}{L_{y}} = \frac{2}{\phi h}\frac{h\delta\phi}{2} =  
\delta.
\end{align}
Since $A$ is row-stochastic,
\begin{align*}
 B(R,[n]\setminus J_+)
 &=\sum_{i\in R,\,j\notin J_+}x_iA_{ij}y_j\\
 &\le\delta A(R,[n]\setminus J_+)
 \le\delta|R|.
\end{align*}
Moreover, by 
$|R|>\alpha|J|$, $\abs{J}\ge m$ and $\alpha = 1+\eta >1$, we have
$|R|>m$.
Then \eqref{eq:seeded-row-mass} implies $B(R,[n])\geq c|R|$.
It follows that
\[
 B(R,J_+)
 =B(R,[n])-B(R,[n]\setminus J_+)
 \ge(c-\delta)|R|.
\]
By \eqref{eq:seeded-B-margins}, $|R|>\alpha|J|$, and the
definition of $g$ in \eqref{eq:seeded-delta-g}, we obtain
\[
 |J_+|=B([n],J_+)
 \ge B(R,J_+) \geq (c-\delta)|R|
 >(c-\delta)\alpha|J|
 =g|J|.
\]
The lemma is proved.
\end{proof}

We now prove \Cref{thm:seeded-multiscale}.

\begin{proof}
We first prove a lower bound for $\kappa$.
Set $h_x\triangleq H_x/\kappa$ and define
\[
 \mathcal R_t\triangleq
 \left\{i\in[n]:x_i\ge\frac{h_x}{L_x^t}\right\}
 \qquad(t\ge0).
\]
The seed assumption gives $R_0\subseteq\mathcal R_0$, so
$|\mathcal R_0|\ge m$.
By \Cref{lem-rtojrplus} and induction,
\[
 |\mathcal R_t|>g\min\{g^{t-1}m,M\}
 \qquad(t\ge1).
\]
Since $g^Km\ge M$, we obtain $|\mathcal R_K|\ge M$ when
$K\ge1$. If $K=0$, then $m=M$ and the same conclusion follows
from $|\mathcal R_0|\ge m$.
Applying \Cref{lem-rtojrplus} to $\mathcal R_K$ with threshold
$h_x/L_x^K$ therefore gives a set $J\subseteq[n]$ such that
\begin{equation}
 |J|>\alpha M,
 \qquad
 y_j\le\frac{2L_x^K}{c\phi H_x}\kappa
 \quad\forall j\in J. \label{eq:seeded-small-columns}
\end{equation}
By column-stochasticity of $B$,
\begin{equation}\label{eq-bnnsetminusjleqnminusalpham}
\begin{aligned}
 B([n],[n]\setminus J)=n-|J|<n-\alpha M.
\end{aligned}
\end{equation}
For the reverse exploration, set $h_y\triangleq H_y/\kappa$ and
\[
 \mathcal C_t\triangleq
 \left\{j\in[n]:y_j\ge\frac{h_y}{L_y^t}\right\}
 \qquad(t\ge0).
\]
The seed assumption gives $C_0\subseteq\mathcal C_0$, and hence
$|\mathcal C_0|\ge m$.
By \Cref{lem-jtorjplus} and induction,
\[
 |\mathcal C_t|>g\min\{g^{t-1}m,M\}
 \qquad(t\ge1).
\]
As above, $g^Km\ge M$ implies $|\mathcal C_K|\ge M$ for
$K\ge1$, and for $K=0$ this follows from $m=M$.
Applying \Cref{lem-jtorjplus} to $\mathcal C_K$ with threshold
$h_y/L_y^K$ therefore gives a set $R\subseteq[n]$ such that
\begin{equation}
 |R|>\alpha M,
 \qquad
 x_i\le\frac{2L_y^K}{\phi H_y}\kappa
 \quad\forall i\in R. \label{eq:seeded-small-rows}
\end{equation}
Combined with   
\eqref{eq:seeded-row-mass}, we obtain
$B(R,[n])>c\alpha M$.  
Therefore, \eqref{eq-bnnsetminusjleqnminusalpham} and \eqref{eq:seeded-omega} give
\begin{align*}
 B(R,J)
 &=B(R,[n])-B(R,[n]\setminus J)\\
 &\ge B(R,[n])-B([n],[n]\setminus J)\\
 &>c\alpha M-(n-\alpha M)=\omega n.
\end{align*}

Define
\begin{equation}
 P\triangleq
 \frac{4L_x^KL_y^K}{c\phi^2H_xH_y}. \label{eq:seeded-P}
\end{equation}
For every $i\in R$ and $j\in J$, 
\eqref{eq:seeded-small-columns}, 
\eqref{eq:seeded-small-rows}, and
\eqref{eq:seeded-P} give
\[
 x_i y_j
 \le\frac{4L_x^KL_y^K}{c\phi^2H_xH_y}\kappa^2
 =P\kappa^2.
\]
Since $B_{ij}=x_iA_{ij}y_j$, $A\ge0$, and $A$ is row-stochastic,
\[
 B(R,J) = \sum_{i\in R,j\in J} x_iA_{ij}y_j \le P\kappa^2A(R,J)
 \le P\kappa^2A(R,[n])
 =P\kappa^2|R|
 \le P\kappa^2n.
\]
Combining this upper bound with $B(R,J)>\omega n$ and using $P,n>0$ gives
\begin{equation}
 \kappa^2>\frac{\omega}{P}. \label{eq:seeded-kappa-lower-bound}
\end{equation}

Next, we bound every product $x_i y_j$.  
By
$\min_{i\in[n]}x_i=\min_{j\in[n]}y_j = \kappa$, we have
\[
 x_i y_j\ge\kappa^2\ge\frac{\omega}{P},\qquad \qquad \forall i,j\in [n].
\]
For every row $i$, row stochasticity of $A$, the bound $y_j\ge\kappa$,
and \eqref{eq:seeded-B-margins} give
\[
 b_+\ge B(i,[n])
 =x_i\sum_{j=1}^n A_{ij}y_j
 \ge x_i\kappa\sum_{j=1}^n A_{ij}
 =x_i\kappa.
\]
Thus $x_i\le b_+/\kappa$.  For every column $j$, the bound
$x_i\ge\kappa$, 
\eqref{eq:seeded-column-bounds}, 
\eqref{eq:seeded-B-margins} give
\[
 1=B([n],j)
 =y_j\sum_{i=1}^n x_iA_{ij}
 \ge y_j\kappa A([n],j)
 \ge y_j\kappa s_{-}.
\]
By the above two inequalities,
we have $b_+\geq x_iy_j\kappa^2 s_{-}$.
By $s_{-},\kappa>0$ and \eqref{eq:seeded-kappa-lower-bound}, we obtain
\[
 x_i y_j
 \le\frac{b_+}{s_{-}\kappa^2}
 <\frac{b_+P}{s_{-}\omega}.
\]
Combining the upper bound with $x_iy_j\ge\omega/P$ proves
\begin{equation}
 \frac{\omega}{P}\le x_i y_j\le\frac{b_+P}{s_{-}\omega},
 \qquad \forall i,j\in[n]. \label{eq:seeded-product-bounds}
\end{equation}
Define
\begin{equation}
 T\triangleq
 \max\left\{1,\frac{P}{\omega},\frac{b_+P}{s_{-}\omega}\right\}. \label{eq:seeded-T}
\end{equation}
Hence, $T\ge P/\omega$ and
$T\ge b_+P/(s_{-}\omega)$.  
Since $P,\omega$ are positive, we have
$1/T\le\omega/P$.  Combining these inequalities with
\eqref{eq:seeded-product-bounds} gives
\begin{equation}
 \frac1T\le x_i y_j\le T
 \qquad \forall i,j\in[n]. \label{eq:seeded-uniform-product-bounds}
\end{equation}
Since
$B_{ij}=x_i y_jA_{ij}$ and $A_{ij}\ge0$,  multiplying
\eqref{eq:seeded-uniform-product-bounds} by $A_{ij}$ proves
\begin{equation}
 \frac{A_{ij}}{T}\le B_{ij}\le T A_{ij}
 \qquad \forall i,j\in[n]. \label{eq:seeded-entrywise-comparison}
\end{equation}
Then, \eqref{eq:seeded-structure-stability} is proved.

It remains to verify \eqref{eq:seeded-distortion-rate}.  Define
\begin{equation}
 \beta\triangleq\frac{\log(L_xL_y)}{\log g},
 \qquad
 C\triangleq
 \max\left\{1,\frac{4b_+L_xL_y}{c\phi^2s_{-}H_xH_y}\right\}. \label{eq:seeded-C-beta}
\end{equation}
Hence, $C\geq 1$.
Furthermore, by $L_x,L_y>1$ and $g>1$, we have $\beta>0$.
By \eqref{eq:seeded-K-loss}, we obtain
\[
 K\le1+\frac{\log(M/m)}{\log g}.
\]
Combined with \eqref{eq:seeded-P} and $M=\lfloor qn\rfloor\le qn$, we obtain
\begin{equation}
 P\le\frac{4L_xL_y}{c\phi^2H_xH_y}
 \left(\frac{qn}{m}\right)^\beta. \label{eq:seeded-P-growth}
\end{equation} 
By \eqref{eq:seeded-omega} and \eqref{eq-def-qalphaomega},
we have
\begin{equation}
 0<\omega\le(1+c)\alpha q-1<1+c - 1 =c. \label{eq:seeded-omega-upper}
\end{equation}
Combined with $c\le1$, we have $0<\omega\leq 1$.
Recall that  $b_+\ge1$ and $0<s_{-}\le1$.
Hence, $b_+/(s_{-} \omega)>0$.
Combined with \eqref{eq:seeded-C-beta} and \eqref{eq:seeded-P-growth}, we have
\begin{align}\label{eq-bpluspoversoneomega-upperbound}
 \frac{b_+P}{s_{-}\omega}
 \le\frac{C}{\omega}\left(\frac{qn}{m}\right)^\beta.
\end{align}
By $b_+\ge1$ and $0<s_{-}\le1$, we also have $b_+/s_{-}\ge1$.
Hence,
\begin{align}\label{eq-poveromega-upperbound}
\frac{P}{\omega}\le \frac{b_+P}{s_{-}\omega}.
\end{align}
Furthermore, $C\ge1$, $0<\omega\le1$, and
$m\le M=\lfloor qn\rfloor\le qn$ give
\begin{align}\label{eq-onele-coveromegaqnovermbeta}
1\le \frac{C}{\omega}\left(\frac{qn}{m}\right)^\beta.
\end{align}
By \eqref{eq:seeded-omega-upper}, \eqref{eq-bpluspoversoneomega-upperbound}, \eqref{eq-poveromega-upperbound}, and \eqref{eq-onele-coveromegaqnovermbeta}
we have 
\begin{equation}
\begin{aligned}
 T&=\max\left\{1,\frac{P}{\omega},
                 \frac{b_+P}{s_-\omega}\right\}\\
 &\le\frac{C}{\omega}\left(\frac{qn}{m}\right)^\beta.
\end{aligned}
\end{equation}
By \eqref{eq:seeded-omega}, we have
\begin{align*}
 T
 &\le\frac{C}{\omega}\left(\frac{qn}{m}\right)^\beta\\
 &=\frac{Cn}{(1+c)(1+\eta)M-n}
   \left(\frac{qn}{m}\right)^\beta\\
 &=\frac{Cn}{(1+c)(1+\eta)\lfloor qn\rfloor-n}
   \left(\frac{qn}{m}\right)^\beta.
\end{align*}
This proves
\eqref{eq:seeded-distortion-rate}.  
The theorem is proved.
\end{proof}

\subsection{From Spectral gap to Buffered robust vertex expansion}\label{sec-spectralgap-brvx}

In this subsection, we show that a sufficiently large spectral gap
implies buffered robust vertex expansion.

Throughout this subsection, fix $q\in(1/2,1)$ and $\eta>0$.  Write
$\alpha\triangleq1+\eta$ and assume $\alpha q<1$.

\begin{lemma}
\label{lem:spectral-weighted-escape}
Let $n\ge2$, and let $A\in\R_{\geq0}^{n\times n}$ have positive row and
column sums.
Define
\begin{align}
 Q_r&\triangleq\diag\bigl(A(1,[n]),\ldots,A(n,[n])\bigr),
 \qquad  Q_c\triangleq\diag\bigl(A([n],1),\ldots,A([n],n)\bigr),\label{eq-defq}\\
 B&\triangleq Q_r^{-1/2}AQ_c^{-1/2},
 \qquad \quad \quad \quad \quad \qquad \quad
 \sigma\triangleq s_2(B).\label{eq-defwb}
\end{align}
For every nonempty $I\subseteq[n]$ and every proper subset
$J\subsetneq[n]$,
\begin{equation}
 \frac{A(I,[n]\setminus J)}{A(I,[n])}
 \ge
 \left(1-\frac{A([n],J)}{A([n],[n])}\right)
 \left(
 1-\sigma
 \sqrt{
 \frac{(A([n],[n])-A(I,[n]))A([n],J)}
 {A(I,[n])(A([n],[n])-A([n],J))}
 }
 \right).
 \label{eq:spectral-exact-forward}
\end{equation}
For every nonempty $J\subseteq[n]$ and every proper subset
$I\subsetneq[n]$,
\begin{equation}
 \frac{A([n]\setminus I,J)}{A([n],J)}
 \ge
 \left(1-\frac{A(I,[n])}{A([n],[n])}\right)
 \left(
 1-\sigma
 \sqrt{
 \frac{A(I,[n])(A([n],[n])-A([n],J))}
 {(A([n],[n])-A(I,[n]))A([n],J)}
 }
 \right).
 \label{eq:spectral-exact-reverse}
\end{equation}
\end{lemma}
\begin{proof}
To prove this lemma, it is sufficient to prove \eqref{eq:spectral-exact-forward}.
The proof of \eqref{eq:spectral-exact-reverse} is analogous.
In the following, we prove \eqref{eq:spectral-exact-forward}.
Let $\boldsymbol 1\in\R^n$ denote the all-ones vector, and
define
\begin{align}\label{eq-def-uv}
 w\triangleq A([n],[n]), \qquad   \qquad  u\triangleq\frac{Q_r^{1/2}\boldsymbol 1}{\sqrt w},
 \qquad  \qquad  
 v\triangleq\frac{Q_c^{1/2}\boldsymbol 1}{\sqrt w}.
\end{align}
By the definitions of $Q_r$, $Q_c$, and $w$, we have
\[
\|u\|_2^2=w^{-1}\sum_{i=1}^n A(i,[n])=\frac{A([n],[n])}{w} = 1,\qquad
\|v\|_2^2=w^{-1}\sum_{j=1}^n A([n],j)=\frac{A([n],[n])}{w} = 1.\]
Moreover,
\begin{align}\label{eq-buu-bvv}
 Bv
 =\frac{Q_r^{-1/2}A\boldsymbol 1}{\sqrt w}
 =\frac{Q_r^{1/2}\boldsymbol 1}{\sqrt w}=u,
 \qquad
 B^{\mathsf T}u
 =\frac{Q_c^{-1/2}A^{\mathsf T}\boldsymbol 1}{\sqrt w}
 =\frac{Q_c^{1/2}\boldsymbol 1}{\sqrt w}=v.
\end{align}
For every
$z\in\R^n$, we have
\begin{align*}
 (Bz)_i
 =\frac{1}{\sqrt{A(i,[n])}}
 \sum_{j=1}^n\frac{A_{ij} z_j}{\sqrt{A([n],j)}}.
\end{align*}
Applying the weighted Cauchy--Schwarz inequality
with the nonnegative weights $A_{ij}$ gives
\begin{align*}
 \|Bz\|_2^2
 =&\sum_{i=1}^n\frac{1}{A(i,[n])}
 \left(\sum_{j=1}^n \frac{A_{ij}z_j}{\sqrt{A([n],j)}}\right)^2 
 \le\sum_{i=1}^n\frac{1}{A(i,[n])}\left(\sum_{j=1}^n
 A_{ij}\right)\left(\sum_{j=1}^n
 \frac{A_{ij}z_j^2}{A([n],j)}\right)\\
 &=\sum_{i=1}^n\sum_{j=1}^n
 \frac{A_{ij}z_j^2}{A([n],j)} =\sum_{j=1}^n\sum_{i=1}^n
 \frac{A_{ij}z_j^2}{A([n],j)}= \sum_{j=1}^n z_j^2=\|z\|_2^2.
\end{align*}
Thus $s_1(B)\le1$.  Since $Bv=u$ and $u,v$ are unit vectors,
$s_1(B)=1$.
Choose orthogonal matrices $U=[\,u\;U_0\,]$ and
$V=[\,v\;V_0\,]$. The identities $Bv=u$ and
$B^{\mathsf T}u=v$ imply
\[
 U^{\mathsf T}BV=
 \begin{pmatrix}1&0\\0&B_0\end{pmatrix},
 \qquad B_0=U_0^{\mathsf T}BV_0.
\]
Since $s_1(B)=1$, orthogonal invariance of singular values gives
$\|B_0\|_2=s_2(B)=\sigma$, including when the top singular
value has multiplicity greater than one. Consequently,
\begin{equation}\label{eq-fzerobgzeroleqsigmafzerogzero}
\begin{aligned}
 |f_0^{\mathsf T}Bg_0|
 \le\sigma\|f_0\|_2\|g_0\|_2
 \qquad\forall f_0\perp u,\ g_0\perp v.
\end{aligned}
\end{equation}

Fix arbitrary sets $I,J\subseteq[n]$, and write
\begin{align}\label{eq-defxy}
 x\triangleq\frac{A(I,[n])}{A([n],[n])},
 \qquad
 y\triangleq\frac{A([n],J)}{A([n],[n])}.
\end{align}
Let $\boldsymbol 1_I$ and $\boldsymbol 1_J$ denote the indicator vectors of
$I$ and $J$, respectively.  Define the weighted indicator vectors
\[
 f\triangleq\frac{Q_r^{1/2}\boldsymbol 1_I}{\sqrt w},
 \qquad
 g\triangleq\frac{Q_c^{1/2}\boldsymbol 1_J}{\sqrt w}.
\]
By \eqref{eq-def-uv} and \eqref{eq-defq}, 
\[
 \|f\|_2^2=\langle f,u\rangle=x,
 \qquad
 \|g\|_2^2=\langle g,v\rangle=y.
\]
Furthermore, by \eqref{eq-defwb},
\begin{equation}\label{eq-fbgeqaoverw}
\begin{aligned}
 f^{\mathsf T}Bg
 &=\frac1w\boldsymbol1_I^{\mathsf T}
   Q_r^{1/2}(Q_r^{-1/2}AQ_c^{-1/2})
   Q_c^{1/2}\boldsymbol1_J\\
 &=\frac{\boldsymbol1_I^{\mathsf T}A\boldsymbol1_J}{w}
 =\frac{A(I,J)}{w}.
\end{aligned}
\end{equation}
Let $f_0\triangleq f-xu$ and $g_0\triangleq g-yv$.  Then
$f_0\perp u$, $g_0\perp v$.
In addition,
\begin{equation}
\begin{aligned}\label{eq-fzerogzeronorem}
\|f_0\|_2^2&=\langle f-xu,f-xu\rangle=\|f\|_2^2-2x\langle f,u\rangle+x^2\|u\|_2^2=x-2x^2+x^2
=x(1-x),\\
\|g_0\|_2^2&=\langle g-yv,g-yv\rangle=\|g\|_2^2-2y\langle g,v\rangle+y^2\|v\|_2^2=y-2y^2+y^2
=y(1-y).
\end{aligned} 
\end{equation}
Since $f_0=f-xu$ and $g_0=g-yv$, we have $f=f_0+xu$ and
$g=g_0+yv$.  Thus,
\begin{align}\label{eq-expansion-fbg}
 f^{\mathsf T}Bg
 =(f_0+xu)^{\mathsf T}B(g_0+yv) =f_0^{\mathsf T}Bg_0
   +y f_0^{\mathsf T}Bv
   +x u^{\mathsf T}Bg_0
   +xy u^{\mathsf T}Bv.
\end{align}
By \eqref{eq-buu-bvv} and
$f_0\perp u$, we have
$f_0^{\mathsf T}Bv=f_0^{\mathsf T}u=0$.  
By \eqref{eq-buu-bvv} and $\|u\|_2=1$, we have
$u^{\mathsf T}Bv=u^{\mathsf T}u=1$.  
Finally,
\eqref{eq-buu-bvv} and $g_0\perp v$ give
$u^{\mathsf T}Bg_0
 =(B^{\mathsf T}u)^{\mathsf T}g_0
 =v^{\mathsf T}g_0
 =0$. 
Substituting these three
identities into \eqref{eq-expansion-fbg} yields
\[
 f^{\mathsf T}Bg=f_0^{\mathsf T}Bg_0+xy.
\]
Combining this identity with
\eqref{eq-fbgeqaoverw} gives
\[
 \frac{A(I,J)}{w}-xy=f_0^{\mathsf T}Bg_0.
\]
Combined with \eqref{eq-fzerobgzeroleqsigmafzerogzero} and \eqref{eq-fzerogzeronorem}, we obtain
\begin{align*}
 \left|\frac{A(I,J)}w-xy\right|
 &=|f_0^{\mathsf T}Bg_0|\\
 &\le\sigma\|f_0\|_2\|g_0\|_2
 =\sigma\sqrt{x(1-x)y(1-y)}.
\end{align*}
Thus,
we have
\begin{align}\label{eq:spectral-mixing}
 \frac{A(I,J)}{w}
 \le xy+\sigma\sqrt{x(1-x)y(1-y)}.
\end{align}

Fix any nonempty $I\subseteq[n]$ and every proper subset
$J\subsetneq[n]$.
Since $A$ has positive row and 
column sums,
we have $A(I,[n])>0$ and
$A([n],[n]\setminus J)>0$.
Combined with \eqref{eq-defxy} and \eqref{eq-defwb},
we have
\begin{align*}
1 - y = 1 - \frac{A([n],J)}{w} = \frac{A([n],[n]\setminus J)}{w} > 0.
\end{align*}
By \eqref{eq-defxy}, we have $A(I,[n])=wx$.
By $A(I,[n])=wx>0$ and \eqref{eq:spectral-mixing},
\begin{align*}
 \frac{A(I,[n]\setminus J)}{A(I,[n])}
 &=\frac{A(I,[n])-A(I,J)}{wx}\\
 &=1-\frac{A(I,J)}{wx}\\
 &\ge1-y-\sigma\sqrt{\frac{(1-x)y(1-y)}x}.
\end{align*}
Combined with $1-y>0$, we have
\begin{equation}
 \frac{A(I,[n]\setminus J)}{A(I,[n])}
 \ge(1-y)
 \left(
 1-\sigma\sqrt{\frac{(1-x)y}{x(1-y)}}
 \right).
\end{equation}
Combined with \eqref{eq-defxy}, \eqref{eq:spectral-exact-forward} is immediate.
The lemma is immediate.
\end{proof}

\begin{theorem}
\label{thm:spectral-weighted-escape}
Assume the conditions and notations in \Cref{lem:spectral-weighted-escape}.
Suppose that $0<r_-\le r_+<\infty$ and
$0<c_-\le c_+<\infty$ satisfy
\begin{equation}
 r_-\le A(i,[n])\le r_+\quad\forall i\in[n],
 \qquad
 c_-\le A([n],j)\le c_+\quad\forall j\in[n]. \label{eq:spectral-margin-bounds}
\end{equation}
Let
\begin{equation}
 \theta\triangleq
 \sqrt{
 \frac{r_-c_-}{r_+c_+}
 \frac{1-\alpha q}{\alpha(1-q)}
 }. \label{eq:spectral-threshold}
\end{equation}
If $\sigma<\theta$, define
\begin{equation}
 \phi\triangleq
 \frac{n\min\{r_-,c_-\}}{A([n],[n])}
 (1-\alpha q)
 \left(1-\frac{\sigma}{\theta}\right)>0. \label{eq:spectral-phi}
\end{equation}
Then, for every nonempty $I\subseteq[n]$ with $|I|\le qn$ and every
$J\subseteq[n]$ with $|J|\le\alpha|I|$,
\begin{equation}
 A(I,[n]\setminus J)\ge\phi A(I,[n]). \label{eq:spectral-weighted-forward}
\end{equation}
For every nonempty $J\subseteq[n]$ with $|J|\le qn$ and every
$I\subseteq[n]$ with $|I|\le\alpha|J|$,
\begin{equation}
 A([n]\setminus I,J)\ge\phi A([n],J). \label{eq:spectral-weighted-reverse}
\end{equation}
\end{theorem}

\begin{proof}
At first, we prove \eqref{eq:spectral-weighted-forward}.
Let
$\varnothing\neq I\subseteq[n]$ and $J\subseteq[n]$ satisfy
$|I|\le qn$ and $|J|\le\alpha|I|$.  
By $\alpha q<1$, we have $|J|\le\alpha qn < n$.
Hence,
$J \subsetneq [n]$.
Thus, $I$ and $J$ satisfy the conditions required to apply
\eqref{eq:spectral-exact-forward}.
Define
\begin{align}\label{eq-defxy-new}
 x\triangleq\frac{A(I,[n])}{A([n],[n])},
 \qquad
 y\triangleq\frac{A([n],J)}{A([n],[n])}.
\end{align}
By \eqref{eq:spectral-exact-forward}, we have
\begin{equation}
 \frac{A(I,[n]\setminus J)}{A(I,[n])}
 \ge(1-y)
 \left(
 1-\sigma\sqrt{\frac{(1-x)y}{x(1-y)}}
 \right). \label{eq:spectral-forward-escape}
\end{equation}

In the next, we provide lower bounds for $1 - y$ and  $1-\sigma\sqrt{((1-x)y)/(x(1-y))}$.
By \eqref{eq-defxy-new}, \eqref{eq:spectral-margin-bounds} and $|J|\le\alpha qn$,
\begin{equation}\label{eq-oneminusy-lowerbound}
\begin{aligned}
 1-y
 &=\frac{A([n],[n]\setminus J)}{A([n],[n])}\\
 &\ge\frac{c_-(n-|J|)}{A([n],[n])}
 \ge\frac{(1-\alpha q)nc_-}{A([n],[n])}.
\end{aligned}
\end{equation}
By \eqref{eq-defxy-new}, \eqref{eq:spectral-margin-bounds}, and $|J|\le\alpha|I|$, we have
\begin{align*}
 \frac{y}{x}
 &=\frac{A([n],J)}{A(I,[n])}
 \le\frac{c_+|J|}{r_-|I|}
 \le\alpha\frac{c_+}{r_-},\\
 \frac{1-x}{1-y}
 &=\frac{A([n]\setminus I,[n])}{A([n],[n]\setminus J)}
 \le\frac{r_+(n-|I|)}{c_-(n-|J|)}
 \le\frac{r_+}{c_-}\frac{n-|I|}{n-\alpha |I|}.
\end{align*}
Hence,
\begin{equation*}
 \frac{(1-x)y}{x(1-y)}
 \le\alpha\frac{r_+c_+}{r_-c_-}
 \frac{n-|I|}{n-\alpha |I|}.
\end{equation*}
The derivative of $(n-|I|)/(n-\alpha |I|)$ about $|I|$ is
$n(\alpha-1)/(n-\alpha |I|)^2$ for $0\leq \abs{I} < n/\alpha$.  
By $\alpha = 1+\eta > 1$,
we have the derivative is always larger than $0$.
Hence, 
we have $(n-|I|)/(n-\alpha |I|)$ is
increasing as $\abs{I}$ increases.
By $|I|\leq qn$, we have
\begin{align}
\frac{n-|I|}{n-\alpha |I|} \leq \frac{n-qn}{n-\alpha qn} = \frac{1-q}{1-\alpha q}.
\end{align}
By \eqref{eq:spectral-threshold},
\begin{equation*}
 \frac{(1-x)y}{x(1-y)}
 \le\alpha\frac{r_+c_+}{r_-c_-}
 \frac{1-q}{1-\alpha q}
 =\theta^{-2}.
\end{equation*}
Hence, by the assumption $\sigma<\theta$, 
\[
 1-\sigma\sqrt{\frac{(1-x)y}{x(1-y)}}
 \ge1-\frac{\sigma}{\theta}>0.
\]
Combined with \eqref{eq:spectral-forward-escape} and \eqref{eq-oneminusy-lowerbound}, we have
\begin{equation}
 \frac{A(I,[n]\setminus J)}{A(I,[n])}
 \ge\frac{(1-\alpha q)n c_-}{A([n],[n])}
 \left(1-\frac{\sigma}{\theta}\right). \label{eq:spectral-forward-bound}
\end{equation}
Combined with \eqref{eq:spectral-phi}, we have
\begin{align}
A(I,[n]\setminus J) \geq \phi A(I,[n]).
\end{align}
Thus, \eqref{eq:spectral-weighted-forward} is proved.

In the following, we prove \eqref{eq:spectral-weighted-reverse}.
Apply
\eqref{eq:spectral-forward-bound} to $A^{\mathsf T}$, with a nonempty
column set $J$ in place of $I$ and a row set $I$ in place of $J$.  The row
and column margins of $A^{\mathsf T}$ are the column and row margins of
$A$, respectively.  Its normalized matrix is
$Q_c^{-1/2}A^{\mathsf T}Q_r^{-1/2}=B^{\mathsf T}$,
which has the same second singular value $\sigma$ as $B$.
The total mass, and the value of $\theta$
in \eqref{eq:spectral-threshold} are unchanged by transposition.  Therefore,
for every $\varnothing\neq J\subseteq[n]$ and $I\subseteq[n]$ satisfying
$|J|\le qn$ and $|I|\le\alpha|J|$, estimate
\eqref{eq:spectral-forward-bound} for $A^{\mathsf T}$ becomes
\begin{equation}
 \frac{A([n]\setminus I,J)}{A([n],J)}
 \ge\frac{(1-\alpha q)n r_-}{A([n],[n])}
 \left(1-\frac{\sigma}{\theta}\right). \label{eq:spectral-reverse-bound}
\end{equation}
Combined with \eqref{eq:spectral-phi}, we have
\begin{align}
A([n]\setminus I,J) \geq \phi A([n],J).
\end{align}
The theorem is proved.
\end{proof}

For a row-stochastic matrix, the row normalization is the identity.  Thus
only the column margins enter the normalized matrix and the spectral
threshold.

\begin{theorem}
\label{thm:sharp-column-balanced-mixing-to-bve}
Let $n\ge2$ and $ K\ge1$.  Let
$A\in\R_{\ge0}^{n\times n}$ be row-stochastic and have positive column
sums.  Define
\[
 Q\triangleq\diag\bigl(A([n],1),\ldots,A([n],n)\bigr),
 \qquad
 \sigma\triangleq s_2(AQ^{-1/2}),
\]
and suppose that
\begin{equation}
 \frac{\max_{j\in[n]}A([n],j)}{\min_{j\in[n]}A([n],j)}
 \le K. \label{eq-maxmin-relation-lambda}
\end{equation}
Define
\begin{equation}
 \theta_K\triangleq
 \sqrt{\frac{1-\alpha q}{ K\alpha(1-q)}}.
 \label{eq:sharp-spectral-threshold}
\end{equation}
Let $\delta\in(0,1)$ and assume $\sigma\le(1-\delta)\theta_K$.
Then, with
\begin{equation}
 \zeta\triangleq
 \frac{\delta(1-\alpha q)}
 { K-( K-1)(1-\alpha q)}.
 \label{eq:sharp-zeta}
\end{equation}
the following two conclusions hold.
For every nonempty $I\subseteq[n]$ with $|I|\le qn$ and every
$J\subseteq[n]$ with $|J|\le\alpha|I|$,
\begin{equation}
 A(I,[n]\setminus J)\ge\zeta|I|.
 \label{eq:sharp-forward-expansion}
\end{equation}
For every nonempty $J\subseteq[n]$ with $|J|\le qn$ and every
$I\subseteq[n]$ with $|I|\le\alpha|J|$,
\begin{equation}
 A([n]\setminus I,J)\ge\zeta|J|.
 \label{eq:sharp-reverse-expansion}
\end{equation}
Consequently, $A$ satisfies
$(q,\eta,\zeta)$-two-sided buffered robust vertex expansion.
\end{theorem}

The following lemma is used in the proof of \Cref{thm:sharp-column-balanced-mixing-to-bve}.
It converts a balance bound on the individual margins into
sharp bounds on the total mass carried by any subset of coordinates.

\begin{lemma}
\label{lem:column-balanced-subset-mass}
Let $n\ge1$, let $K\ge1$, and let $c_1,\ldots,c_n>0$ satisfy
\begin{align}\label{eq-ratiobound}
 \sum_{j=1}^n c_j=n,
 \qquad
 \frac{\max_{j\in[n]}c_j}{\min_{j\in[n]}c_j}\le K.
\end{align}
Then, for every $U\subseteq[n]$,
\begin{equation}
 \frac{|U|}{ K n-( K-1)|U|}
 \le \frac1n\sum_{j\in U}c_j
 \le \frac{ K|U|}{n+( K-1)|U|}.
 \label{eq:column-balanced-subset-mass}
\end{equation}
\end{lemma}

\begin{proof}
If $U=\varnothing$ or $U=[n]$, both bounds in
\eqref{eq:column-balanced-subset-mass} follow directly from
$\sum_{j=1}^n c_j=n$.  Suppose that $0<|U|<n$.
By \eqref{eq-ratiobound}, 
we have
\[
 \frac{n-\sum_{j\in U}c_j}{n-|U|} \leq \max_{j\in[n]}c_j = \frac{\max_{j\in[n]}c_j}{\min_{j\in[n]}c_j}\cdot \min_{j\in[n]}c_j \leq 
  K\frac{\sum_{j\in U}c_j}{|U|}.
\]
Hence, 
\[
|U| \left(n-\sum_{j\in U}c_j\right)\leq  K \left(n-\abs{U}\right)  \sum_{j\in U}c_j.
\]
Therefore,
\[
 n|U|
 \le
 \bigl( K n-( K-1)|U|\bigr)
 \sum_{j\in U}c_j.
\]
Dividing by the positive quantity
$n\bigl( K n-( K-1)|U|\bigr)$ proves the lower bound in
\eqref{eq:column-balanced-subset-mass}.

Conversely, by \eqref{eq-ratiobound}, 
we also have
\[
 \frac{\sum_{j\in U}c_j}{|U|} \leq \max_{j\in[n]}c_j = \frac{\max_{j\in[n]}c_j}{\min_{j\in[n]}c_j}\cdot \min_{j\in[n]}c_j 
 \le K\frac{n-\sum_{j\in U}c_j}{n-|U|}.
\]
Hence, 
\[
 (n-|U|)\sum_{j\in U}c_j
 \le K|U|\left(n-\sum_{j\in U}c_j\right).
\]
Thus,
$\sum_{j\in U}c_j$ gives
\[
 \bigl(n+( K-1)|U|\bigr)
 \sum_{j\in U}c_j
 \le K n|U|.
\]
Dividing by the positive quantity
$n\bigl(n+( K-1)|U|\bigr)$ proves the upper bound in
\eqref{eq:column-balanced-subset-mass}.
\end{proof}

Now we can prove \Cref{thm:sharp-column-balanced-mixing-to-bve}.

\begin{proof}[Proof of \Cref{thm:sharp-column-balanced-mixing-to-bve}]
We first prove \eqref{eq:sharp-forward-expansion}.  
Since $A$ is row-stochastic, its row-sum diagonal matrix is the identity.
Thus,  the matrix $B$ in \Cref{lem:spectral-weighted-escape} is $AQ^{-1/2}$, and 
$s_2(B)$ in that lemma is the $\sigma$ defined in this theorem.
Fix a nonempty set
$I\subseteq[n]$ with $|I|\le qn$ and a set $J\subseteq[n]$ with
$|J|\le\alpha|I|$.  
By $\alpha q<1$,
we have
$|J|\le\alpha qn<n$.
Hence, $J\subsetneq[n]$.  
Thus the pair $(I,J)$ satisfies
all the conditions required to apply \eqref{eq:spectral-exact-forward}.
Using $A(I,[n])=|I|$ and $A([n],[n])=n$ in that inequality gives
\begin{align}\label{eq-ratioajnsetminusjoversizei-lowerbound}
 \frac{A(I,[n]\setminus J)}{|I|}
 \ge
 \left(1-\frac{A([n],J)}{n}\right)
 \left(
 1-\sigma
 \sqrt{
 \frac{(n-|I|)A([n],J)}
 {|I|(n-A([n],J))}
 }
 \right).
\end{align}
Furthermore, since $A$ is row-stochastic, we have $\sum_{j\in [n]} A([n],j) = \sum_{i\in [n]} A(i,[n]) = n$.
Applying the lower bound in Lemma~\ref{lem:column-balanced-subset-mass} with
$U=[n]\setminus J$, we have
\begin{equation}
 \frac{n - \abs{J}}{ K n-( K-1)(n - \abs{J})}
 \le \frac{A([n],[n]\setminus J)}{n}.
\end{equation}
Since 
\[\frac{A([n],[n]\setminus J)}{n} = \frac{A([n],[n])-A([n],J)}{n} = \frac{n-A([n],J)}{n} = 1 - \frac{A([n],J)}{n},\]
we have 
\[
 1-\frac{A([n],J)}{n}
 \ge \frac{n - \abs{J}}{ K n-( K-1)(n - \abs{J})}\ge
 \frac{n-|J|}{n+( K-1)|J|} = \frac{1 - \abs{J}/n}{1 + ( K -1)\abs{J}/n}.
\]
The function $a\mapsto(1-a)/(1+( K-1)a)$ is decreasing whenever
$0\le a<1$.  
By $|J|/n\le\alpha q$, we obtain
\begin{equation}
 1-\frac{A([n],J)}{n}
 \ge\frac{1-\alpha q}{1+(K-1)\alpha q}=
 \frac{1-\alpha q}
 { K-( K-1)(1-\alpha q)}.
 \label{eq:sharp-forward-main-factor}
\end{equation}
Applying the upper bound in Lemma~\ref{lem:column-balanced-subset-mass} with
$U=J$, we have
\[
\frac{A([n],J)}{n}
 \le \frac{ K|J|}{n+( K-1)|J|}.
\]
Hence,
\[
 \frac{A([n],J)}{n-A([n],J)}
 \le\frac{ K|J|}{n+( K - 1)\abs{J}- K|J|} = \frac{ K|J|}{n-|J|}.
\]
Therefore,
\begin{align*}
 \frac{(n-|I|)A([n],J)}
 {|I|(n-A([n],J))}\le
 \frac{ K(n-|I|)|J|}{|I|(n-|J|)}.
\end{align*}
Because $|J|\le\alpha|I|$, $|I|\le qn$, and $\alpha>1$, it follows that
\begin{align*}
 \frac{ K(n-|I|)|J|}{|I|(n-|J|)} \le
 \frac{ K\alpha(n-|I|)}{n-\alpha|I|}\le  K\alpha
 \frac{1-q}{1-\alpha q}.
\end{align*}
Thus, we have
\[
 \frac{(n-|I|)A([n],J)}
 {|I|(n-A([n],J))}
 \le  K\alpha
 \frac{1-q}{1-\alpha q} = \theta_K^{-2},
\]
where the last equality is by \eqref{eq:sharp-spectral-threshold}.
Combining
$\sigma\le(1-\delta)\theta_K$ gives
\[
 1-\sigma
 \sqrt{
 \frac{(n-|I|)A([n],J)}
 {|I|(n-A([n],J))}
 }
 \ge1-\frac{\sigma}{\theta_K}\ge\delta.
\]
Substituting this inequality and
\eqref{eq:sharp-forward-main-factor} into
\eqref{eq-ratioajnsetminusjoversizei-lowerbound},
we have
\begin{align*}
 \frac{A(I,[n]\setminus J)}{|I|}
 \ge
 \frac{\delta(1-\alpha q)}
 { K-( K-1)(1-\alpha q)}.
\end{align*}
Hence, 
\eqref{eq:sharp-forward-expansion} is immediate by \eqref{eq:sharp-zeta}.

We next prove \eqref{eq:sharp-reverse-expansion}.  Fix a nonempty set
$J\subseteq[n]$ with $|J|\le qn$ and a set $I\subseteq[n]$ with
$|I|\le\alpha|J|$.  By $\alpha q<1$,
we have 
$|I|\le\alpha qn<n$.
Hence, $I\subsetneq[n]$.  
Thus the pair $(I,J)$ satisfies
all the conditions required to apply \eqref{eq:spectral-exact-reverse}.
Using $A(I,[n])=|I|$ and $A([n],[n])=n$ in that inequality gives
\begin{equation*}
 \frac{A([n]\setminus I,J)}{A([n],J)}
 \ge
 \left(1-\frac{\abs{I}}{n}\right)
 \left(
 1-\sigma
 \sqrt{
 \frac{A(I,[n])(A([n],[n])-A([n],J))}
 {(A([n],[n])-A(I,[n]))A([n],J)}
 }
 \right).
\end{equation*}
Hence, multiplying both sides by $A([n],J)/|J|>0$ gives
\begin{align}\label{eq:spectral-exact-reverse-ansetminusijoverj}
 \frac{A([n]\setminus I,J)}{|J|}
 \ge
 \frac{(n-|I|)A([n],J)}{n|J|}
 \left(
 1-\sigma
 \sqrt{
 \frac{|I|(n-A([n],J))}
 {(n-|I|)A([n],J)}
 }
 \right).
\end{align}
Applying the
lower bound in Lemma~\ref{lem:column-balanced-subset-mass} with $U=J$, we have
\begin{align}\label{eq-lemcolumnmass-uequalj}
 \frac{|J|}{ K n-( K-1)\abs{J}}
 \le
 \frac{A([n],J)}{n}.
\end{align}
Hence, multiplying both sides by $(n-\abs{I})/|J|>0$ gives
\[
 \frac{n -|I|}{ K n-( K-1)\abs{J}}
 \le
 \frac{(n -|I|)A([n],J)}{n|J|}.
\]
Since $|I|\le\alpha|J|$, we therefore have
\[
 \frac{(n-|I|)A([n],J)}{n|J|}
 \ge \frac{n -\alpha|J|}{ K n-( K-1)\abs{J}} =
 \frac{1-\alpha|J|/n}
 { K-( K-1)|J|/n} = \frac{1}{ K} - \frac{(\alpha - 1 + 1/ K)|J|/n}
 { K-( K-1)|J|/n}.
\]
By $\alpha > 1$ and $|J|\le qn$, we have
\[
\frac{(n-|I|)A([n],J)}{n|J|}
 \ge \frac{1}{ K} - \frac{(\alpha - 1 + 1/ K)|J|/n}
 { K-( K-1)|J|/n} \geq \frac{1}{ K} - \frac{(\alpha - 1 + 1/ K)q}
 { K-( K-1)q}   =
 \frac{1-\alpha q}{ K-( K-1)q}.
\]
Furthermore, $\alpha>1$ and $q>1/2$ imply
$1-\alpha q<1-q<q$.  Since $ K \ge 1$, we conclude that
\begin{equation}
 \frac{(n-|I|)A([n],J)}{n|J|}
 \ge
 \frac{1-\alpha q}
 { K-( K-1)(1-\alpha q)}.
 \label{eq:sharp-reverse-main-factor}
\end{equation}
By \eqref{eq-lemcolumnmass-uequalj}, we also have
\[
 \frac{n-A([n],J)}{A([n],J)}
 \le\frac{ K(n-|J|)}{|J|}.
\]
Because $|I|\le\alpha|J|$ and the function $a\mapsto a/(n-a)$ is
increasing whenever $0\le a<n$, we obtain
\begin{align*}
 \frac{|I|(n-A([n],J))}
 {(n-|I|)A([n],J)}
 &\le
 \frac{ K|I|(n-|J|)}{(n-|I|)|J|} \le 
 \frac{ K\alpha |J|(n-|J|)}{(n-\alpha |J|)|J|} 
 =
 \frac{ K\alpha(n-|J|)}{n-\alpha|J|}.
\end{align*}
Combined with $\alpha > 1$, $|J|\le qn$, and $\alpha q < 1$, we have
\begin{align*}
 \frac{|I|(n-A([n],J))}
 {(n-|I|)A([n],J)}
 &\le
 \frac{ K\alpha(n-|J|)}{n-\alpha|J|} \le  \frac{ K\alpha(n-qn)}{n-\alpha qn } = 
 \frac{ K\alpha(1-q)}{1-\alpha q}.
\end{align*}
Hence, \eqref{eq:sharp-spectral-threshold} gives
\begin{align*}
 \frac{|I|(n-A([n],J))}
 {(n-|I|)A([n],J)}
\le 
 \frac{ K\alpha(1-q)}{1-\alpha q}
 =\theta_K^{-2}.
\end{align*}
Combined with $\sigma\le(1-\delta)\theta_K$, we obtain
\[
 1-\sigma
 \sqrt{
 \frac{|I|(n-A([n],J))}
 {(n-|I|)A([n],J)}
 }
 \ge1-\frac{\sigma}{\theta_K}\ge\delta.
\]
Combining this inequality with \eqref{eq:sharp-reverse-main-factor} and
\eqref{eq:spectral-exact-reverse-ansetminusijoverj} gives 
\begin{align}
\frac{A([n]\setminus I,J)}{|J|}
 \ge \frac{\delta(1-\alpha q)}
 { K-( K-1)(1-\alpha q)}.
\end{align}
Hence, 
\eqref{eq:sharp-reverse-expansion} is immediate by \eqref{eq:sharp-zeta}.
The lemma is proved.
\end{proof}

\newpage

%% file: Spectral_Condition.tex
\section{Spectral Conditions for general matrices}\label{sec-spectralconditions}

In this section, we prove Theorems \ref{cor-k-sigmaa}
and \ref{thm:polynomial-scaling-gap-seeded} .

\subsection{Proof of \Cref{thm:polynomial-scaling-gap-seeded}}

\begin{proof}[Proof of \Cref{thm:polynomial-scaling-gap-seeded}]
We first establish \eqref{eq-distrotion-t} and then verify
$\sigma_{\perp}<1$. The iteration bound
\eqref{eq-errorbound-maintheorem} will follow from
\Cref{thm:endpoint-gap-global-sinkhorn}.

Define
\[
\begin{aligned}
 Q&=\operatorname{diag}\bigl(A([n],1),\ldots,A([n],n)\bigr),\\
 e&=n^{-1/2}\mathbf1,\qquad f=n^{-1/2}Q^{1/2}\mathbf1.
\end{aligned}
\]
We have $A^\top\mathbf1=Q\mathbf1$.
Since $A$ is row-stochastic, $\sum_j A([n],j)=n$. Thus
\[
\|e\|_2^2=1,
\qquad
\|f\|_2^2
=\frac1n\mathbf1^\top Q\mathbf1
=\frac1n A([n],[n])
=1.
\]
Moreover, $A\mathbf1=\mathbf1$ and
$A^\top\mathbf1=Q\mathbf1$ imply
\[
Mf=
\frac1{\sqrt n}AQ^{-1/2} Q^{1/2}\mathbf1 = \frac1{\sqrt n}A\mathbf1=e,
\qquad
M^\top e
=\frac1{\sqrt n}Q^{-1/2}A^\top\mathbf1
=\frac1{\sqrt n}Q^{1/2}\mathbf1=f.
\]
Since $B$ is doubly stochastic, we also have
$Be=B^\top e=e$.

We first verify that $s_1(M)=s_1(B)=1$.
For every $z\in\mathbb R^n$, the weighted Cauchy--Schwarz
inequality and row-stochasticity of $A$ give
\begin{align*}
\|Mz\|_2^2
&=\sum_i
  \left(\sum_j A_{ij}\frac{z_j}{\sqrt{A([n],j)}}\right)^2
  = \sum_i
  \left(\sum_j \sqrt{A_{ij}}\frac{\sqrt{A_{ij}} z_j}{\sqrt{A([n],j)}}\right)^2
\le
  \sum_i\left(\sum_jA_{ij}\right)
  \left(\sum_j\frac{A_{ij} z_j^2}{A([n],j)}\right)\\
&=\sum_j\frac{z_j^2}{A([n],j)}\sum_iA_{ij} = \sum_j\frac{z_j^2}{A([n],j)}A([n],j)
=\|z\|_2^2.
\end{align*}
Hence $\|M\|_2\le1$. Since $Mf=e$ and both $e$ and $f$
are unit vectors, we have $s_1(M)=1$.
Likewise, double stochasticity of $B$ gives
\[
\|Bz\|_2^2
=\sum_i\left(\sum_jB_{ij}z_j\right)^2 = \sum_i
  \left(\sum_j \sqrt{B_{ij}}\sqrt{B_{ij}} z_j\right)^2
\le
  \sum_i\left(\sum_jB_{ij}\right)
  \left(\sum_jB_{ij} z_j^2\right)\\
\le\sum_{i,j}B_{ij}z_j^2
=\|z\|_2^2.
\]
Together with $Be=e$, this yields $s_1(B)=1$.

We then show that $s_2(M)=\sigma$ and $s_2(B)=\sigma_{\perp}$.
The identities $Mf=e$ and $M^\top e=f$ show that
$e$ and $f$ form a pair of left and right singular vectors
of $M$ corresponding to the singular value $1$.
We may therefore choose a singular value decomposition of $M$
of the form
\[
M=ef^\top+\sum_{k=2}^n s_k(M)u_kv_k^\top,
\]
where $(e,u_2,\ldots,u_n)$ and $(f,v_2,\ldots,v_n)$
are orthonormal bases. Consequently,
\begin{equation}\label{eq:product-gap-centered-M}
\|M-ef^\top\|_2=s_2(M)=\sigma.
\end{equation}
Similarly, $Be=B^\top e=e$ shows that the
left and right singular vectors
of $B^\top$ corresponding to the singular value $1$ can both be chosen as $e$.
Hence,
\begin{equation}\label{eq:product-gap-centered-B}
\|B^\top-ee^\top\|_2=s_2(B^\top)=s_2(B)=\sigma_{\perp}.
\end{equation}

Set
\begin{align}\label{eq-def-pefr}
P=B^\top A,\qquad E=M-ef^\top,\qquad
F=B^\top-ee^\top,\qquad
R=FE.
\end{align}
We next prove 
\begin{align}\label{eq:seeded-reduction-spectrum}
s_2(PQ^{-1/2})
\le\sigma\sigma_{\perp}
=\frac{1-\Delta}{\sqrt K}.
\end{align}
Using $e^\top e= \|e\|_2^2 = 1$, $  f^\top f=\|f\|_2^2 = 1$, $Mf=e$, $M^\top e=f$, and $Be=B^\top e=e$, we obtain
\begin{align*}
Ef&=Mf-e(f^\top f)=0,
\qquad
e^\top E=e^\top M-(e^\top e)f^\top=0,\\
Fe&=B^\top e-e(e^\top e)=0,
\qquad
e^\top F=(Be)^\top-(e^\top e)e^\top=0.
\end{align*}
It follows that
\begin{equation}\label{eq-pqminusoneovertwo-eftopr}
\begin{aligned}
 PQ^{-1/2}
 &=B^\top M=(ee^\top+F)(ef^\top+E)\\
 &=ef^\top+FE=ef^\top+R.
\end{aligned}
\end{equation}
Furthermore, by $R=FE$, $Ef = 0$ and $e^\top F = 0$, we have
\begin{align}\label{eq-efzero-etoprzero}
Rf=0,\qquad e^\top R=0.
\end{align}
Choose orthogonal matrices
\[
U=[\,e\;\;U_0\,],\qquad V=[\,f\;\;V_0\,],
\]
where the columns of $U_0$ and $V_0$ form orthonormal bases
of $e^\perp$ and $f^\perp$, respectively.
Writing $R_0=U_0^\top R V_0$, by \eqref{eq-pqminusoneovertwo-eftopr} and \eqref{eq-efzero-etoprzero} we have
\begin{align}\label{eq-block-decomposition-r-pqminusoneovertwo}
U^\top R V=
\begin{pmatrix}
0&0\\
0&R_0
\end{pmatrix},
\qquad
U^\top(PQ^{-1/2})V=
\begin{pmatrix}
1&0\\
0&R_0
\end{pmatrix}.
\end{align}
Since orthogonal multiplication preserves singular values, we have
$\|R_0\|_2=\|R\|_2$.
By submultiplicativity of the operator norm and
\eqref{eq:product-gap-centered-M}--\eqref{eq-def-pefr},
\[
\|R_0\|_2
=\|R\|_2
\le\|F\|_2\|E\|_2
=\sigma_{\perp}\sigma
\le1.
\]
Thus, \eqref{eq-block-decomposition-r-pqminusoneovertwo} gives
$s_2(PQ^{-1/2})
=\|R_0\|_2
\le\sigma\sigma_{\perp}$.
Finally, the definition
$\Delta=1-\sqrt K\,\sigma\sigma_{\perp}$ gives
$\sigma\sigma_{\perp}=(1-\Delta)/\sqrt K$.
Hence, \eqref{eq:seeded-reduction-spectrum} is proved.

Set
\begin{align}\label{eq-def-qetadeltaphi}
 q\triangleq\frac34,\qquad
 \eta\triangleq\frac{\Delta}{8},\qquad
 \delta\triangleq\frac{\Delta}{2},\qquad
\zeta\triangleq\frac{5\Delta}{64K}.
\end{align}
Since $\Delta=1-\sqrt K\,\sigma\sigma_{\perp}>0$, we have
$0<\Delta\le1$. Together with $K\ge1$, this gives
$\eta,\delta,\zeta\in(0,1)$.
Also, $q\in(1/2,1)$ and $q(1+\eta)<1$.
Define 
\[
 \theta_K =\sqrt{\frac{1-(1+\eta)q}{K(1+\eta)(1-q)}}.
\]
By \eqref{eq-def-qetadeltaphi}, we have
\[
 \theta_K =\frac1{\sqrt K}\sqrt{\frac{1-3\eta}{1+\eta}}
 \ge\frac{1-\Delta/2}{\sqrt K}.
\]
Hence, by \eqref{eq:seeded-reduction-spectrum},
\[
 (1-\delta)\theta_K
 \ge\frac{(1-\Delta/2)^2}{\sqrt K} > \frac{1-\Delta}{\sqrt K}
 \ge s_2(PQ^{-1/2}).
\]
By \Cref{thm:sharp-column-balanced-mixing-to-bve}, the matrix $P$
has $(q,\eta,\zeta_*)$-two-sided buffered robust vertex expansion,
where
\begin{align*}
 \zeta_*
 &\triangleq\frac{\delta(1-(1+\eta)q)}
 {K-(K-1)(1-(1+\eta)q)}\\
 &\ge\frac{\delta(1-(1+\eta)q)}K
 \ge\frac{5\Delta}{64K}=\zeta.
\end{align*}
Thus $P$ also has $(q,\eta,\zeta)$-two-sided buffered robust
vertex expansion.

Define
\[
 \tau\triangleq\frac{\eta}{2K},\qquad
 \widehat A\triangleq(1-\tau)I+\tau P,
 \qquad \widehat B\triangleq\widehat A^\top.
\]
We show all hypotheses of \Cref{thm:seeded-multiscale} are now satisfied by $\widehat A$ and $\widehat B$.
Recall $P=B^\top A$.
Since $A$ is row-stochastic and $B$ is doubly stochastic,
\begin{equation}\label{eq-byinversex}
\begin{aligned}
 P\mathbf1&=\mathbf1,\\
 \mathbf1^\top P&=\mathbf1^\top A
 =\bigl(A([n],1),\ldots,A([n],n)\bigr),\\
 BY^{-1}\mathbf1&=XA\mathbf1=X\mathbf1.
\end{aligned}
\end{equation}
Thus $\widehat A$ is row-stochastic.
Since the column sums of $A$ have average $1$ and ratio $K\ge1$,
\[
 \frac1K\le A([n],j)\le K
 \qquad\forall j\in[n].
\]
The same bounds hold for the column sums of $P$.
Consequently,
\begin{equation}\label{eq-range-acolumn}
\begin{aligned}
 1-\frac\eta2
 &\le1-\tau<\widehat A([n],j)\\
 &\le1-\tau+\tau K
 \le1+\frac\eta2
 \qquad\forall j\in[n].
\end{aligned}
\end{equation}
Furthermore, since $P$ has
$(q,\eta,\zeta)$-two-sided buffered robust vertex expansion and $\widehat A\ge\tau P$ entrywise, we have that $\widehat A$ has
$(q,\eta,\tau\zeta)$-two-sided buffered robust vertex expansion.
In addition, by $\widehat B = \widehat A^\top$,
we also have $\widehat B$ is column-stochastic and has row sums in $[1-\eta/2,1+\eta/2]$. 
Moreover,
\begin{align*}
 Y^{-1}P
 &=Y^{-1}B^\top A=A^\top XA\\
 &=A^\top BY^{-1}=P^\top Y^{-1}.
\end{align*}
Thus $Y^{-1}PY=P^\top$, and hence
\begin{equation}\label{eq-yinverseayeqb}
\begin{aligned}
 Y^{-1}\widehat A Y
 &=(1-\tau)I+\tau Y^{-1}PY\\
 &=(1-\tau)I+\tau P^\top
 =\widehat A^\top=\widehat B.
\end{aligned}
\end{equation}
Let 
\begin{align}\label{eq-def-hatxhaty}
 a=\sqrt{\min_{j}Y_j\max_{j}Y_j}, \qquad\widehat X=aY^{-1},\qquad \widehat Y=a^{-1}Y,\qquad
 \kappa=\sqrt{\min_{j}Y_j/\max_{j}Y_j}.
\end{align}
Then, \eqref{eq-yinverseayeqb} gives $\widehat B=\widehat X\widehat A\widehat Y$.
Moreover, 
the minimum diagonal entries of both $\widehat X$ and $\widehat Y$
equal $\kappa$, and the maximum diagonal entries of both equal
$1/\kappa$. 
Write $\widehat x_i=(\widehat X)_{ii}$ and
$\widehat y_j=(\widehat Y)_{jj}$.
Thus, there are sets $R_0,C_0\subseteq[n]$ such
that $|R_0|,|C_0|\ge 1$ and
\begin{equation}
 \widehat{x}_i\ge \frac{1}{\kappa}\quad\forall i\in R_0,
 \qquad
 \widehat{y}_j\ge \frac{1}{\kappa}\quad\forall j\in C_0.
\end{equation}

Use the following parameters in Theorem~\ref{thm:seeded-multiscale}:
\[
 m=1,\qquad 
 c=s_-=1-\eta/2,\qquad
 s_+=b_+=1+\eta/2,\qquad H_x=H_y=1,\qquad \phi = \tau\zeta.
\]
We have \eqref{eq:seeded-forward-bve}, \eqref{eq:seeded-reverse-bve}, and \eqref{eq:seeded-seed-assumption} are satisfied.
By \eqref{eq-range-acolumn}, we have
$s_{-}\leq \widehat A([n],j)\leq s_{+}$ for each $j\in [n]$.
Hence, \eqref{eq:seeded-column-bounds} is satisfied.
Similarly, we also have $\widehat  B([n],j)=1, \widehat  B(j,[n])\le b_+$ for each $j\in [n]$.
Moreover,
\[c\leq \widehat B(i,[n])\leq 1+\eta/2 < c^{-1},\qquad \forall\, i\in [n]\]
Hence, 
\[ c|I|\le\widehat  B(I,[n])\le c^{-1}|I|,\qquad \forall\, \text{nonempty } I\subseteq [n] \]
Thus, \eqref{eq:seeded-B-margins} and \eqref{eq:seeded-row-mass} are satisfied.
Furthermore, recall $\eta\in (0,1)$ and $c=1-\eta/2$. We have
\[
 c(1+\eta)=1+\frac{\eta(1-\eta)}2>1.
\]
For every $n\ge2$, we have $\lfloor3n/4\rfloor\ge n/2$.
Hence,
\begin{equation*}
 (1+c)(1+\eta)\lfloor qn\rfloor-n
 \ge \left(2 - \frac{\eta}{2}\right) \left(1 + \eta\right)\frac{n}{2} - n
 =\frac{(3\eta-\eta^2)n}{4}
 \ge\frac{\eta n}{2}>0.
\end{equation*}
Recall $q(1+\eta)<1$.
Hence, \eqref{eq:seeded-parameter-assumptions} is satisfied.
In summary, all hypotheses of \Cref{thm:seeded-multiscale} are now satisfied.
It follows that there are $C\ge1$ and $\beta>0$, depending only on
$K$ and $\Delta$, and a number $T\ge1$ such that
\begin{equation}\label{eqw-hatxhaty-boundedbyt}
 T\le\frac{2C}{\eta}(qn)^\beta,
 \qquad
 T^{-1}\le\widehat x_{i}\widehat y_{j}\le T,
 \qquad \forall i,j\in[n].
\end{equation}
Furthermore, by \eqref{eq-def-hatxhaty} we have
\[
\widehat x_{i}\widehat y_{j} = \frac{y_j}{y_i},\qquad \forall i,j\in[n].
\]
Finally, since $BY^{-1}\mathbf1=(x_1,\ldots,x_n)^\top$ by \eqref{eq-byinversex},
we have
\[
x_i y_j
=\sum_{k=1}^n B_{ik}\frac{y_j}{y_k}
=\sum_{k=1}^n B_{ik}\widehat x_k\widehat y_j, \qquad 
\forall i,j\in[n].
\]
By \eqref{eqw-hatxhaty-boundedbyt} and row-stochasticity of $B$,
\[
 T^{-1}\le x_i y_j\le T\qquad\forall i,j\in[n].
\]
Set $C_0=\max\{1,2Cq^\beta/\eta\}$.
This proves \eqref{eq-distrotion-t}.

It remains to verify $\sigma_{\perp}<1$ before applying the global
convergence theorem. Suppose instead that $s_2(B)=1$.
Choose a nonzero right singular vector $z\perp\mathbf1$ of $B$
with singular value $1$, and set $u=Bz$.
Double stochasticity gives $u\perp\mathbf1$ and
$\|u\|_2=\|z\|_2$. Also,
\[
 \sum_i\left(\sum_jB_{ij}z_j^2
                  -\left(\sum_jB_{ij}z_j\right)^2\right)
 =\|z\|_2^2-\|Bz\|_2^2=0.
\]
Each summand is a nonnegative weighted variance. Equality therefore
implies $z_j=u_i$ whenever $B_{ij}>0$.
Since $A$ and $B$ have the same support, it follows that
\[
 Az=u,\qquad A^\top u=Qz.
\]
Set $v=Q^{1/2}z\ne0$. Then
\begin{align*}
 Mv&=u,\qquad M^\top u=v,\\
 f^\top v
 &=\frac{\mathbf1^\top Qz}{\sqrt n}
 =\frac{\mathbf1^\top Az}{\sqrt n}\\
 &=\frac{\mathbf1^\top u}{\sqrt n}=0.
\end{align*}
Thus $M$ has a nontrivial singular value $1$, so $\sigma=1$.
This contradicts
$\Delta=1-\sqrt K\,\sigma\sigma_{\perp}>0$, because $K\ge1$.
Hence $\sigma_{\perp}<1$.

Finally, $\log T=O_{K,\Delta}(\log n)$, so
\Cref{thm:endpoint-gap-global-sinkhorn} gives the claimed number
of full column--row cycles. Its columnwise error bound
$\varepsilon/n$, together with exact row-stochasticity, implies
\eqref{eq-errorbound-maintheorem}.
\end{proof}

\subsection{Proof of \Cref{cor-k-sigmaa}}

\Cref{cor-k-sigmaa} is immediate by \Cref{thm:polynomial-scaling-gap-seeded} and the following lemma.

\begin{lemma}
\label{lem:input-spectral-strict-hall}
Let $n\ge2$ and let $A\in\mathbb R_{\ge0}^{n\times n}$
be row-stochastic with positive column sums. Define
\[
 K=\frac{\max_j c_j(A)}{\min_j c_j(A)},
 \qquad
 M=A\operatorname{diag}(c_1(A),\ldots,c_n(A))^{-1/2},
 \qquad
 \sigma=s_2(M).
\]
If $\sqrt K\,\sigma<1$, then
$|N_A(I)|>|I|$ for each $\varnothing\ne I\subsetneq[n]$.
Consequently, there exist positive diagonal matrices $X,Y$
such that $XAY$ is doubly stochastic.
\end{lemma}

The following lemma is used in the proof of \Cref{lem:input-spectral-strict-hall}.

\begin{lemma}
\label{prop:strict-hall}
Let $n\ge2$, $q\in(1/2,1)$, and $\eta,\phi>0$. Assume
$q(1+\eta)<1$.  Let $A\in\R_{\geq0}^{n\times n}$ be row-stochastic.  If
$A$ satisfies $(q,\eta,\phi)$-two-sided buffered robust vertex expansion,
then
\begin{equation}
 |N_A(I)|>|I| \qquad \forall\varnothing\neq I\subsetneq[n]. \label{eq:strict-hall}
\end{equation}
Consequently, 
there exist positive diagonal
matrices $X$ and $Y$ such that $XAY$ is doubly stochastic.
\end{lemma}

Now we can prove \Cref{lem:input-spectral-strict-hall}, which completes the proof of \Cref{cor-k-sigmaa}.
\begin{proof}
Fix $q=3/4$. Since
\[
 \lim_{\alpha\downarrow1}
 \sqrt{\frac{1-\alpha q}{K\alpha(1-q)}}
 =\frac1{\sqrt K}>\sigma,
\]
we may choose $\alpha\in(1,4/3)$ such that
\[
 \sigma<\theta_K
 \triangleq
 \sqrt{\frac{1-\alpha q}{K\alpha(1-q)}}.
\]
Set
\[
 \eta=\alpha-1,
 \qquad
 \delta=\frac12\left(1-\frac{\sigma}{\theta_K}\right).
\]
Then $\eta>0$, $q(1+\eta)=\alpha q<1$, and
$\delta\in(0,1)$. Moreover,
\[
 \sigma<(1-\delta)\theta_K
 =\frac{\theta_K+\sigma}{2}.
\]
Thus all hypotheses of
\Cref{thm:sharp-column-balanced-mixing-to-bve}
are satisfied. It follows that $A$ satisfies
$(q,\eta,\zeta)$-two-sided buffered robust vertex expansion
for some $\zeta>0$.
Applying \Cref{prop:strict-hall} gives
\[
 |N_A(I)|>|I|
 \qquad
 \forall\,\varnothing\ne I\subsetneq[n],
\]
and the existence of positive diagonal matrices $X,Y$
such that $XAY$ is doubly stochastic.
\end{proof}

\newpage

%% file: Lowerbounds.tex
\section{Lower bounds for general matrices.}\label{sec-lowerbounds}

In this section, we prove Propositions \ref{prop:product-threshold-counterexample}, \ref{prop:critical-spectral-counterexample}, \ref{prop:bounded-log-iterations} and \Cref{cor:fixed-gap-logradius-tight}.
The following corollary further shows that the framework of \Cref{prop:bounded-log-iterations} encompasses
several important matrix-scaling algorithms, including the
box-constrained Newton scaling method of \cite{cohen2017matrix},
\textsc{Scaling1}, and the two-stage implementation of
\textsc{Scaling2} followed by \textsc{Scaling3} in
\cite{allen2017much}.
Consequently, each of these algorithms requires $\Omega(n)$
outer iterations to achieve total marginal $\ell_1$ error at most
$n^{-2}$ on the family constructed in
\Cref{prop:product-threshold-counterexample}.

\begin{corollary}
\label{cor:three-scaling-algorithms}
Let $A_n$ be the family of positive matrices in Proposition~\ref{prop:product-threshold-counterexample}, and set
\[
R_n=\log\!\left(1+\frac{2^n}{3}\right)
    +\log\!\left(1-\frac{2+\sqrt2}{n^2}\right).
\]
Consider the following exact-arithmetic update rules with the
initializations specified below. Count one accepted update of the
logarithmic scaling vector as one outer iteration. Suppose that the
returned scaling $M$ satisfies $\Delta(M)\le n^{-2}$,
where $\Delta(M)$ denotes the total marginal $\ell_1$ error.
For all sufficiently large positive integers $n$ divisible by $12$:
\begin{enumerate}[label=(\roman*),leftmargin=*]
\item The box-constrained Newton scaling method in \cite{cohen2017matrix}, initialized at zero logarithmic row and column scalings, requires at least $R_n/2$ iterations.
\item Scaling1 in \cite{allen2017much}, initialized at zero logarithmic column scaling, requires at least $88R_n/5$ iterations.
\item The Scaling2 followed by Scaling3 implementation in \cite{allen2017much}, with Scaling2 initialized at zero and its output used as the initial vector of Scaling3, requires at least $8R_n/5$ iterations in total across both stages.

\end{enumerate}
In particular, all three outer-iteration counts are $\Omega(n)$.
\end{corollary}

These bounds count accepted outer updates from the stated initializations.
They exclude computations within the linear system, oracle, and
multiplicative weights subroutines. A running-time bound requires
separate accounting for these costs and for input precision.
For another initialization or update rule, the hypotheses of
Proposition~\ref{prop:bounded-log-iterations} must be checked separately.

\subsection{Proof of \Cref{prop:product-threshold-counterexample}}

\begin{proof}[Proof of \Cref{prop:product-threshold-counterexample}]
Write $n=12m$ and set
\begin{align}\label{eq-def-taupq}
 \tau=2^{-n},\qquad p=\frac23-\tau,\qquad q=\tau.
\end{align}
In particular, $0<q<p<1$ and $p+q=2/3$.
For $a,b\in(0,1)$, define the $6m\times6m$ matrix
\begin{equation}\label{eq:ptc-block-family}
 F(a,b)=
 \begin{pmatrix}
 \dfrac{a}{4m}\mathbf{1}_{3m\times4m}
 &\dfrac{1-a}{2m}\mathbf{1}_{3m\times2m}\\[4pt]
 \dfrac{b}{4m}\mathbf{1}_{3m\times4m}
 &\dfrac{1-b}{2m}\mathbf{1}_{3m\times2m}
 \end{pmatrix},
\end{equation}
where $\mathbf{1}_{r\times s}$ denotes the $r\times s$ all-ones matrix.
Thus the two row blocks have equal size $3m$, whereas the column blocks
have sizes $4m$ and $2m$.
We have $F(a,b)$ is row-stochastic for all $a,b\in (0,1)$.
Let
\[
 A_0=F(p,q),\qquad C_0=F(1-q,1-p),\qquad
 H=\frac12
 \begin{pmatrix}
 1+1/\sqrt2&1-1/\sqrt2\\
 1-1/\sqrt2&1+1/\sqrt2
 \end{pmatrix},
\]
and define
\begin{equation}\label{eq:ptc-construction}
 A=A_0\otimes H,\qquad C=C_0\otimes H.
\end{equation}
All these matrices are strictly positive and row-stochastic.
The matrix $H$ is doubly stochastic.

Recall $p+q=2/3$.
The column sums of $A_0$ on its two column blocks are
\[
 \frac34(p+q)= \frac12,
 \qquad \frac32(2-p-q) = 2.
\]
By  $A=A_0\otimes H$, $A$ is row-stochastic, and $H$ is doubly stochastic, we have
$A$ is
row-stochastic.
Since $H$ is doubly stochastic, all its column sums equal $1$.
Thus the column sums of $A=A_0\otimes H$ are those of $A_0$,
each repeated twice. 
Formally, 
\begin{align}\label{eq-relation-columnsum}
(A([n],1),\ldots,A([n],n)) = \left(A_0([6m],1),\cdots,A_0([6m],6m)\right)\otimes (1,1).
\end{align}
Thus, the column sums of $A$ are $1/2$ and $2$.
Therefore,
\begin{align}\label{eq-keqfour}
 K=\frac{\max_j A([n],j)}{\min_j A([n],j)}=4.
\end{align}
Similarly, 
The column sums of $C_0$ are
\[
 \frac34(2-p-q)=1,\qquad \frac32(p+q)=1.
\]
Consequently, $C_0$ is is doubly stochastic.
By  $C=C_0\otimes H$ and $H$ is doubly stochastic, we have $C$ is doubly stochastic.
Define the positive diagonal matrices
\begin{align*}
 X_0&=\operatorname{diag}\left(
 \frac{1-q}{p}I_{3m},\frac{1-p}{q}I_{3m}\right),\\
 Y_0&=\operatorname{diag}\left(
 I_{4m},\frac{pq}{(1-p)(1-q)}I_{2m}\right).
\end{align*}
A direct block multiplication gives $X_0A_0Y_0=C_0$.
Let $X=X_0\otimes I_2$ and $Y=Y_0\otimes I_2$.
By \eqref{eq:ptc-construction},
\begin{equation}
\begin{aligned}
C
&=C_0\otimes H=(X_0A_0Y_0)\otimes(I_2HI_2)\\
&=(X_0\otimes I_2)(A_0\otimes H)(Y_0\otimes I_2)
 =XAY.
\end{aligned}
\end{equation}
In this construction, we use the explicit factors $X$ and $Y$
without imposing $\min_i x_i=\min_j y_j$.
Since $A$ is strictly positive, $C$ is its unique doubly stochastic scaling.

Since $A=A_0\otimes H$ and $C=C_0\otimes H$, the common positive
factor from $H$ cancels in each ratio of corresponding entries.
Thus the ratios $C_{ij}/A_{ij}$ take exactly the same values as
the ratios $(C_0)_{rs}/(A_0)_{rs}$.
Hence, the four possible values of $C_{ij}/A_{ij}$ are
\begin{align}\label{eq-ratios}
 \frac{1-q}{p},\qquad \frac{q}{1-p},\qquad
 \frac{1-p}{q},\qquad \frac{p}{1-q}.
\end{align}
Recall that $0<q<p<1$ and $p+q=2/3<1$. In particular,
$1-p-q>0$. 
Thus,
\[
\frac{1-q}{p}
=1+\frac{1-p-q}{p}>1,
\qquad
\frac{1-p}{q}
=1+\frac{1-p-q}{q}>1.
\]
Furthermore,
\begin{align*}
\frac{1-p}{q}-\frac{1-q}{p}=\frac{p-p^2-q+q^2}{pq}=\frac{(p-q)(1-p-q)}{pq}>0.
\end{align*}
Consequently,
\[
\frac{1-p}{q}>\frac{1-q}{p}>1.
\]
It follows from \eqref{eq-ratios} that the maximum and minimum
values of $C_{ij}/A_{ij}$ are $(1-p)/q$ and $q/(1-p)$,
respectively. Consequently,
\[
\max_{i,j}\frac{C_{ij}}{A_{ij}}
=
\max_{i,j}\frac{A_{ij}}{C_{ij}}
=
\frac{1-p}{q}.
\]
Moreover, by \eqref{eq-def-taupq},
\[
 \frac{1-p}{q}=\frac{1/3+\tau}{\tau}=1+\frac{2^n}{3}.
\]
Hence, \eqref{eq:ptc-exact-distortion} is proved.

Let $Q_0=\operatorname{diag}\left(A_0([6m],1),\cdots,A_0([6m],6m)\right)$ and $B_0=A_0Q_0^{-1/2}$.
Define
\[
U=
\begin{pmatrix}
\dfrac{\mathbf1_{3m}}{\sqrt{3m}} & \mathbf0_{3m}\\[6pt]
\mathbf0_{3m} & \dfrac{\mathbf1_{3m}}{\sqrt{3m}}
\end{pmatrix},
\qquad
V=
\begin{pmatrix}
\dfrac{\mathbf1_{4m}}{\sqrt{4m}} & \mathbf0_{4m}\\[6pt]
\mathbf0_{2m} & \dfrac{\mathbf1_{2m}}{\sqrt{2m}}
\end{pmatrix}
\in\mathbb R^{6m\times2},
\]
where $\mathbf1_r$ and $\mathbf0_r$ denote the all-ones and
all-zeros column vectors in $\mathbb R^r$, respectively.
Thus $U^{\top}U=V^{\top}V=I_2$.
By construction, $U^\top U=V^\top V=I_2$.
Since $B_0$ and $C_0$ are constant on each of the four blocks,
\[
B_0=U\widehat B_0V^\top,
\qquad
C_0=U\widehat C_0V^\top,
\]
where
\[
\widehat B_0=U^\top B_0V
=
\begin{pmatrix}
\sqrt{3/2}\,p & \dfrac{\sqrt3}{2}(1-p)\\[4pt]
\sqrt{3/2}\,q & \dfrac{\sqrt3}{2}(1-q)
\end{pmatrix},
\qquad 
\widehat C_0=U^\top C_0V
=
\begin{pmatrix}
\dfrac{\sqrt3}{2}(1-q) & \sqrt{3/2}\,q\\[4pt]
\dfrac{\sqrt3}{2}(1-p) & \sqrt{3/2}\,p
\end{pmatrix}.
\]
In both $\widehat B_0$ and $\widehat C_0$, the entries in the
first and second columns equal the corresponding block constants
of the original matrix multiplied by $\sqrt{3m\cdot4m}$ and
$\sqrt{3m\cdot2m}$, respectively.

Since $V$ has two orthonormal columns,
$\dim\ker(V^\top)=6m-2$. 
For every $z\in\ker(V^\top)$, we have $V^\top z=0$, and hence
\[
B_0z=U\widehat B_0V^\top z=0,
\qquad
C_0z=U\widehat C_0V^\top z=0.
\]
On its orthogonal complement, the column space of $V$, we have
\[
B_0(Vx)=U\widehat B_0x,
\qquad
C_0(Vx)=U\widehat C_0x
\qquad \forall x\in\mathbb R^2.
\]
With respect to the orthonormal bases given by the columns
of $V$ and $U$, the maps
\[
B_0:\operatorname{col}(V)\longrightarrow\operatorname{col}(U),
\qquad
C_0:\operatorname{col}(V)\longrightarrow\operatorname{col}(U)
\]
are represented by $\widehat B_0$ and $\widehat C_0$,
respectively. Since singular values are invariant under
orthonormal changes of basis, these two maps have the same
singular values as $\widehat B_0$ and $\widehat C_0$,
respectively.
Consequently, the singular values of $B_0$ are the two singular
values of $\widehat B_0$ together with $6m-2$ additional zeros.
Similarly, the singular values of $C_0$ are the two singular
values of $\widehat C_0$ together with $6m-2$ additional zeros.
All singular values are counted with multiplicity.

By \Cref{lem-singlevalue} and $B_0=A_0Q_0^{-1/2}$, we have $s_1(B_0)=1$.
Since $C_0$ is doubly stochastic, we also have $s_1(C_0)=1$.
Moreover,
\[
 \det\widehat B_0=\det\widehat C_0
 =\frac{3}{2\sqrt2}(p-q)
 =\frac{1-3\tau}{\sqrt2}>0.
\]
The product of the singular values of a square matrix is the absolute
value of its determinant. Thus
\[
 s_2(B_0)=\frac{\det\widehat B_0}{s_1(B_0)}=\frac{1-3\tau}{\sqrt2}.
\]
Recall that
$Q_0=\operatorname{diag}\left(A_0([6m],1),\cdots,A_0([6m],6m)\right)$ and $Q=\operatorname{diag}(A([n],1),\ldots,A([n],n))$.
By \eqref{eq-relation-columnsum}
we have $Q=Q_0\otimes I_2$.
Using $B=AQ^{-1/2}$, $A=A_0\otimes H$ and $B_0=A_0Q_0^{-1/2}$, we have
\begin{align*}
B=AQ^{-1/2}=(A_0\otimes H)(Q_0^{-1/2}\otimes I_2)=(A_0Q_0^{-1/2})\otimes(HI_2)=(A_0Q_0^{-1/2})\otimes H=B_0\otimes H.
\end{align*}
The singular values of $H$ are $1$ and $1/\sqrt2$, so the four nonzero
singular values of $B$, in decreasing order, are
\begin{align}\label{eq-singlevalue-bc}
 1,\qquad \frac1{\sqrt2},\qquad
 \frac{1-3\tau}{\sqrt2},\qquad \frac{1-3\tau}{2}.
\end{align}
Similarly, we also have
\[
 s_2(C_0)=\frac{\det\widehat C_0}{s_1(C_0)}=\frac{1-3\tau}{\sqrt2}.
\]
Combined with $C = C_0\otimes H$, we obtain that the four nonzero
singular values of $C$ are the values in \eqref{eq-singlevalue-bc}.
Thus, we have 
$ s_2(B)=s_2(C)=1/\sqrt{2}$.
Therefore, \eqref{eq-keqfour} gives $\sqrt K\,s_2(B)s_2(C)=1$.
Hence, \eqref{eq:ptc-spectral-equality} is proved.

We claim that 
\[
 A^{(2t)}=F(p_t,q_t)\otimes H,
 \qquad \qquad 0<p_t,q_t<1,
\]
where
\begin{equation}\label{eq:ptc-sk-recurrence}
p_{t+1}=f_{\ell_t}(p_t),\qquad q_{t+1}=f_{\ell_t}(q_t),\qquad \ell_t=\frac{2(2-(p_t+q_t))}{p_t+q_t},\qquad
 f_\ell(u)=\frac{\ell u}{1-u+\ell u}.
\end{equation}
For $t=0$, the claim follows from
\[
A^{(0)}=A=F(p,q)\otimes H,
\qquad p_0=p,\quad q_0=q.
\]
Suppose that the claim holds for some $t\ge0$.
The column sums of $F(p_t,q_t)$ on its first and second
column blocks are, respectively,
\begin{equation}
\begin{aligned}\label{eq-def-donetdtwot}
d_{1,t}
&\triangleq 3m\frac{p_t}{4m}+3m\frac{q_t}{4m}
=\frac34(p_t+q_t),\\
d_{2,t}
&\triangleq 3m\frac{1-p_t}{2m}+3m\frac{1-q_t}{2m}
=\frac32(2-p_t-q_t).
\end{aligned}
\end{equation}
Both $d_{1,t}$ and $d_{2,t}$ are positive because $0<p_t,q_t<1$.
Set
\[
D_t=\operatorname{diag}
\left(d_{1,t}I_{4m},\,d_{2,t}I_{2m}\right).
\]
Since $H$ is column-stochastic, the column sums of
$A^{(2t)}=F(p_t,q_t)\otimes H$ are those of $F(p_t,q_t)$,
each repeated twice. 
Thus column normalization gives
\[
A^{(2t+1)}
=A^{(2t)}(D_t^{-1}\otimes I_2)
=G_t\otimes H,
\]
where
\[
G_t=F(p_t,q_t)D_t^{-1}
=
\begin{pmatrix}
\dfrac{p_t}{4m\,d_{1,t}}\mathbf1_{3m\times4m}
&
\dfrac{1-p_t}{2m\,d_{2,t}}\mathbf1_{3m\times2m}\\[6pt]
\dfrac{q_t}{4m\,d_{1,t}}\mathbf1_{3m\times4m}
&
\dfrac{1-q_t}{2m\,d_{2,t}}\mathbf1_{3m\times2m}
\end{pmatrix}.
\]

The row sums of $G_t$ on its first and second row blocks are
\begin{align*}
r_{1,t}
&=4m\frac{p_t}{4m\,d_{1,t}}
  +2m\frac{1-p_t}{2m\,d_{2,t}}
=\frac{p_t}{d_{1,t}}+\frac{1-p_t}{d_{2,t}},\\
r_{2,t}
&=4m\frac{q_t}{4m\,d_{1,t}}
  +2m\frac{1-q_t}{2m\,d_{2,t}}
=\frac{q_t}{d_{1,t}}+\frac{1-q_t}{d_{2,t}}.
\end{align*}
Set
\[
E_t=\operatorname{diag}
\left(r_{1,t}I_{3m},\,r_{2,t}I_{3m}\right).
\]
Since $H$ is row-stochastic, the row sums of
$A^{(2t+1)}=G_t\otimes H$ are those of $G_t$,
each repeated twice. Consequently, the subsequent row
normalization gives
\[
A^{(2(t+1))}
=(E_t^{-1}\otimes I_2)A^{(2t+1)}
=(E_t^{-1}G_t)\otimes H.
\]

Define
\[
p_{t+1}=\frac{p_t/d_{1,t}}{r_{1,t}},
\qquad
q_{t+1}=\frac{q_t/d_{1,t}}{r_{2,t}}.
\]
The definitions of $r_{1,t}$ and $r_{2,t}$ imply
\[
1-p_{t+1}
=\frac{(1-p_t)/d_{2,t}}{r_{1,t}},
\qquad
1-q_{t+1}
=\frac{(1-q_t)/d_{2,t}}{r_{2,t}}.
\]
All four quantities are strictly positive. 
Hence, it is easy to verify that
\[
E_t^{-1}G_t
=
\begin{pmatrix}
\dfrac{p_{t+1}}{4m}\mathbf1_{3m\times4m}
&
\dfrac{1-p_{t+1}}{2m}\mathbf1_{3m\times2m}\\[6pt]
\dfrac{q_{t+1}}{4m}\mathbf1_{3m\times4m}
&
\dfrac{1-q_{t+1}}{2m}\mathbf1_{3m\times2m}
\end{pmatrix}
=F(p_{t+1},q_{t+1}),
\]
with $0<p_{t+1},q_{t+1}<1$. Hence
\[
A^{(2(t+1))}=F(p_{t+1},q_{t+1})\otimes H.
\]

Finally, by \eqref{eq:ptc-sk-recurrence} and \eqref{eq-def-donetdtwot}, we obtain
$\ell_t=d_{2,t}/d_{1,t}$.
Combined with \eqref{eq:ptc-sk-recurrence}, we have
\begin{align*}
p_{t+1}=\frac{p_t/d_{1,t}}
{p_t/d_{1,t}+(1-p_t)/d_{2,t}}=\frac{d_{2,t}p_t}
{d_{2,t}p_t+d_{1,t}(1-p_t)}=\frac{\ell_t p_t}{1-p_t+\ell_t p_t}
=f_{\ell_t}(p_t).
\end{align*}
Similarly,
\[
q_{t+1}
=\frac{q_t/d_{1,t}}
{q_t/d_{1,t}+(1-q_t)/d_{2,t}}
=\frac{\ell_t q_t}{1-q_t+\ell_t q_t}
=f_{\ell_t}(q_t).
\]
This completes the induction.

Furthermore, we claim that, for every $t\ge0$,
\begin{equation}\label{eq:ptc-invariant-interval}
\frac23\le p_t+q_t<\frac43,
\qquad
1<\ell_t\le4.
\end{equation}
We prove this by induction. Initially, $p_0+q_0=2/3$, and
\[
\ell_0=\frac{2(2-(p_0+q_0))}{p_0+q_0}=4.
\]
Hence the claimed bounds hold initially.
Suppose that the bounds hold at time $t$.
Since $\ell_t>1$, the function $f_{\ell_t}$ is strictly
concave on $(0,1)$ and satisfies $f_{\ell_t}(u)>u$ there.
Recall $0<q_t<p_t<1$.
we obtain
\[
p_{t+1}+q_{t+1}
=f_{\ell_t}(p_t)+f_{\ell_t}(q_t)
>p_t+q_t
\ge\frac23.
\]
On the other hand, strict concavity of $f_{\ell_t}$ and $p_t\ne q_t$ give
\begin{align*}
\frac{p_{t+1}+q_{t+1}}2=\frac{f_{\ell_t}(p_t)+f_{\ell_t}(q_t)}2<f_{\ell_t}\left(\frac{p_t+q_t}{2}\right)=\frac{2-(p_t+q_t)}{3-\frac32(p_t+q_t)}=\frac23,
\end{align*}
where the second equality uses
$\ell_t=2(2-(p_t+q_t))/(p_t+q_t)$.
Thus $2/3\le p_{t+1}+q_{t+1}<4/3$, which also implies
\[
1<\ell_{t+1}
=\frac{4}{p_{t+1}+q_{t+1}}-2
\le4.
\]
This completes the induction.
Hence, \eqref{eq:ptc-invariant-interval} is proved.

Recall that the first $4m$ columns of $F(p_t,q_t)$ have sum
$d_{1,t}$, while the remaining $2m$ columns have sum $d_{2,t}$.
Each column sum is repeated twice in $A^{(2t)}$.
Thus, since $n=12m$, the first $8m=2n/3$ columns of $A^{(2t)}$
have sum $d_{1,t}$, and the remaining $4m=n/3$ columns have sum
$d_{2,t}$.
Therefore,
\begin{align}\label{eq-sumerrorofat}
\sum_{j=1}^n\left|A^{(2t)}([n],j)-1\right| =  \frac{2n}{3}\left|d_{1,t} - 1\right|+ \frac{n}{3}\left|d_{2,t} - 1\right|.
\end{align}
Moreover,
by \eqref{eq-def-donetdtwot} and \eqref{eq:ptc-invariant-interval},
we have
\[
\frac12\le d_{1,t}<1,
\qquad
1<d_{2,t}\le2.
\]
Therefore,
\[
|d_{1,t}-1|=1-d_{1,t},
\qquad
|d_{2,t}-1|=d_{2,t}-1.
\]
In addition, \eqref{eq-def-donetdtwot} gives
\begin{align}\label{eq-relation-dtwominusone-doneminusone}
d_{2,t}-1=\frac32\bigl(2-(p_t+q_t)\bigr)-1=2-\frac32(p_t+q_t)=2\left(1-\frac34(p_t+q_t)\right)=2(1-d_{1,t}).
\end{align}
Thus,
\begin{align}\label{eq-maxerrorat-relation}
\max_{j\in [n]}\left|A^{(2t)}([n],j)-1\right|=\max\{1-d_{1,t},\,d_{2,t}-1\}=d_{2,t}-1=2-\frac32(p_t+q_t).
\end{align}
Moreover, by \eqref{eq-sumerrorofat} and \eqref{eq-relation-dtwominusone-doneminusone}, we obtain
\begin{equation}
\begin{aligned}\label{eq-sumerrorofat-1}
\sum_{j=1}^n\left|A^{(2t)}([n],j)-1\right| &= \frac{2n}{3}(1-d_{1,t})
  +\frac n3(d_{2,t}-1)=\frac{2n}{3}\cdot\frac{d_{2,t}-1}{2}
  +\frac n3(d_{2,t}-1)
\\&=\frac{2n}{3}(d_{2,t}-1).
\end{aligned}
\end{equation}

By \eqref{eq:ptc-sk-recurrence},
\[
q_{t+1}
=\frac{\ell_t q_t}{1-q_t+\ell_t q_t}.
\]
Consequently,
\begin{align*}
1-q_{t+1}=1-\frac{\ell_t q_t}{1-q_t+\ell_t q_t}=\frac{1-q_t+\ell_t q_t-\ell_t q_t}
{1-q_t+\ell_t q_t}=\frac{1-q_t}{1-q_t+\ell_t q_t}.
\end{align*}
Since $0<q_{t+1}<1$, we have $1-q_{t+1}>0$.
Hence,
\begin{align*}
\frac{q_{t+1}}{1-q_{t+1}}=
\frac{\dfrac{\ell_t q_t}{1-q_t+\ell_t q_t}}
{\dfrac{1-q_t}{1-q_t+\ell_t q_t}}=\ell_t\frac{q_t}{1-q_t}.
\end{align*}
Iterating this identity from time $0$ to time $t-1$, we obtain
\[
\frac{q_t}{1-q_t}
=
\left(\prod_{s=0}^{t-1}\ell_s\right)
\frac{q_0}{1-q_0},\qquad \forall t\ge1.
\]
By \eqref{eq:ptc-invariant-interval},
\[
\prod_{s=0}^{t-1}\ell_s\le4^t.
\]
Using $q_0=q = \tau$, we conclude that
\[
\frac{q_t}{1-q_t}
\le4^t\frac{q_0}{1-q_0}
=4^t\frac{\tau}{1-\tau}.
\]
One can verify that
the same inequality holds with equality at $t=0$.
Hence, 
\begin{align}\label{eq-qtoveroneminusqtlefourttauoveroneminustau}
 \frac{q_t}{1-q_t}\le4^t\frac{\tau}{1-\tau},\qquad \forall t\ge 0.
\end{align}
Given that $A^{(2t)}$ is row-stochastic, and that
\[\sum_{j\in [n]}\left|A^{(2t)}([n],j)-1\right| + \sum_{i\in [n]}\left|A^{(2t)}(i,[n])-1\right|\le n^{-2},\]
we have 
\[
\sum_{j\in [n]}\left|A^{(2t)}([n],j)-1\right|\leq  n^{-2}.\]
By \eqref{eq-sumerrorofat-1}, we have
\[
d_{2,t}-1
=\frac{3}{2n}
\sum_{j=1}^n\left|A^{(2t)}([n],j)-1\right|
\le\frac{3}{2n^3}.
\]
By \eqref{eq-def-donetdtwot} and $p_t\le1$, we obtain
\begin{align*}
q_t=2-p_t-\frac23d_{2,t}\ge1-\frac23d_{2,t}=\frac13-\frac23(d_{2,t}-1)\ge\frac13-\frac1{n^3}\ge\frac16,
\end{align*}
where the last inequality uses $n\ge12$.
Since $0<q_t<1$, it follows that
\[
\frac{q_t}{1-q_t}
\ge\frac{1/6}{1-1/6}
=\frac15.
\]
Combining this \eqref{eq-qtoveroneminusqtlefourttauoveroneminustau},
we obtain
\[
4^t\frac{\tau}{1-\tau}\ge\frac15.
\]
Since $\tau=2^{-n}$, this implies
\[
4^t
\ge\frac{1-\tau}{5\tau}
=\frac{2^n-1}{5}.
\]
Taking logarithms to base $4$, we conclude that
\[
t
\ge\log_4\left(\frac{2^n-1}{5}\right)
\ge\frac n2-\log_4 10,
\]
where the last inequality follows from $2^n-1\ge2^{n-1}$.
We next consider the intervening odd iterations. Set
$s_t=p_t+q_t$. By the formulas for $r_{1,t}$ and $r_{2,t}$ above,
\[
 r_{1,t}-1
 =\frac{(p_t-q_t)(4-3s_t)}{3s_t(2-s_t)},
 \qquad
 r_{2,t}-1=-(r_{1,t}-1).
\]
The matrix $A^{(2t+1)}=G_t\otimes H$ is column-stochastic.
Each of its two row sums is repeated $n/2$ times. Hence its total
marginal error is
\[
 \begin{aligned}
 &\sum_{j=1}^n\left|A^{(2t+1)}([n],j)-1\right|
 +\sum_{i=1}^n\left|A^{(2t+1)}(i,[n])-1\right|\\
 &\qquad=n\left|r_{1,t}-1\right|.
 \end{aligned}
\]
Suppose that $q_t\le1/6$. Since $\ell_t>1$, the recurrence gives
$p_t\ge p_0=2/3-\tau$. Thus, using $n\ge12$,
\[
 p_t-q_t\ge\frac12-\tau>\frac13,
 \qquad
 s_t<\frac76,
 \qquad
 4-3s_t>\frac12.
\]
Moreover, $0<s_t<2$ implies
$0<3s_t(2-s_t)\le3$. It follows that
\[
 n\left|r_{1,t}-1\right|\ge\frac{n}{18}>n^{-2}.
\]
Consequently, an odd iterate with total marginal error at most
$n^{-2}$ must satisfy $q_t>1/6$. Applying
\eqref{eq-qtoveroneminusqtlefourttauoveroneminustau} then gives
\[
 4^t\frac{\tau}{1-\tau}
 \ge\frac{q_t}{1-q_t}>\frac15,
 \qquad
 t>\log_4\!\left(\frac{2^n-1}{5}\right)=\Omega(n).
\]
Thus the same asymptotic lower bound holds for odd iterations.
In summary,
$T(n^{-2})=\Omega(n)$.
The proposition is proved.
\end{proof}

\subsection{Proof of \Cref{cor:fixed-gap-logradius-tight}}

\begin{proof}[Proof of \Cref{cor:fixed-gap-logradius-tight}]
The spectral identity follows from
\Cref{prop:product-threshold-counterexample}.
Since $A_n$ is strictly positive, every feasible pair $(u,v)$
representing $C_n$ satisfies
\[
 u_i-v_j=\log\frac{(C_n)_{ij}}{(A_n)_{ij}}
 \qquad\text{for all }i,j.
\]
The two exact distortion identities in that proposition therefore give
\[
 \max_{i,j}|u_i-v_j|=L_n.
\]
It follows that $L_n\le2\max\{\|u\|_\infty,\|v\|_\infty\}$
for every feasible pair, proving $D_n\ge L_n/2$.

For the upper bound, take any feasible pair $(u,v)$ and consider the
$2n$ coordinates of $(u,v)$. Coordinates from opposite groups differ
by at most $L_n$. Coordinates in the same group differ by at most
$2L_n$, by comparison with any coordinate from the other group.
Thus the diameter of all $2n$ coordinates is at most $2L_n$.
If $m$ and $M$ are their minimum and maximum, replace both vectors by
\[
 \widetilde u=u-\frac{M+m}{2}\mathbf1,
 \qquad
 \widetilde v=v-\frac{M+m}{2}\mathbf1.
\]
This common shift preserves $u_i-v_j$ and hence preserves $C_n$,
while
\[
 \max\{\|\widetilde u\|_\infty,
         \|\widetilde v\|_\infty\}
 =\frac{M-m}{2}\le L_n.
\]
This centering is used only to bound the minimum defining $D_n$;
actual diagonal factors can be returned to the usual reciprocal
normalization without changing any products.
Therefore $D_n\le L_n$. Since $L_n=\Theta(n)$, we obtain
$D_n=\Theta(n)$.

Applying \Cref{thm:global-sk-logradius-gap} with
$D_n=\Theta(n)$, $g=1/2$, and $\varepsilon=n^{-2}$
gives $O(n)$ full column--row cycles.
Each cycle consists of two normalization steps, so
$T(n^{-2})=O(n)$.
Combining this with the lower bound
$T(n^{-2})=\Omega(n)$ from
\Cref{prop:product-threshold-counterexample} proves
\eqref{eq:fixed-gap-logradius-matching-time}.
\end{proof}



\subsection{Proof of \Cref{prop:critical-spectral-counterexample}}

The following lemma is used in the proof of \Cref{prop:critical-spectral-counterexample}.

\begin{lemma}\label{lem-singlevalue}
Let $A\in\mathbb{R}_{\ge0}^{n\times n}$ be row-stochastic
with positive column sums, and define
\[
 B=A\operatorname{diag}\bigl(A([n],1),\ldots,A([n],n)\bigr)^{-1/2}.
\]
Then $s_1(B) = 1$. 
\end{lemma}
\begin{proof}
For any $z=(z_1,\cdots,z_n)^\top\in\mathbb R^n$, we have
\[
(Bz)_i=\sum_{j=1}^n A_{ij}\frac{z_j}{\sqrt{A([n],j)}}.
\]
Since $A_{ij}\ge0$, the Cauchy--Schwarz inequality gives,
for each $i\in[n]$,
\begin{align*}
\left(\sum_{j=1}^n A_{ij}\frac{z_j}{\sqrt{A([n],j)}}\right)^2
&=
\left(
\sum_{j=1}^n
\sqrt{A_{ij}}\,
\frac{\sqrt{A_{ij}}z_j}{\sqrt{A([n],j)}}
\right)^2\le
\left(\sum_{j=1}^n A_{ij}\right)
\left(\sum_{j=1}^n A_{ij}\frac{z_j^2}{A([n],j)}\right)=
\sum_{j=1}^n A_{ij}\frac{z_j^2}{A([n],j)},
\end{align*}
where the last equality uses the row-stochasticity of $A$.
Summing over $i$ and using $\sum_{i=1}^n A_{ij}=A([n],j)$, we obtain
\begin{align*}
\|Bz\|_2^2&=\sum_{i=1}^n
\left(\sum_{j=1}^n A_{ij}\frac{z_j}{\sqrt{A([n],j)}}\right)^2\le\sum_{i=1}^n\sum_{j=1}^n A_{ij}\frac{z_j^2}{A([n],j)}=\sum_{j=1}^n\frac{z_j^2}{A([n],j)}
\left(\sum_{i=1}^n A_{ij}\right)
\\&=\sum_{j=1}^n\frac{z_j^2A([n],j)}{A([n],j)}=\sum_{j=1}^nz_j^2
=\|z\|_2^2.
\end{align*}
Thus $s_1(B)\leq 1$.
Moreover, let $Q = \diag\left(A([n],1),\cdots,A([n],n)\right)$.
By $B = AQ^{-1/2}$ and $A$ is row-stochastic, we obtain
\[
B\left(\sqrt{A([n],1)},\cdots,\sqrt{A([n],n)}\right)^{\top}= AQ^{-1/2}Q^{1/2}\mathbf1 = \mathbf1.
\]
Moreover, by $A$ is row-stochastic, we have 
\begin{align}
\left\|\left(\sqrt{A([n],1)},\cdots,\sqrt{A([n],n)}\right)^{\top}\right\|^2_2  = \sum_{j\in [n]}\left(\sqrt{A([n],j)}\right)^2 = \sum_{j\in [n]}A([n],j) = \sum_{i\in [n]}A(i,[n]) = n.
\end{align}
Thus, 
\[s_1(B)
\ge\frac{\left\|B\left(\sqrt{A([n],1)},\cdots,\sqrt{A([n],n)}\right)^{\top}\right\|_2}{\left\|\left(\sqrt{A([n],1)},\cdots,\sqrt{A([n],n)}\right)^{\top}\right\|_2}
=\frac{\left\|\mathbf1\right\|_2}{\sqrt n}
=1.\]
Combined with $s_1(B)\leq 1$, we have $s_1(B)=1$.
\end{proof}

Now we can prove \Cref{prop:critical-spectral-counterexample}.
\begin{proof}[Proof of \Cref{prop:critical-spectral-counterexample}]
Write $n=4m$, and set $\varepsilon=2^{-n}$ and $\rho=K^{-1/2}$.
Define
\[
 P=\frac{1}{K+1}
 \begin{pmatrix}
 K+1-\varepsilon^2 & \varepsilon^2\\
 K-1+\varepsilon^2 & 2-\varepsilon^2
 \end{pmatrix},
 \qquad
 H=\frac12
 \begin{pmatrix}
 1+\rho & 1-\rho\\
 1-\rho & 1+\rho
 \end{pmatrix},
\]
and let
\[
 U_m=\frac{1}{m}\mathbf{1}_m\mathbf{1}_m^{\top},
 \qquad A=P\otimes H\otimes U_m.
\]
All three factors are strictly positive and row-stochastic, while $H$
and $U_m$ are also column-stochastic. Consequently, $A$ is also strictly positive
and row-stochastic.

\noindent\underline{Proof of \Cref{first-prop-critical-spectral-counterexample}.}
The column sums of $P$ are
\[
 d_1=\frac{2K}{K+1},\qquad d_2=\frac{2}{K+1}.
\]
Thus the first $n/2$ columns of $A$ have sum $d_1$, and the remaining
$n/2$ columns have sum $d_2$. In particular, their ratio is exactly $K$.
Thus, \eqref{eq-k-exactratio} is proved.

\medskip
\noindent\underline{Proof of \Cref{second-prop-critical-spectral-counterexample}.}
Let
\[
 D=\operatorname{diag}(d_1,d_2),\qquad M=PD^{-1/2}.
\]
Then $\operatorname{diag}(A([n],1),\ldots,A([n],n))=D\otimes I_{2m}$.
We obtain
\begin{align*}
B
&=A\operatorname{diag}\bigl(A([n],1),\ldots,A([n],n)\bigr)^{-1/2}\\
&=\bigl(P\otimes(H\otimes U_m)\bigr)
   (D^{-1/2}\otimes I_{2m})\\
&=(PD^{-1/2})\otimes\bigl((H\otimes U_m)I_{2m}\bigr)\\
&=M\otimes H\otimes U_m.
\end{align*}
By \Cref{lem-singlevalue} and $M=PD^{-1/2}$, we have $s_1(M)=1$.
Since the product of the two singular values of $M$ equals
$|\det M|$ and $s_1(M)=1$, we obtain
\[
 s_2(M)
 =\frac{|\det P|}{\sqrt{d_1d_2}}
 =\frac{2(1-\varepsilon^2)/(K+1)}{2\sqrt{K}/(K+1)}
 =(1-\varepsilon^2)\rho.
\]
The singular values of $H$ are $1,\rho$, and the only nonzero singular
value of $U_m$ is $1$. Therefore, by $B=M\otimes H\otimes U_m$,
the four nonzero singular values of
$B$, in decreasing order, are
\[
 1,\quad \rho,\quad (1-\varepsilon^2)\rho,
 \quad (1-\varepsilon^2)\rho^2.
\]
In particular, $s_2(B)=\rho=K^{-1/2}$ exactly.
Hence, \eqref{eq-stwobk-exactratio} is proved.

\medskip
\noindent\underline{Proof of \Cref{third-prop-critical-spectral-counterexample}.}
Set
\begin{align}\label{eq-defsh}
 h=\sqrt{\frac{P_{12}P_{21}}{P_{11}P_{22}}}
 =\varepsilon
 \sqrt{\frac{K-1+\varepsilon^2}
 {(K+1-\varepsilon^2)(2-\varepsilon^2)}},
 \qquad
 S=\frac{1}{1+h}
 \begin{pmatrix}1&h\\h&1\end{pmatrix}.
\end{align}
Also define
\[
 L=\frac{1}{(1+h)}\operatorname{diag}\left(
 \frac{1}{P_{11}},\frac{h}{P_{21}}\right),
 \qquad
 R=\operatorname{diag}\left(1,\frac{hP_{11}}{P_{12}}\right).
\]
We have 
\[
\begin{aligned}
LPR
&=\frac1{1+h}
\begin{pmatrix}
\dfrac{P_{11}}{P_{11}}&
\dfrac{P_{12}}{P_{11}}\cdot\dfrac{hP_{11}}{P_{12}}\\[8pt]
\dfrac{hP_{21}}{P_{21}}&
\dfrac{hP_{22}}{P_{21}}\cdot\dfrac{hP_{11}}{P_{12}}
\end{pmatrix}\\
&=\frac1{1+h}
\begin{pmatrix}
1&h\\[4pt]
h&\dfrac{h^2P_{11}P_{22}}{P_{12}P_{21}}
\end{pmatrix}.
\end{aligned}
\]
By the definitions of $P$ and $h$, one can also verify
$h^2P_{11}P_{22}=P_{12}P_{21}$.
Hence,
\[
\begin{aligned}
LPR=
\frac1{1+h}
\begin{pmatrix}
1&h\\[4pt]
h&1
\end{pmatrix} =S.
\end{aligned}
\]
Hence, with $X=L\otimes I_{2m}$ and $Y=R\otimes I_{2m}$, we have
\begin{equation}\label{eq-ceqshu}
\begin{aligned}
C=XAY
&=(L\otimes I_{2m})(P\otimes H\otimes U_m)
  (R\otimes I_{2m})\\
&=(LPR)\otimes(H\otimes U_m)\\
&=S\otimes H\otimes U_m.
\end{aligned}
\end{equation}
In this construction, we use the explicit factors $X$ and $Y$
without imposing $\min_i x_i=\min_j y_j$.
Moreover, one can verify that $S$ is doubly stochastic.
Recall that $H$ and $U_m$ are doubly stochastic,
we have $C$ is doubly stochastic. 
Since $A$ is strictly positive, its doubly
stochastic scaling is unique.
Hence, $C$ is the unique doubly
stochastic scaling of $A$.

By $A=P\otimes H\otimes U_m$ and $C=S\otimes H\otimes U_m$,
we have
\[
 A=\begin{pmatrix}P_{11}\left(H\otimes U_m\right)&P_{12}\left(H\otimes U_m\right)\\P_{21}\left(H\otimes U_m\right)&P_{22}\left(H\otimes U_m\right)\end{pmatrix},
 \qquad
 C=\frac{1}{1+h}\begin{pmatrix}H\otimes U_m&h\left(H\otimes U_m\right)\\h\left(H\otimes U_m\right)&H\otimes U_m\end{pmatrix}.
\]
Recall that $A$ is positive.
Thus the ratio $C_{ij}/A_{ij}$ is constant on each of the four blocks,
with respective values
\[
 r_{11}=\frac{1}{(1+h)P_{11}},\quad
 r_{12}=\frac{h}{(1+h)P_{12}},\quad
 r_{21}=\frac{h}{(1+h)P_{21}},\quad
 r_{22}=\frac{1}{(1+h)P_{22}}.
\]
By $P_{21} = (K-1+\varepsilon^2)/(K+1) > \varepsilon^2/(K+1) = P_{12}$, we have
$r_{21}< r_{12}$.
Similarly, we also have $P_{11}>P_{22}$.
Hence, $r_{11}< r_{22}$.
Moreover,  
\begin{align}\label{eq-r12r22-ratio}
\frac{r_{12}}{r_{22}} = \frac{P_{22}h}{P_{12}} = \frac{P_{22}}{P_{12}}\cdot \sqrt{\frac{P_{12}P_{21}}{P_{11}P_{22}}} = \sqrt{\frac{P_{21}P_{22}}{P_{11}P_{12}}} = \sqrt{\frac{(K-1+\varepsilon^2)(2-\varepsilon^2)}{(K+1-\varepsilon^2)\varepsilon^2}}.
\end{align}
Since
\begin{align}\label{eq-kminusoneepsilonsquaretwominuisepsilonsquarelarger}
(K-1+\varepsilon^2)(2-\varepsilon^2)
-(K+1-\varepsilon^2)\varepsilon^2=2(K-1)(1-\varepsilon^2)>0,
\end{align}
\eqref{eq-r12r22-ratio} gives $r_{12}>r_{22}$. 
In summary, $r_{12}>\max\{r_{11},r_{21},r_{22}\}$.
Similarly,
\begin{align}
\frac{r_{21}}{r_{11}} = \frac{P_{11}h}{P_{21}} = \frac{P_{11}}{P_{21}}\cdot \sqrt{\frac{P_{12}P_{21}}{P_{11}P_{22}}} = \sqrt{\frac{P_{11}P_{12}}{P_{21}P_{22}}} = \sqrt{\frac{(K+1-\varepsilon^2)\varepsilon^2}{(K-1+\varepsilon^2)(2-\varepsilon^2)}}.
\end{align}
By \eqref{eq-kminusoneepsilonsquaretwominuisepsilonsquarelarger},
We have $r_{21}<r_{11}$. 
In summary, $r_{21}<\min\{r_{11},r_{12},r_{22}\}$.
Consequently, 
\[
 \max_{i,j}\frac{C_{ij}}{A_{ij}}= r_{12}
 =\frac{h}{(1+h)P_{12}},
 \qquad
 \max_{i,j}\frac{A_{ij}}{C_{ij}} = \frac{1}{r_{21}}
 =\frac{(1+h)P_{21}}{h}.
\]

Fix $K>1$. 
By \eqref{eq-defsh} and $\varepsilon=2^{-n}$, we have
\[
\lim_{n\rightarrow \infty} \frac{h}{\varepsilon} =\lim_{n\rightarrow \infty}
 \sqrt{\frac{K-1+\varepsilon^2}
 {(K+1-\varepsilon^2)(2-\varepsilon^2)}}=\sqrt{\frac{K-1}{2(K+1)}}>0.
\]
Using $P_{12}=\varepsilon^2/(K+1)$,
we have 
\[
 \lim_{n\rightarrow \infty} \varepsilon\max_{i,j}\frac{C_{ij}}{A_{ij}}
 =\lim_{n\rightarrow \infty}\frac{\varepsilon h}{(1+h)P_{12}} = \lim_{n\rightarrow \infty}\frac{(K+1) h}{(1+h)\varepsilon} = \sqrt{\frac{K^2-1}{2}}.
\]
Similarly, using $P_{21}=(K-1+\varepsilon^2)/(K+1)$, we have
\begin{align*}
 \lim_{n\to\infty}\varepsilon\max_{i,j}\frac{A_{ij}}{C_{ij}}
 &=\lim_{n\to\infty}\frac{\varepsilon(1+h)P_{21}}{h}\\
 &=\lim_{n\to\infty}
   \frac{\varepsilon(1+h)(K-1+\varepsilon^2)}{(K+1)h}\\
 &=\sqrt{\frac{2(K-1)}{K+1}}.
\end{align*}
By $\varepsilon^{-1}=2^n$,
we obtain
\begin{align*}
 \max_{i,j}\frac{C_{ij}}{A_{ij}}
 &=\left(\sqrt{\frac{K^2-1}{2}}+o_K(1)\right)2^n,\\
 \max_{i,j}\frac{A_{ij}}{C_{ij}}
 &=\left(\sqrt{\frac{2(K-1)}{K+1}}+o_K(1)\right)2^n.
\end{align*}
Hence, \eqref{eq-cvsa-exactratio} is proved.

\medskip
\noindent\underline{Proof of \Cref{fourth-prop-critical-spectral-counterexample}.}
Define 
$F\triangleq H\otimes U_m.$
Thus, $F$ is a doubly stochastic matrix of order $n/2$ and  $A=P\otimes F$.
For a strictly positive square matrix $Z$, let
\[
 \mathcal{R}(Z)
 \triangleq \operatorname{diag}(Z\mathbf{1})^{-1}Z,
 \qquad
 \mathcal{C}(Z)
 \triangleq Z\operatorname{diag}(Z^\top\mathbf{1})^{-1}
\]
denote its row and column normalizations, respectively.
Then we have
\[
 A^{(0)}=A,\qquad
 A^{(2k+1)}=\mathcal{C}\left(A^{(2k)}\right),\qquad
 A^{(2k+2)}=\mathcal{R}\left(A^{(2k+1)}\right)
 \qquad \forall\, k\ge0.
\]

Since every row and column sum of $F$ equals $1$, for every strictly
positive $2\times2$ matrix $Z$ we have
\[
 \mathcal{R}(Z\otimes F)=\mathcal{R}(Z)\otimes F,
 \qquad
 \mathcal{C}(Z\otimes F)=\mathcal{C}(Z)\otimes F.
\]
Indeed, each row sum of $Z$ is repeated $n/2$ times among the row
sums of $Z\otimes F$, and the same holds for column sums.
Consequently, the iterates have the form
\[
 A^{(2k)}=P_k\otimes F,
 \qquad
 A^{(2k+1)}=\widehat P_k\otimes F,
\]
where
\[
 P_0=P,\qquad
 \widehat P_k=\mathcal{C}(P_k),\qquad
 P_{k+1}=\mathcal{R}(\widehat P_k).
\]
The same repetition of row and column sums also gives
\begin{equation}\label{eq:critical-error-tensor}
 \begin{aligned}
 &\sum_{i=1}^{n}\left|r_i(Z\otimes F)-1\right|
 +
 \sum_{j=1}^{n}\left|c_j(Z\otimes F)-1\right|=
 \frac n2
 \left(
 \sum_{a=1}^{2}|r_a(Z)-1|
 +
 \sum_{b=1}^{2}|c_b(Z)-1|
 \right).
 \end{aligned}
\end{equation}

Each $P_k$ is strictly positive and row-stochastic, so we may write
\[
 P_k=
 \begin{pmatrix}
  1-u_k & u_k\\
  v_k & 1-v_k
 \end{pmatrix},
 \qquad 0<u_k,v_k<1.
\]
Initially,
\[
 u_0=\frac{\varepsilon^2}{K+1},
 \qquad
 v_0=\frac{K-1+\varepsilon^2}{K+1}.
\]
Set
$\delta_k\triangleq v_k-u_k$.
The column sums of $P_k$ are $1+\delta_k$ and $1-\delta_k$.
In particular, $|\delta_k|<1$, and
\begin{equation}\label{eq:critical-column-normalized}
 \widehat P_k=
 \begin{pmatrix}
  \dfrac{1-u_k}{1+\delta_k}
  &
  \dfrac{u_k}{1-\delta_k}
  \\[6pt]
  \dfrac{v_k}{1+\delta_k}
  &
  \dfrac{1-v_k}{1-\delta_k}
 \end{pmatrix}.
\end{equation}
Normalizing the second row of this matrix gives
\[
 v_{k+1}
 =
 \frac{v_k/(1+\delta_k)}
 {v_k/(1+\delta_k)+(1-v_k)/(1-\delta_k)}
 =
 \frac{v_k(1-\delta_k)}
 {1+(1-2v_k)\delta_k}.
\]
The denominator is positive, since it equals
\[
 v_k(1-\delta_k)+(1-v_k)(1+\delta_k)>0.
\]
Moreover, using $\delta_k=v_k-u_k$, we obtain
\begin{align*}
 v_{k+1}-\frac{v_k}{1+2v_k}
 &=
 \frac{
 v_k\bigl((1-\delta_k)(1+2v_k)
          -1-(1-2v_k)\delta_k\bigr)
 }{
 (1+2v_k)\bigl(1+(1-2v_k)\delta_k\bigr)
 }\\
 &=
 \frac{2u_kv_k}{
 (1+2v_k)\bigl(1+(1-2v_k)\delta_k\bigr)
 }
 \ge0.
\end{align*}
It follows that
\[
 \frac1{v_{k+1}}\le\frac1{v_k}+2.
\]
Iterating this inequality, and defining
\[
 L_k\triangleq v_0^{-1}+2k,
\]
we obtain
\begin{equation}\label{eq:critical-v-lower}
 v_k\ge\frac1{L_k}.
\end{equation}

We next bound $u_k$ using the invariant cross-ratio.
Multiplication by positive diagonal matrices preserves the quantity
$Z_{12}Z_{21}/(Z_{11}Z_{22})$, because each row and column scaling
factor occurs once in both the numerator and the denominator.
Thus, with $h$ as defined in \eqref{eq-defsh},
\[
 \frac{u_kv_k}{(1-u_k)(1-v_k)}=h^2
 \qquad\text{for every }k\ge0.
\]
Since $(1-u_k)(1-v_k)\le1$, this implies
\[
 u_kv_k\le h^2.
\]
Combining this inequality with \eqref{eq:critical-v-lower} gives
\[
 u_k\le\frac{h^2}{v_k}\le h^2L_k,
\]
and hence
\begin{equation}\label{eq:critical-delta-lower}
 \delta_k=v_k-u_k
 \ge\frac1{L_k}-h^2L_k
 =\frac{1-h^2L_k^2}{L_k}.
\end{equation}

To make this estimate uniform over a sufficiently long range of
iterations, observe that
\[
 h^2
 =
 \varepsilon^2
 \frac{K-1+\varepsilon^2}
 {(K+1-\varepsilon^2)(2-\varepsilon^2)}
 \le\varepsilon^2
 =4^{-n}.
\]
Here the inequality follows from $0<\varepsilon^2<1$, which gives
\[
 K-1+\varepsilon^2<K,
 \qquad
 (K+1-\varepsilon^2)(2-\varepsilon^2)>K.
\]
Also,
\[
 v_0^{-1}
 =\frac{K+1}{K-1+\varepsilon^2}
 \le\frac{K+1}{K-1}.
\]
Therefore, for every integer $0\le k\le n^4$,
\[
 h^2L_k^2
 \le
 4^{-n}
 \left(\frac{K+1}{K-1}+2n^4\right)^2.
\]
Since $K>1$ is fixed, the right-hand side tends to zero as
$n\to\infty$.
Thus, for all sufficiently large $n$, depending only on $K$,
\[
 h^2L_k^2\le\frac12
 \qquad \forall\, 0\le k\le n^4.
\]
By \eqref{eq:critical-delta-lower}, it follows that
\begin{equation}\label{eq:critical-delta-uniform}
 \delta_k\ge\frac1{2L_k}>0
 \qquad \forall\, 0\le k\le n^4.
\end{equation}

We now translate this bound into an error bound at both even and
odd iterations.
At iteration $2k$, the matrix $P_k$ is row-stochastic and has column
sums $1+\delta_k$ and $1-\delta_k$.
Using \eqref{eq:critical-error-tensor} and
\eqref{eq:critical-delta-uniform}, we obtain
\begin{equation}\label{eq:critical-even-error}
 \begin{aligned}
 &\sum_{i=1}^{n}\left|r_i(A^{(2k)})-1\right|
 +
 \sum_{j=1}^{n}\left|c_j(A^{(2k)})-1\right|=
 \frac n2\,(2\delta_k)
 =n\delta_k
 \ge\frac{n}{2L_k},
 \qquad \forall\ 0\le k\le n^4.
 \end{aligned}
\end{equation}

For the odd iterations, we first establish a lower bound on
$\det P_k$.
Every $2\times2$ matrix occurring in the normalization sequence
$P_0,\widehat P_0,P_1,\widehat P_1,\ldots$
is either row-stochastic or column-stochastic, and therefore has
total sum $2$.
If $z_1,z_2>0$ are the two margins being normalized in the next
iteration, then
\[
 z_1+z_2=2,\qquad z_1z_2\le1.
\]
That normalization divides the determinant by $z_1z_2$.
Since
\[
 \det P_0=\frac{2(1-\varepsilon^2)}{K+1}>0,
\]
all subsequent determinants remain positive and cannot decrease.
Consequently,
\begin{equation}\label{eq:critical-det-lower}
 1-u_k-v_k
 =\det P_k
 \ge\det P_0
 =\frac{2(1-\varepsilon^2)}{K+1}
 \ge\frac1{K+1}.
\end{equation}
The last inequality holds because $n\ge4$ and
$\varepsilon^2=4^{-n}\le1/2$.

By \eqref{eq:critical-column-normalized}, the row sums of
$\widehat P_k$ are $1-\zeta_k$ and $1+\zeta_k$, where
\[
 \zeta_k
 \triangleq
 \frac{\delta_k(1-u_k-v_k)}{1-\delta_k^2}.
\]
Indeed, subtracting $1$ from the first row sum gives
\[
 \frac{1-u_k}{1+\delta_k}
 +\frac{u_k}{1-\delta_k}-1
 =
 -\frac{\delta_k(1-u_k-v_k)}{1-\delta_k^2},
\]
and the second row sum follows because $\widehat P_k$ has total
sum $2$.
For $0\le k\le n^4$, equations
\eqref{eq:critical-delta-uniform} and
\eqref{eq:critical-det-lower}, together with
$0<1-\delta_k^2\le1$, imply
\[
 \zeta_k\ge\frac{\delta_k}{K+1}>0.
\]
Since $\widehat P_k$ is column-stochastic, we conclude that
\begin{equation}\label{eq:critical-odd-error}
 \begin{aligned}
 &\sum_{i=1}^{n}\left|r_i(A^{(2k+1)})-1\right|
 +
 \sum_{j=1}^{n}\left|c_j(A^{(2k+1)})-1\right|=
 \frac n2\,(2\zeta_k)
 \ge\frac{n\delta_k}{K+1}
 \ge\frac{n}{2(K+1)L_k}
 \qquad (0\le k\le n^4).
 \end{aligned}
\end{equation}

Combining \eqref{eq:critical-even-error} and
\eqref{eq:critical-odd-error} gives a bound valid at every iteration
in this range.
Indeed, for $0\le t\le2n^4+1$, let $k=\lfloor t/2\rfloor$.
Then
\[
 L_k=v_0^{-1}+2k
 \le\frac{K+1}{K-1}+t,
\]
and therefore
\begin{equation}\label{eq:critical-all-iterations-error}
 \sum_{i=1}^{n}\left|r_i(A^{(t)})-1\right|
 +
 \sum_{j=1}^{n}\left|c_j(A^{(t)})-1\right|
 \ge
 \frac{n}{
 2(K+1)\left(\dfrac{K+1}{K-1}+t\right)
 }.
\end{equation}
In particular, if
\[
 \sum_{i=1}^{n}\left|r_i(A^{(t)})-1\right|
 +
 \sum_{j=1}^{n}\left|c_j(A^{(t)})-1\right|
 \le n^{-2}
\]
at such an iteration, then
\[
 \frac{n}{
 2(K+1)\left(\dfrac{K+1}{K-1}+t\right)
 }
 \le n^{-2}.
\]
Rearranging gives
\begin{equation}\label{eq:critical-sk-iteration-lower}
 t
 \ge
 \frac{n^3}{2(K+1)}
 -\frac{K+1}{K-1}.
\end{equation}
If instead $t>2n^4+1$, the same lower bound holds automatically.

Thus, for all sufficiently large $n$, every iterate whose total
marginal $\ell_1$ error is at most $n^{-2}$ satisfies
\eqref{eq:critical-sk-iteration-lower}.
Since $K>1$ is fixed, this proves the claimed
$\Omega_K(n^3)$ iteration lower bound.

\end{proof}

\subsection{Lower bounds for bounded logarithmic updates}

\begin{proposition}
\label{prop:bounded-log-iterations}
Let $n\ge2$ and let $A,C\in\mathbb R_{>0}^{n\times n}$. Suppose that $C$ is doubly stochastic and $C=XAY$ for positive diagonal matrices $X,Y$. Let $s_2(C)$ be the second largest singular value of $C$, and assume
\[
s_2(C)\le1-\gamma,\qquad 0<\gamma\le1.
\]
For positive matrices $P,Q$ of the same size, and for a positive square matrix $M$, define
\begin{align}\label{eq-defdpq-deltam}
d_{\log}(P,Q)\triangleq\max_{i,j}\left|\log\frac{P_{ij}}{Q_{ij}}\right|,
\qquad
\Delta(M)\triangleq\|M\mathbf1-\mathbf1\|_1+
\|M^\top\mathbf1-\mathbf1\|_1.
\end{align}
Let $T\ge0$ be an integer and let $M^{(0)},\ldots,M^{(T)}$ be positive matrices such that $M^{(0)}=A$ and $M^{(t)}=X_tAY_t$ for positive diagonal matrices $X_t,Y_t$. Suppose that, for some constant $h>0$,
\begin{align}\label{eq-deltadlog-upperbound}
d_{\log}\left(M^{(t+1)},M^{(t)}\right)\le h
\quad \forall\ 0\le t<T.
\end{align}
If $0\le\varepsilon<\gamma$ and $\Delta\left(M^{(T)}\right)\le\varepsilon$, then
\begin{align}\label{eq-tlowerbound}
T\ge\frac{d_{\log}(C,A)+\log(1-\varepsilon/\gamma)}{h}.
\end{align}
In particular, consider the family of positive matrices $A_n$ in Proposition~\ref{prop:product-threshold-counterexample}.
We have
every such sequence attaining $\Delta\left(M^{(T)}\right)\le n^{-2}$ with an absolute constant $h$ satisfies
\begin{align}\label{eq-tlowerbound-main}
T\ge\frac{\log (1+2^n/3)+\log(1-1/((1-1/\sqrt2) n^2))}{h}
=\Omega(n).
\end{align}
\end{proposition}

\begin{proof}[Proof of \Cref{prop:bounded-log-iterations}]
We first prove 
\begin{align}\label{eq-dlogmc}
d_{\log}(M,C)\le-\log \left(1-\Delta(M)/\gamma\right) 
\end{align}
for every positive diagonal scaling $M$ of $A$ with $\Delta(M) < \gamma$. Since $M$ and $C$ are positive diagonal scalings of the same positive matrix, there exist $u,v\in\mathbb R^n$ such that
\begin{align}\label{eq-relation-mc}
M_{ij}=C_{ij}e^{u_i-v_j}.
\end{align}
Without loss of generality, we further assume that 
$z=(u,v)\in\mathbb R^{2n}$ is orthogonal to $\mathbf1_{2n}$.
Since if $z=(u,v)$ is not orthogonal to $\mathbf1_{2n}$,
one can
set $\widetilde u=u+a\mathbf1_n$ and
$\widetilde v=v+a\mathbf1_n$ where 
\[
a=-\frac{\sum_{i=1}^n u_i+\sum_{j=1}^n v_j}{2n}.
\]
Thus, we have $\sum_i\widetilde u_i+\sum_j\widetilde v_j=0$ and then $\widetilde z=(\widetilde u, \widetilde v)$ is orthogonal to $\mathbf1_{2n}$.
Furthermore, since
\[
\widetilde u_i-\widetilde v_j
=(u_i+a)-(v_j+a)=u_i-v_j,
\]
these new vectors represent the same matrix.
Formally,
$M_{ij}=C_{ij}e^{\widetilde u_i-\widetilde v_j}$.
In the following, we assume $z=(u,v)\in\mathbb R^{2n}$ is orthogonal to $\mathbf1_{2n}$.

Define the smooth function
\[
F(u,v)=\sum_{i,j}C_{ij}e^{u_i-v_j}
-\sum_i u_i+\sum_j v_j.
\]
For each fixed $i$, only the terms with row index $i$ in the
exponential sum depend on $u_i$. Since
\[
\frac{\partial}{\partial u_i}e^{u_i-v_j}
=e^{u_i-v_j},
\]
we obtain
\[
\begin{aligned}
\frac{\partial F}{\partial u_i}(u,v)
&=
\frac{\partial}{\partial u_i}
\left(
  \sum_{j=1}^n C_{ij}e^{u_i-v_j}-u_i
\right)=\sum_{j=1}^n C_{ij}e^{u_i-v_j}-1=
\sum_{j=1}^n M_{ij}-1=
\bigl(M\mathbf1_n-\mathbf1_n\bigr)_i.
\end{aligned}
\]
Thus the first $n$ coordinates of the gradient are the
row-sum errors:
\[
\nabla_u F(u,v)=M\mathbf1_n-\mathbf1_n.
\]
Similarly, the last $n$ coordinates are the negatives of the
column-sum errors:
\[
\nabla_v F(u,v)=\mathbf1_n-M^\top\mathbf1_n.
\]
Therefore,
\begin{align}\label{eq-nabldafz}
\nabla F(z)
=
\begin{pmatrix}
\nabla_u F(u,v)\\
\nabla_v F(u,v)
\end{pmatrix}
=
\begin{pmatrix}
M\mathbf1_n-\mathbf1_n\\
\mathbf1_n-M^\top\mathbf1_n
\end{pmatrix}.
\end{align}

Finally, by the definition of the vector $\ell_1$ norm,
\begin{equation}
\begin{aligned}\label{eq-nabldafz-lonenorm}
\|\nabla F(z)\|_1
&=
\sum_{i=1}^n
\left|\frac{\partial F}{\partial u_i}(u,v)\right|
+
\sum_{j=1}^n
\left|\frac{\partial F}{\partial v_j}(u,v)\right|\\
&=
\sum_{i=1}^n
\left|\sum_{j=1}^n M_{ij}-1\right|
+
\sum_{j=1}^n
\left|1-\sum_{i=1}^n M_{ij}\right|\\
&=
\|M\mathbf1_n-\mathbf1_n\|_1
+
\|M^\top\mathbf1_n-\mathbf1_n\|_1\\
&=\Delta(M).
\end{aligned}
\end{equation}
By \eqref{eq-nabldafz},
we have
\[
\begin{aligned}
\mathbf1_{2n}^\top\nabla F(z)=
\sum_{i=1}^n\left(\sum_{j=1}^n M_{ij}-1\right)
+
\sum_{j=1}^n\left(1-\sum_{i=1}^n M_{ij}\right)=
\sum_{i=1}^n\sum_{j=1}^n M_{ij}-n
+n-\sum_{j=1}^n\sum_{i=1}^n M_{ij}=0.
\end{aligned}
\]
Thus,
the sum of the coordinates of $\nabla F(z)$ is zero. For any real vector $g$ with coordinate sum zero, the sum of its positive coordinates and the sum of the absolute values of its negative coordinates both equal $\|g\|_1/2$. The sum of the squares of nonnegative numbers is at most the square of their sum. Applying this separately to the positive and negative coordinates proves
$\|g\|_2^2\le2(\|g\|_1/2)^2$.
Consequently, by \eqref{eq-nabldafz-lonenorm},
\begin{align}\label{eq-nabldafz-upperbound}
\|\nabla F(z)\|_2\le\frac{\Delta(M)}{\sqrt2}.
\end{align}
If
$u=v=0$, we have $e^{u_i-v_j}=1$ for all $i,j$.
Hence, $M = C$. 
By \eqref{eq-nabldafz}, we have
Thus,
\[
\nabla F(0)
=
\begin{pmatrix}
C\mathbf1_n-\mathbf1_n\\
\mathbf1_n-C^\top\mathbf1_n
\end{pmatrix}.
\]
Since $C$ is doubly stochastic, its row and column sums are $1$.
Thus,
\begin{align}\label{eq-nabdafzero-eqzero}
\nabla F(0)
=
\begin{pmatrix}
C\mathbf1_n-\mathbf1_n\\
\mathbf1_n-C^\top\mathbf1_n
\end{pmatrix}
=
\begin{pmatrix}
0_n\\
0_n
\end{pmatrix}.
\end{align}

We compute the Hessian
\[
\nabla^2F(u,v)
=
\begin{pmatrix}
H_{uu}(u,v)&H_{uv}(u,v)\\
H_{vu}(u,v)&H_{vv}(u,v)
\end{pmatrix}.
\]
Let $\delta_{ab}$ denote the Kronecker delta:
$\delta_{ab}=1$ if $a=b$, and $\delta_{ab}=0$ otherwise.
For the $u,u$ block, we have
\[
\begin{aligned}
\bigl(H_{uu}(u,v)\bigr)_{ik}
=
\frac{\partial}{\partial u_k}
\left(\frac{\partial F}{\partial u_i}\right)=
\frac{\partial}{\partial u_k}
\left(\sum_{j=1}^n C_{ij}e^{u_i-v_j}-1\right)=
\delta_{ik}\sum_{j=1}^n C_{ij}e^{u_i-v_j}.
\end{aligned}
\]
Thus this block is diagonal, and its $i$th diagonal entry
is the $i$th row sum of the matrix with entries
$C_{ij}e^{u_i-v_j}$.

For the $v,v$ block, we similarly obtain
\[
\begin{aligned}
\bigl(H_{vv}(u,v)\bigr)_{j\ell}=
\frac{\partial}{\partial v_\ell}
\left(\frac{\partial F}{\partial v_j}\right)=
\frac{\partial}{\partial v_\ell}
\left(1-\sum_{i=1}^n C_{ij}e^{u_i-v_j}\right)=
\delta_{j\ell}\sum_{i=1}^n C_{ij}e^{u_i-v_j}.
\end{aligned}
\]
Hence,
this block is diagonal, with diagonal entries equal to
the corresponding column sums.

For the upper-right block,
\[
\begin{aligned}
\bigl(H_{uv}(u,v)\bigr)_{ij}=
\frac{\partial}{\partial v_j}
\left(\frac{\partial F}{\partial u_i}\right)=
\frac{\partial}{\partial v_j}
\left(\sum_{q=1}^n C_{iq}e^{u_i-v_q}-1\right)=-C_{ij}e^{u_i-v_j}.
\end{aligned}
\]
For the lower-left block,
\[
\begin{aligned}
\bigl(H_{vu}(u,v)\bigr)_{ji}
=
\frac{\partial}{\partial u_i}
\left(\frac{\partial F}{\partial v_j}\right)=
\frac{\partial}{\partial u_i}
\left(1-\sum_{p=1}^n C_{pj}e^{u_p-v_j}\right)=-C_{ij}e^{u_i-v_j}.
\end{aligned}
\]
Hence, $H_{vu}=H_{uv}^\top$.
Consequently, at a point where
$M_{ij}=C_{ij}e^{u_i-v_j}$, the Hessian is
\begin{align}\label{eq-nabda2fuv}
\nabla^2F(u,v)
=
\begin{pmatrix}
\operatorname{diag}(M\mathbf1_n)&-M\\
-M^\top&\operatorname{diag}(M^\top\mathbf1_n)
\end{pmatrix}.
\end{align}
For a vector $a$, $\operatorname{diag}(a)$ denotes the diagonal matrix
whose diagonal entries are the components of $a$.

Recall that  $M = C$ if $u=v=0$.
Since $C$ is doubly stochastic, we have
\begin{equation}\label{eq-defhzero}
\begin{aligned}
H_0\triangleq\nabla^2F(0)
&=
\begin{pmatrix}
\operatorname{diag}(C\mathbf1_n)&-C\\
-C^\top&\operatorname{diag}(C^\top\mathbf1_n)
\end{pmatrix}=
\begin{pmatrix}
I_n&-C\\
-C^\top&I_n
\end{pmatrix},
\end{aligned}
\end{equation}

We determine the kernel of $H_0$ and establish a lower bound
for its quadratic form on $\mathbf1_{2n}^{\perp}$.
Write $s_k\triangleq s_k(C)$. Since $C$ is doubly stochastic,
\[
1=s_1\ge s_2\ge\cdots\ge s_n\ge0.
\]
Choose orthonormal left and right singular vectors
$p_1,\ldots,p_n$ and $q_1,\ldots,q_n$ satisfying
\[
Cq_k=s_kp_k,
\qquad
C^\top p_k=s_kq_k
\quad (1\le k\le n).
\]
Double stochasticity also gives
\[
C\frac{\mathbf1_n}{\sqrt n}
=\frac{\mathbf1_n}{\sqrt n},
\qquad
C^\top\frac{\mathbf1_n}{\sqrt n}
=\frac{\mathbf1_n}{\sqrt n},
\]
so we may choose
\[
p_1=q_1=\frac{\mathbf1_n}{\sqrt n}.
\]
For each $k$, define
\[
a_k\triangleq\frac1{\sqrt2}
\begin{pmatrix}p_k\\q_k\end{pmatrix},
\qquad
b_k\triangleq\frac1{\sqrt2}
\begin{pmatrix}p_k\\-q_k\end{pmatrix}.
\]
Using the singular-vector identities, we obtain
\[
\begin{aligned}
H_0a_k
=\frac1{\sqrt2}
\begin{pmatrix}
I_n&-C\\
-C^\top&I_n
\end{pmatrix}
\begin{pmatrix}p_k\\q_k\end{pmatrix}=
\frac1{\sqrt2}
\begin{pmatrix}
p_k-Cq_k\\
-C^\top p_k+q_k
\end{pmatrix}=
\frac1{\sqrt2}
\begin{pmatrix}
(1-s_k)p_k\\
(1-s_k)q_k
\end{pmatrix}
=(1-s_k)a_k.
\end{aligned}
\]
Similarly,
\[
\begin{aligned}
H_0b_k
=\frac1{\sqrt2}
\begin{pmatrix}
I_n&-C\\
-C^\top&I_n
\end{pmatrix}
\begin{pmatrix}p_k\\-q_k\end{pmatrix}=
\frac1{\sqrt2}
\begin{pmatrix}
p_k+Cq_k\\
-C^\top p_k-q_k
\end{pmatrix}=
\frac1{\sqrt2}
\begin{pmatrix}
(1+s_k)p_k\\
-(1+s_k)q_k
\end{pmatrix}
=(1+s_k)b_k.
\end{aligned}
\]

Moreover, orthonormality of the singular-vector bases gives
\[
\begin{aligned}
a_k^\top a_\ell
=b_k^\top b_\ell
&=
\frac12
\left(p_k^\top p_\ell+q_k^\top q_\ell\right)
=\delta_{k\ell},\\
a_k^\top b_\ell
&=
\frac12
\left(p_k^\top p_\ell-q_k^\top q_\ell\right)
=0.
\end{aligned}
\]
Thus the $2n$ vectors
$a_1,\ldots,a_n,b_1,\ldots,b_n$
form an orthonormal eigenbasis of $H_0$, with corresponding
eigenvalues $1-s_k$ and $1+s_k$.
Since $s_1=1$ and $p_1=q_1=\mathbf1_n/\sqrt n$,
\[
a_1=\frac{\mathbf1_{2n}}{\sqrt{2n}},
\qquad
H_0a_1=0.
\]
For every $k\ge2$, the assumption $s_2\le1-\gamma$ gives
\[
1-s_k\ge1-s_2\ge\gamma>0.
\]
Furthermore, since $s_k\ge0$ and $0<\gamma\le1$,
\[
1+s_k\ge1\ge\gamma>0
\quad (1\le k\le n).
\]
Hence exactly one vector in this eigenbasis, namely $a_1$,
has eigenvalue zero; all other eigenvalues are at least
$\gamma$. Consequently,
\[
\ker H_0
=
\operatorname{span}\{a_1\}
=
\operatorname{span}\{\mathbf1_{2n}\}.
\]

For any vector $x\perp\mathbf1_{2n}$, since
$a_1=\mathbf1_{2n}/\sqrt{2n}$, we have $a_1^\top x=0$.
Therefore the expansion of $x$ in the above eigenbasis
contains no $a_1$ component:
\[
x
=
\sum_{k=2}^n\alpha_k a_k
+
\sum_{k=1}^n\beta_k b_k,
\qquad
\alpha_k=a_k^\top x,
\quad
\beta_k=b_k^\top x.
\]
By orthonormality,
\[
\|x\|_2^2
=
\sum_{k=2}^n\alpha_k^2
+
\sum_{k=1}^n\beta_k^2.
\]
Using the eigenvalue identities and the lower bounds above,
we obtain
\begin{equation}\label{eq-xtophzeroxlargegammaxnorm}
\begin{aligned}
x^\top H_0x
&=
\sum_{k=2}^n(1-s_k)\alpha_k^2
+
\sum_{k=1}^n(1+s_k)\beta_k^2\ge
\gamma\sum_{k=2}^n\alpha_k^2
+
\gamma\sum_{k=1}^n\beta_k^2=\gamma\|x\|_2^2,\qquad \forall \ x\perp\mathbf1_{2n}.
\end{aligned}
\end{equation}

If $u=v=0$, then $M=C$ and \eqref{eq-dlogmc} is immediate. Otherwise write $z=rw$ with $r=\|z\|_2>0$ and $\|w\|_2=1$. 
Since $z\perp\mathbf1_{2n}$, we have $w\perp\mathbf1_{2n}$. 
Let $e_1,\ldots,e_{2n}$ denote the standard basis
of $\mathbb R^{2n}$. 
For each pair $i,j\in [n]$, we have
\begin{align}\label{eq-expansion-prodeinplusj}
(e_i-e_{n+j})(e_i-e_{n+j})^\top
=e_ie_i^\top+e_{n+j}e_{n+j}^\top
-e_ie_{n+j}^\top-e_{n+j}e_i^\top.
\end{align}
Thus $(e_i-e_{n+j})(e_i-e_{n+j})^\top$ has entries $1$ at positions $(i,i)$ and
$(n+j,n+j)$, entries $-1$ at positions $(i,n+j)$ and $(n+j,i)$,
and zero entries elsewhere. 
Let $u = (tw_1,tw_2,\cdots,tw_n) , v = (tw_{n+1},tw_{n+2},\cdots,tw_{2n})$.
By \eqref{eq-nabda2fuv} and \eqref{eq-relation-mc},
\begin{align}
\nabla^2F(tw)
=\sum_{i,j}C_{ij}e^{t(w_i-w_{n+j})}
(e_i-e_{n+j})(e_i-e_{n+j})^\top.
\end{align}
Letting $u=v=0$, we have 
\begin{align}
H_0\triangleq\nabla^2F(0)=\sum_{i,j}C_{ij}
(e_i-e_{n+j})(e_i-e_{n+j})^\top.
\end{align}
Since $\|w\|_2=1$, the Cauchy--Schwarz inequality gives
\[
\begin{aligned}
|w_i-w_{n+j}|=|(e_i-e_{n+j})^\top w|\le\|e_i-e_{n+j}\|_2\,\|w\|_2
=\sqrt2.
\end{aligned}
\]
Consequently, for every
$t\ge0$,
\[
t(w_i-w_{n+j})\ge-\sqrt2t,
\qquad
e^{t(w_i-w_{n+j})}\ge e^{-\sqrt2t}.
\]
Because $C_{ij}>0$, the coefficients
$\alpha_{ij}(t)\triangleq
C_{ij}\left(e^{t(w_i-w_{n+j})}-e^{-\sqrt2t}\right)$
are nonnegative. Subtracting the two outer-product representations
therefore yields
\[
\begin{aligned}
\nabla^2F(tw)-e^{-\sqrt2t}H_0
&=\sum_{i,j}C_{ij}
 \left(e^{t(w_i-w_{n+j})}-e^{-\sqrt2t}\right)(e_i-e_{n+j})(e_i-e_{n+j})^\top\\
 &=\sum_{i,j}\alpha_{ij}(t)(e_i-e_{n+j})(e_i-e_{n+j})^\top.
\end{aligned}
\]
This is a nonnegative linear combination of positive semidefinite
matrices. Explicitly, for every $x\in\mathbb R^{2n}$, by \eqref{eq-expansion-prodeinplusj} we have
\[
\begin{aligned}
x^\top\left(\nabla^2F(tw)-e^{-\sqrt2t}H_0\right)x
&=\sum_{i,j}\alpha_{ij}(t)\,x^\top (e_i-e_{n+j})(e_i-e_{n+j})^\top x=\sum_{i,j}\alpha_{ij}(t)(x_i-x_{n+j})^2\ge0.
\end{aligned}
\]
Thus the difference is positive semidefinite, proving
\begin{align}\label{eq-nabdatwoftwrelationesquarttwothzero}
\nabla^2F(tw)\succeq e^{-\sqrt2t}H_0,\qquad 
\forall\ t\ge 0.
\end{align}

By the Cauchy--Schwarz inequality, $z=rw$, and $\|w\|_2 = 1$, we obtain
\[
\begin{aligned}
w^\top\nabla F(rw)\le\bigl|w^\top\nabla F(rw)\bigr|\le\|w\|_2\,\|\nabla F(rw)\|_2=\|\nabla F(z)\|_2.
\end{aligned}
\]
Combined with \eqref{eq-nabldafz-upperbound},
we have
\[
\begin{aligned}
w^\top\nabla F(rw)\leq \frac{\Delta(M)}{\sqrt2}.
\end{aligned}
\]
Combined with \eqref{eq-nabdafzero-eqzero},
we have 
\begin{align}\label{eq-deltawtopnabdafrw-upperbound}
w^\top\nabla F(rw) - w^\top\nabla F(0)  \leq \frac{\Delta(M)}{\sqrt2}.
\end{align}
In addition, by the fundamental theorem of calculus,
\[
w^\top\nabla F(rw) - w^\top\nabla F(0)
=\int_0^r w^\top\nabla^2F(tw)w\,dt.
\]
For each $t\in[0,r]$, \eqref{eq-nabdatwoftwrelationesquarttwothzero} gives
\[
w^\top\nabla^2F(tw)w
\ge e^{-\sqrt2t}w^\top H_0w.
\]
Because $w\perp\mathbf1_{2n}$, by \eqref{eq-xtophzeroxlargegammaxnorm} and using
$\|w\|_2=1$, we obtain
\[
w^\top H_0w\ge\gamma\|w\|_2^2=\gamma.
\]
Combining these two inequalities yields the pointwise bound
\[
w^\top\nabla^2F(tw)w\ge\gamma e^{-\sqrt2t}
\qquad \forall\ 0\le t\le r.
\]
Integrating over $[0,r]$ gives
\[
\int_0^r w^\top\nabla^2F(tw)w\,dt
\ge\gamma\int_0^r e^{-\sqrt2t}\,dt =\gamma\left[-\frac1{\sqrt2}e^{-\sqrt2t}\right]_0^r
=\frac{\gamma}{\sqrt2}(1-e^{-\sqrt2r}).
\]
Hence, \eqref{eq-deltawtopnabdafrw-upperbound} implies
\[
\begin{aligned}
\frac{\Delta(M)}{\sqrt2}\geq
\frac{\gamma}{\sqrt2}(1-e^{-\sqrt2r}).
\end{aligned}
\]
It follows that 
\begin{align}\label{eq-expminussqrttworlowerbound}
e^{-\sqrt2r}\ge1-\Delta(M)/\gamma>0.
\end{align}
Moreover, since $z=(u,v)\in\mathbb R^{2n}$, for every $i,j$ we have
\[
u_i^2+v_j^2
\le\sum_{k=1}^n u_k^2+\sum_{\ell=1}^n v_\ell^2
=\|z\|_2^2.
\]
Applying the Cauchy--Schwarz inequality in $\mathbb R^2$ yields
\[
\begin{aligned}
|u_i-v_j|
&=|1\cdot u_i+(-1)\cdot v_j|\le\sqrt{1^2+(-1)^2}\,\sqrt{u_i^2+v_j^2}=\sqrt2\,\sqrt{u_i^2+v_j^2}\le\sqrt2\|z\|_2.
\end{aligned}
\]
Combining it with $\|z\|_2=r$ and \eqref{eq-expminussqrttworlowerbound} proves
\[
\max_{i,j}|u_i-v_j|
\le\sqrt2r\le-\log\left(1-\Delta(M)/\gamma\right).
\]
Moreover, by the definition of $d_{\log}$ in \eqref{eq-defdpq-deltam} and \eqref{eq-relation-mc}, we have
\[
d_{\log}(M,C)
\triangleq \max_{1\le i,j\le n}
\left|\log\frac{M_{ij}}{C_{ij}}\right| = \max_{1\le i,j\le n}\left|\log\frac{C_{ij}e^{u_i-v_j}}{C_{ij}}\right| = \max_{1\le i,j\le n}\left|u_i-v_j\right| .
\]
This establishes \eqref{eq-dlogmc}.

Recall that $\Delta\left(M^{T}\right) \leq \varepsilon < \gamma$.
By \eqref{eq-dlogmc}, we obtain 
\begin{align}\label{eqw-dlogcmt-upperbound}
d_{\log}\left(C,M^{(T)}\right) \le-\log(1-\varepsilon/\gamma).
\end{align}
For any strictly positive matrices $P,Q,R$ of the same size,
\[
\begin{aligned}
d_{\log}(P,R)
&=\max_{i,j}\left|\log\frac{P_{ij}}{R_{ij}}\right|=\max_{i,j}\left|
 \log\frac{P_{ij}}{Q_{ij}}+\log\frac{Q_{ij}}{R_{ij}}
 \right|\le\max_{i,j}\left(
 \left|\log\frac{P_{ij}}{Q_{ij}}\right|
 +\left|\log\frac{Q_{ij}}{R_{ij}}\right|
 \right)\\
&\le\max_{i,j}\left|\log\frac{P_{ij}}{Q_{ij}}\right|
 +\max_{i,j}\left|\log\frac{Q_{ij}}{R_{ij}}\right|=d_{\log}(P,Q)+d_{\log}(Q,R).
\end{aligned}
\]
Applying this inequality, \eqref{eqw-dlogcmt-upperbound}, and \eqref{eq-deltadlog-upperbound} gives
\[
\begin{aligned}
d_{\log}(C,A)
&\le d_{\log}\left(C,M^{(T)}\right)+d_{\log}\left(M^{(T)},A\right)\le-\log(1-\varepsilon/\gamma)+
\sum_{t=0}^{T-1}d_{\log}(M^{(t+1)},M^{(t)})\\
&\le-\log(1-\varepsilon/\gamma)+Th.
\end{aligned}
\]
Thus, \eqref{eq-tlowerbound} is immediate.

For the specified family of positive matrices $A_n$ in Proposition~\ref{prop:product-threshold-counterexample},
we have 
\begin{align*}
n\geq 12,\qquad s_2(C)=\frac1{\sqrt2},\qquad
\max_{i,j}\frac{C_{ij}}{A_{ij}}
=\max_{i,j}\frac{A_{ij}}{C_{ij}} = 1+\frac{2^n}{3}.
\end{align*}
Let $\gamma = 1-1/\sqrt2$ and $\varepsilon = 1/n^2$. 
Since $n\ge12$ and $\gamma>1/4$, we have $\varepsilon = n^{-2}\le1/144<\gamma$.
By the definition of $d_{\log}$ in \eqref{eq-defdpq-deltam} and \eqref{eq-dlogmc}, we have
\begin{align*}
T&\ge \frac{\max_{i,j}\left|\log C_{ij}/A_{ij} \right|}{h} + \frac{\log(1-1/((1-1/\sqrt2) n^2))}{h} \\
&= \frac{\log (1+2^n/3)+\log(1-1/((1-1/\sqrt2) n^2))}{h}
=\Omega(n).
\end{align*}
Thus, \eqref{eq-tlowerbound-main} is proved.
The proposition is established. 
\end{proof}

\subsection{Proof of \Cref{cor:three-scaling-algorithms}}

\begin{proof}[Proof of \Cref{cor:three-scaling-algorithms}]
We verify that the matrix sequence generated by each algorithm
satisfies the hypotheses of
Proposition~\ref{prop:bounded-log-iterations}.

For the box-constrained Newton scaling method in \cite{cohen2017matrix}, use logarithmic variables $(x,y)$ with associated matrix
\[
M(x,y)_{ij}=A_{ij}e^{x_i-y_j}.
\]
Let $\left(x^{(t)},y^{(t)}\right)$ denote the logarithmic scaling variables
after $t$ iterations, and set
\[
M^{(t)}\triangleq M\left(x^{(t)},y^{(t)}\right).
\]
The matrix-scaling implementation is initialized at $\left(x^{(0)},y^{(0)}\right)=(0,0)$, as specified in \cite[Appendix A.4, proof of Theorem 4.6]{cohen2017matrix}. Thus its initial matrix is $A$. By \cite[Definition 3.3]{cohen2017matrix} , the $k$-oracle returns a vector $d$ with $\|d\|_\infty\le k$. 
The update step in (A.1) of
\cite[Appendix A.1]{cohen2017matrix} is $d/k$.
Thus, every logarithmic coordinate changes by at most $1$. 
The coordinate bounds for each update are
\[
\|x^{(t+1)}-x^{(t)}\|_\infty\le1,
\qquad
\|y^{(t+1)}-y^{(t)}\|_\infty\le1.
\]
Consequently, for every pair of consecutive iterates and every $i,j$,
\[
\begin{aligned}
\left|\log\frac{M^{(t+1)}_{ij}}{M^{(t)}_{ij}}\right|
&=\left|\log
 \frac{M\left(x^{(t+1)},y^{(t+1)}\right)_{ij}}
      {M\left(x^{(t)},y^{(t)}\right)_{ij}}\right|=\left|
 \left(x_i^{(t+1)}-x_i^{(t)}\right)
 -\left(y_j^{(t+1)}-y_j^{(t)}\right)
 \right|
 \\&\le \left|x_i^{(t+1)}-x_i^{(t)}\right|
 + \left|y_j^{(t+1)}-y_j^{(t)}\right|\le2.
\end{aligned}
\]
Taking the maximum over $i,j$ gives
$d_{\log}\left(M^{(t+1)},M^{(t)}\right)\le2$.
The regularized objective used in the scaling reduction does not change this update rule or the associated matrix.
Applying Proposition~\ref{prop:bounded-log-iterations} with $h=2$
shows that the box-constrained Newton method in \cite{cohen2017matrix} requires
at least $R_n/2$ iterations.

For each implementation in \cite{allen2017much}, define the row-normalized
scaling map by
\begin{align}\label{eq-def-mz}
\left(M(z)\right)_{ij}\triangleq
\frac{A_{ij}e^{z_j}}{\sum_k A_{ik}e^{z_k}},
\qquad \forall\ z\in\mathbb R^n.
\end{align}
For each implementation, let $z^{(t)}$ denote the logarithmic
column vector after $t$ iterations. 
For simplicity, let $M^{(t)}$ denote $M\left(z^{(t)}\right)$.
Since $A$ is row-stochastic, the initialization $z^{(0)}=0$
gives $M^{(0)}=A$.

We estimate the logarithmic change in each matrix entry.
Suppose that, for some $b\ge0$,
\[
\left\|z^{(t+1)}-z^{(t)}\right\|_\infty\le b.
\]
Then, for every row $i$,
\[
e^{-b}\sum_kA_{ik}e^{z_k^{(t)}}
\le
\sum_kA_{ik}e^{z_k^{(t+1)}}
\le
e^b\sum_kA_{ik}e^{z_k^{(t)}}.
\]
Using \eqref{eq-def-mz}, the changes in the logarithms of both the
numerator and the denominator are at most $b$ in absolute
value. Therefore,
\begin{equation}\label{eq-dlogm-upperbound}
\begin{aligned}
d_{\log}\left(M^{(t+1)},M^{(t)}\right)
&=
d_{\log}\left(M\left(z^{(t+1)}\right),M\left(z^{(t)}\right)\right)\le 2b.
\end{aligned}
\end{equation}
Furthermore, 
the row-maximum normalization used in \cite{allen2017much} does not change
the row-normalized matrices. Indeed, for any positive
diagonal matrix $S$ and any $z\in\mathbb R^n$, replacing
$A$ by $SA$ in \eqref{eq-def-mz} gives
\[
\frac{(SA)_{ij}e^{z_j}}
     {\sum_k(SA)_{ik}e^{z_k}}
=
\frac{S_{ii}A_{ij}e^{z_j}}
     {S_{ii}\sum_kA_{ik}e^{z_k}}
=
\left(M(z)\right)_{ij}.
\]
Moreover, for any nonnegative matrix
$B\in\mathbb R_{\ge0}^{n\times n}$ with strictly positive
row sums, define
\[
f_B(z)\triangleq
\sum_i\log\left(\sum_jB_{ij}e^{z_j}\right)-\sum_jz_j,
\qquad z\in\mathbb R^n.
\]
The corresponding objective for $SA$ satisfies
\[
\begin{aligned}
f_{SA}(z)
&=
\sum_i\log\left(S_{ii}\sum_jA_{ij}e^{z_j}\right)
-\sum_jz_j = f_A(z)+\sum_i\log S_{ii}.
\end{aligned}
\]
The added term is independent of $z$.
Thus,
\[
\nabla f_{SA}(z)=\nabla f_A(z),
\qquad
\nabla^2f_{SA}(z)=\nabla^2f_A(z).
\]
Thus this preliminary row normalization leaves the logarithmic
update rules, and hence the associated matrix sequences,
unchanged.

For Scaling1, Algorithm 4 in \cite{allen2017much}  starts
at $z^{(0)}=0$ and accepts
$z^{(t+1)}=z^{(t)}+\delta^{(t)}/6.6$.
Definition 5.2 gives $\|\alpha^{(t)}\|_\infty\le1/32$,
and Lemma 6.1 gives
\[
 \|\delta^{(t)}-\alpha^{(t)}\|_\infty
 \le\frac1{32}+\frac{1}{8K}.
\]
Here $K\ge1$ is the subroutine accuracy parameter. Hence
\[
 \|\delta^{(t)}\|_\infty
 \le\frac1{16}+\frac{1}{8K}.
\]
We have
\[
\left\|z^{(t+1)}-z^{(t)}\right\|_\infty
\le\frac{1/16+1/(8K)}{6.6}
\le\frac{3/16}{33/5}
=\frac5{176}.
\]
By \eqref{eq-dlogm-upperbound}, 
\begin{equation*}
\begin{aligned}
d_{\log}\left(M^{(t+1)},M^{(t)}\right) \leq \frac{5}{88}.
\end{aligned}
\end{equation*}
Proposition~\ref{prop:bounded-log-iterations} with $h=5/88$ proves that the Scaling1 algorithm in \cite{allen2017much} requires
at least $88R_n/5$ iterations.

For Scaling2, Algorithm 5 in \cite{allen2017much} also starts
at $z^{(0)}=0$ and accepts
$z^{(t+1)}=z^{(t)}+\delta^{(t)}$.
Each coordinate of the update direction $\delta^{(t)}$
is constrained to an interval specified in Definition 5.2.
By Fact 5.3, these constraints imply
$\|\delta^{(t)}\|_\infty\le1/16$.
By \eqref{eq-dlogm-upperbound}, 
\begin{equation}\label{eq-scaletwo-uppperbound}
\begin{aligned}
d_{\log}\left(M^{(t+1)},M^{(t)}\right) \leq \frac{1}{8}.
\end{aligned}
\end{equation}
For Scaling3, Algorithm 6 accepts
$z^{(t+1)}=z^{(t)}+\delta^{(t)}/6.6$, with $t$ continuing from
the preceding Scaling2 stage. Here the reference vector $\alpha^{(t)}$
and the parameter $K$ satisfy $\|\alpha^{(t)}\|_\infty\le1/32$
and $K\ge1$. By Lemma 8.4,
\[
\|\delta^{(t)}-\alpha^{(t)}\|_\infty\le\frac1{32}+\frac2K.
\]
Consequently,
\[
\left\|z^{(t+1)}-z^{(t)}\right\|_\infty
\le\frac{1/16+2/K}{6.6}
\le\frac{33/16}{33/5}
=\frac5{16},
\]
By \eqref{eq-dlogm-upperbound}, 
\begin{equation}\label{eq-scalethree-uppperbound}
\begin{aligned}
d_{\log}\left(M^{(t+1)},M^{(t)}\right) \leq \frac{5}{8}.
\end{aligned}
\end{equation} 
The implementation in \cite[Corollary 8.2(2)]{allen2017much}
uses the output vector of Scaling2 as the initial vector of Scaling3.
Hence, by \eqref{eq-scaletwo-uppperbound} and \eqref{eq-scalethree-uppperbound}, the combined matrix sequence starts at $A$ and satisfies
\[
d_{\log}\left(M^{(t+1)},M^{(t)}\right)\le\frac58
\]
at every iteration.
Applying Proposition~\ref{prop:bounded-log-iterations}
with $h=5/8$ shows that the two stages together require
at least $8R_n/5$ outer iterations to attain
the prescribed accuracy.

Finally, one can verify $R_n=\Theta(n)$. 
Hence, all three iteration counts are $\Omega(n)$.
\end{proof}

\newpage

%% file: Dense_Matrix.tex
\section{Applications on dense matrices}\label{sec-upperbound}
In this section, we prove Theorems \ref{cor:dense-seed} and \ref{thm-tight-dense}.

\subsection{Upper bounds for dense matrices}

The following lemmas are used in the proof of \Cref{cor:dense-seed}.

\begin{lemma}\label{lem-spectragap-relation}
Let $t>0$ and $\gamma,\rho\in(0,1]$ be constants, and let $n\ge2$ be an integer.
Let \(A\in [0,1]^{n\times n}\) be a $(\gamma,\rho)$-dense matrix.
Define
\begin{align}\label{eq-def-qone-qtwo-aprime}
Q_1 = \diag(r_1(A),\dots,r_n(A)),\qquad Q_2 = \diag(c_1(A),\dots,c_n(A)),\qquad A' = Q^{-1/2}_1 A Q^{-1/2}_2.
\end{align}
Let $s_1(A')\geq s_2(A') \geq \cdots \geq s_n(A')$ be the singular values of $A'$, counted with multiplicity. 
Let \(B\in [0,1]^{n\times n}\) be a matrix with
\begin{align}\label{eq-rib-bound-cib-eqone}
&r_i\left(B\right)\in [\rho\gamma ,1/(\rho\gamma)],\quad c_i\left(B\right) = 1, \qquad \forall\  i\in [n].\\
&\frac{A_{i,j}}{tn} \leq B_{i,j}\leq 
\frac{tA_{i,j}}{n\rho^3 \gamma^3},
\qquad  \qquad \qquad  \quad \ \  
\forall\  i,j\in [n].\label{eq-relation-aijbij}
\end{align}
Let
\begin{align}\label{eq-def-qd}
Q=\diag\left(r_1(B),\dots,r_n(B)\right),
\qquad
D \triangleq B^{\top}Q^{-1}B.
\end{align}
Then 
\[
1-\lambda_2(D) \geq \frac{(\rho\gamma)^5}{t^2} \cdot \left(1 - \left(s_2(A')\right)^2\right).\]
\end{lemma}

\begin{lemma}
\label{lem:diffuse-seed}
Let $n\ge2$ and let $\kappa,\mu,s_-,s_+,b_-,b_+>0$.
Let $A\in\R_{\ge0}^{n\times n}$ be row-stochastic, with
\[
 s_-\le A([n],j)\le s_+\quad\forall j\in[n],
 \qquad A_{ij}\le\mu\quad\forall i,j\in[n].
\]
Let $B=XAY$ be column-stochastic, where
$X=\diag(x_1,\ldots,x_n)$ and $Y=\diag(y_1,\ldots,y_n)$
are positive diagonal matrices satisfying
\[
 \min_i x_i=\min_j y_j=\kappa,
 \qquad b_-\le B(i,[n])\le b_+\quad\forall i\in[n].
\]
Define
\[
 R_0\triangleq\left\{i\in[n]:x_i\ge\frac{1}{2s_+\kappa}\right\},
 \qquad
 C_0\triangleq\left\{j\in[n]:y_j\ge\frac{b_-}{2\kappa}\right\}.
\]
Then
\begin{equation}\label{eq:diffuse-seed-sizes}
 |R_0|\ge\frac{1}{2b_+\mu},
 \qquad
 |C_0|\ge\frac{b_-s_-}{2\mu}.
\end{equation}
\end{lemma}

\begin{proof}
Choose $j_0\in[n]$ with $y_{j_0}=\kappa$.
For $i\notin R_0$, we have $x_i<1/(2s_+\kappa)$, and hence
\[
 B([n]\setminus R_0,j_0)
 =\kappa\sum_{i\notin R_0}x_iA_{ij_0}
 \le\frac{A([n],j_0)}{2s_+}\le\frac12.
\]
Since $B([n],j_0)=1$, it follows that $B(R_0,j_0)\ge1/2$.
For every $i$, row-stochasticity of $A$ and $y_j\ge\kappa$ give
\[
 \kappa x_i
 =\kappa x_i\sum_j A_{ij}
 \le x_i\sum_j A_{ij}y_j
 =B(i,[n])\le b_+.
\]
Thus $B(i,j_0)=\kappa x_iA_{ij_0}\le b_+\mu$, so
\[
 \frac12\le B(R_0,j_0)\le |R_0|b_+\mu.
\]
This proves the first bound in \eqref{eq:diffuse-seed-sizes}.

For the second bound, choose $i_0\in[n]$ with $x_{i_0}=\kappa$.
For $j\notin C_0$, we have $y_j<b_-/(2\kappa)$, so
\[
 B(i_0,[n]\setminus C_0)
 =\kappa\sum_{j\notin C_0}A_{i_0j}y_j
 \le\frac{b_-}{2}\sum_{j\notin C_0}A_{i_0j}
 \le\frac{b_-}{2}.
\]
As $B(i_0,[n])\ge b_-$, we obtain $B(i_0,C_0)\ge b_-/2$.
For each $j\in[n]$, column-stochasticity of $B$ gives
\[
 1=B([n],j)
 =y_j\sum_i x_iA_{ij}
 \ge y_j\kappa\sum_i A_{ij}
 \ge y_j\kappa s_-.
\]
Consequently, $B(i_0,j)=\kappa A_{i_0j}y_j\le\mu/s_-$, and hence
\[
 \frac{b_-}{2}\le B(i_0,C_0)\le |C_0|\frac{\mu}{s_-}.
\]
This proves the second bound in \eqref{eq:diffuse-seed-sizes}.
\end{proof}

\begin{lemma}
\label{lem:dense-buffered-expansion}
Let $n\ge2$, $\gamma\in(1/2,1]$, and $\rho\in(0,1]$.
Let $D\in[0,1]^{n\times n}$ be $(\gamma,\rho)$-dense, and define
\[
 A_{ij}=\frac{D_{ij}}{D(i,[n])}.
\]
For any $q\in(1/2,1)$ and $\alpha>1$ satisfying $\alpha q<\gamma$,
the matrix $A$ satisfies $(q,\alpha-1,\phi)$-two-sided buffered
robust vertex expansion, where
\[
 \phi=\rho(\gamma-\alpha q)>0.
\]
\end{lemma}

\begin{proof}
Density gives $0<D(i,[n])\le n$ for every $i$, so $A$ is well-defined
and row-stochastic. Whenever $D_{ij}\ge\rho$, we have $A_{ij}\ge\rho/n$.
Let $\varnothing\ne I\subseteq[n]$ satisfy $|I|\le qn$, and let
$J\subseteq[n]$ satisfy $|J|\le\alpha|I|$.
Each row indexed by $I$ has at least
\[
 \lceil\gamma n\rceil-|J|\ge(\gamma-\alpha q)n
\]
entries of $D$ outside $J$ that are at least $\rho$.
Therefore,
\[
 A(I,[n]\setminus J)
 \ge |I|(\gamma-\alpha q)n\frac{\rho}{n}
 =\phi|I|.
\]
Similarly, if $\varnothing\ne J\subseteq[n]$ satisfies $|J|\le qn$
and $I\subseteq[n]$ satisfies $|I|\le\alpha|J|$, each column indexed
by $J$ has at least $(\gamma-\alpha q)n$ entries outside $I$ that
are at least $\rho$. Each such entry has $A_{ij}\ge\rho/n$, so
\[
 A([n]\setminus I,J)
 \ge |J|(\gamma-\alpha q)n\frac{\rho}{n}
 =\phi|J|.
\]
Since $\alpha q<\gamma\le1$, the required parameter condition
$q(1+(\alpha-1))<1$ also holds.
\end{proof}

\begin{lemma}[Spectral gap of dense matrices]
\label{lem:dense-normalized-spectral-gap}
Let $n\ge2$, $\gamma\in(1/2,1]$, and $\rho\in(0,1]$.
Suppose that $D\in[0,1]^{n\times n}$ is $(\gamma,\rho)$-dense.
Define
\[
 R=\operatorname{diag}(D\mathbf1),\qquad
 C=\operatorname{diag}(D^\top\mathbf1),\qquad
 E=R^{-1/2}DC^{-1/2}.
\]
Then
\[
 s_2(E)\le\sqrt{1-(2\gamma-1)\rho^2}<1.
\]
\end{lemma}

\begin{proof}
Set $\beta=(2\gamma-1)\rho^2$ and
$P=R^{-1}DC^{-1}D^\top$.
The matrix $P$ is row-stochastic and similar to $EE^\top$.
Any two rows of $D$ share at least $(2\gamma-1)n$ positions
at which both entries are at least $\rho$.
Since every row and column sum of $D$ is at most $n$,
\[
 P_{ik}
 =\frac1{r_i}\sum_{j=1}^n\frac{D_{ij}D_{kj}}{c_j}
 \ge\frac{\beta}{n}
 \qquad\forall i,k\in[n].
\]
Consequently, the Dobrushin coefficient satisfies
\[
 \tau_1(P)
 =1-\min_{i,k}\sum_{j=1}^n\min\{P_{ij},P_{kj}\}
 \le1-\beta.
\]
The Dobrushin eigenvalue bound
\cite{AkelbekKirkland2009} therefore yields
\[
 s_2(E)^2=\lambda_2(P)\le\tau_1(P)\le1-\beta,
\]
as required.
\end{proof}

Now we can prove \Cref{cor:dense-seed}.

\begin{proof}[Proof of \Cref{cor:dense-seed}]
Write $d\triangleq\gamma\rho$ and let $A=\operatorname{diag}(r)^{-1}D$.
Since $D$ is $(\gamma,\rho)$-dense and belongs to $[0,1]^{n\times n}$,
\begin{align}\label{eq-range-ricj-1}
 dn\le r_i(D),c_j(D)\le n\qquad\forall i,j\in[n].
\end{align}
In particular, $D$ has positive row and column sums, $A$ is row-stochastic,
and $dn^2\le D([n],[n])\le n^2$.
We prove each item using its corresponding choice of parameters below.
\begin{itemize}
\item For \Cref{item-sigmaleqgammarhominustau}, the assumption
$0\le\sigma<d-\tau$ implies $0<\tau<d$. Set
\[
 q=\frac34,\qquad
 \theta=d-\frac{\tau}{2},\qquad
 \alpha=\frac{4d^2}{3d^2+\theta^2},\qquad
 \phi=d^2(1-\alpha q)\frac{\tau}{2\theta}.
\]
Then $1<\alpha<4/3$, $\alpha q<1$, and
\[
 d\sqrt{\frac{1-\alpha q}{\alpha(1-q)}}=\theta>d-\tau>\sigma,
 \qquad
 1-\frac{\sigma}{\theta}>\frac{\tau}{2\theta}>0.
\]
Thus $0<\phi<1$, and $q,\alpha,\phi$ depend only on
$\gamma,\rho,\tau$, with no dependence on how close $\sigma$ is to $d-\tau$.
\item For \Cref{item-gammalargeroneovertwo}, set
\[
 q=\frac{1+2\gamma}{4},\qquad
 \alpha=\frac{q+\gamma}{2q},\qquad
 \phi=\frac{\rho(\gamma-q)}{2}.
\]
Since $\gamma>1/2$, we have
\[
 \frac12<q<\gamma\le1,\qquad
 \alpha>1,\qquad
 \alpha q=\frac{q+\gamma}{2}<\gamma,\qquad
 0<\phi=\rho(\gamma-\alpha q)<1.
\]
These parameters depend only on $\gamma,\rho$.
\end{itemize}

We next verify $(q,\alpha-1,\phi)$-two-sided buffered robust vertex
expansion for each of the two choices.
For \Cref{item-gammalargeroneovertwo}, this follows directly from
\Cref{lem:dense-buffered-expansion}.
For \Cref{item-sigmaleqgammarhominustau}, apply
\Cref{thm:spectral-weighted-escape} to the original matrix $D$, taking
\[
 r_-=c_-=dn,\qquad r_+=c_+=n.
\]
Its spectral threshold is the $\theta$ chosen above.
For every nonempty $I\subseteq[n]$ with $|I|\le qn$ and every
$J\subseteq[n]$ with $|J|\le\alpha|I|$, we obtain
\begin{equation}\label{eq-djnsetminusjgeqphidinoverprhogamma-1}
\begin{aligned}
 D(I,[n]\setminus J)
 &\ge\frac{dn^2}{D([n],[n])}(1-\alpha q)
       \left(1-\frac{\sigma}{\theta}\right)D(I,[n])\\
 &\ge d(1-\alpha q)\frac{\tau}{2\theta}D(I,[n])
 =\frac{\phi}{d}D(I,[n]).
\end{aligned}
\end{equation}
Row-normalization and \eqref{eq-range-ricj-1} give
\begin{align}\label{eq-aijgedijovern-1}
 A_{ij}\ge\frac{D_{ij}}{n}\qquad\forall i,j\in[n].
\end{align}
The same marginal bounds imply
\begin{align}\label{eq-dingeqrhogammaabsi}
 D(I,[n])\ge dn|I|,\qquad D([n],J)\ge dn|J|
 \qquad\forall I,J\subseteq[n].
\end{align}
Consequently,
\[
 A(I,[n]\setminus J)
 \ge\frac{D(I,[n]\setminus J)}{n}
 \ge\frac{\phi}{dn}D(I,[n])
 \ge\phi|I|.
\]
For every nonempty $J\subseteq[n]$ with $|J|\le qn$ and every
$I\subseteq[n]$ with $|I|\le\alpha|J|$, the reverse estimate
\eqref{eq:spectral-weighted-reverse} similarly yields
\[
 D([n]\setminus I,J)\ge\frac{\phi}{d}D([n],J),
 \qquad
 A([n]\setminus I,J)
 \ge\frac{\phi}{dn}D([n],J)\ge\phi|J|.
\]
Thus the required two-sided buffered robust vertex expansion holds
in both cases.

By \Cref{prop:strict-hall}, there is a doubly stochastic scaling
$B=XAY$, where $X=\diag(x_1,\ldots,x_n)$ and
$Y=\diag(y_1,\ldots,y_n)$ are positive diagonal matrices.
Normalize these factors so that
\[
 \kappa\triangleq\min_i x_i=\min_j y_j>0.
\]
In particular, $B(i,[n])=B([n],i)=1$ for each $i\in[n]$.
Moreover, \eqref{eq-range-ricj-1} and $D\in[0,1]^{n\times n}$ imply
\[
 A_{ij}=\frac{D_{ij}}{r_i(D)}
 \le\frac{1}{\gamma\rho n}
 \qquad \forall i,j\in[n].
\]
Therefore, 
\[
 A([n],j)=\sum_{i=1}^n A_{ij}
 \le n\cdot \frac{1}{\gamma\rho n}
 \le(\gamma\rho)^{-1}.
\]
Fix a column $j$.  Density gives at least $\gamma n$ indices $i$ for which
$D_{ij}\ge\rho$.  For each such index, \eqref{eq-range-ricj-1} implies
$A_{ij}\ge\rho/n$.  Summing these terms gives
$A([n],j)\ge\gamma\rho$.  
In summary, we have
\begin{equation}
 A(i,[n])=1\quad\forall i\in[n],
 \qquad
 \gamma\rho\le A([n],j)\le(\gamma\rho)^{-1}\quad\forall j\in[n],
 \qquad
 \max_{i,j\in[n]}A_{ij}\le\frac{1}{\gamma\rho n}. \label{eq:dense-standardized-bounds}
\end{equation}
Apply \Cref{lem:diffuse-seed} with
$s_-=d$, $s_+=d^{-1}$, $b_-=b_+=1$, and $\mu=(dn)^{-1}$.
Define
\[
 R_0\triangleq\left\{i\in[n]:x_i\ge\frac{d}{2\kappa}\right\},
 \qquad
 C_0\triangleq\left\{j\in[n]:y_j\ge\frac{1}{2\kappa}\right\}.
\]
Then
\[
 |R_0|\ge\frac{1}{2\mu}=\frac{dn}{2},
 \qquad
 |C_0|\ge\frac{s_-}{2\mu}=\frac{d^2n}{2}.
\]
Let
\[
 m=\min\left\{\lfloor qn\rfloor,
                 \left\lceil\frac{d^2n}{2}\right\rceil\right\}.
\]
Because $0<d\le1$ and both seed cardinalities are integers,
$|R_0|,|C_0|\ge m$.
Set $\eta=\alpha-1$ and take the parameter $c=1$ in
\Cref{thm:seeded-multiscale}. We have $0<\eta<1$, since
$\alpha>1$, $q>1/2$, and $q\alpha<1$.
Moreover, $0<\phi<1$, $c\alpha=\alpha>1$, and, for all sufficiently
large $n$,
\[
 (1+c)\alpha\lfloor qn\rfloor
 =2\alpha\lfloor qn\rfloor>n.
\]
Since $B$ is doubly stochastic, the row-mass condition holds with
$c=1$. All the remaining hypotheses of \Cref{thm:seeded-multiscale}
are satisfied with
\[
 b_+=1,\qquad s_-=d,\qquad s_+=d^{-1},\qquad
 H_x=\frac d2,\qquad H_y=\frac12.
\]
For all sufficiently large $n$, $m=\Theta(n)$ and
\[
 2\alpha\lfloor qn\rfloor-n=\Theta(n),
\]
with positive constants determined by the corresponding parameter choice.
Therefore \eqref{eq:seeded-distortion-rate} gives a constant $T\ge1$
such that
\[
 T^{-1}\le x_i y_j\le T,\qquad
 T^{-1}A_{ij}\le B_{ij}\le T A_{ij}
 \qquad\forall i,j\in[n].
\]
Here $T$ depends only on $\gamma,\rho,\tau$ for
\Cref{item-sigmaleqgammarhominustau}, and only on $\gamma,\rho$ for
\Cref{item-gammalargeroneovertwo}.
To apply \Cref{lem-spectragap-relation}, use the original matrix $D$
as its input matrix, rather than the row-normalized matrix $A$.
Indeed, \eqref{eq-range-ricj-1} and the preceding comparison give
\[
 \frac{D_{ij}}{Tn}\le B_{ij}
 \le\frac{T D_{ij}}{dn}
 \le\frac{T D_{ij}}{d^3n}
 \qquad\forall i,j\in[n].
\]
Moreover, $B\in[0,1]^{n\times n}$ and all its row and column sums
equal $1\in[d,d^{-1}]$. Thus the lemma applies with $t=T$.
Its normalized input is precisely the matrix $E$ in \Cref{cor:dense-seed},
while its matrix $Q$ is the identity,
so its two-step walk is $B^\top B$. Consequently,
\[
 1-s_2(B)^2
 =1-\lambda_2(B^\top B)
 \ge\frac{d^5}{T^2}(1-\sigma^2).
\]
For \Cref{item-sigmaleqgammarhominustau}, $\sigma<d-\tau$ gives
\[
 1-s_2(B)^2
 \ge\frac{d^5}{T^2}\bigl(1-(d-\tau)^2\bigr)>0.
\]
This is a positive constant depending only on $\gamma,\rho,\tau$.
By \Cref{thm:endpoint-gap-global-sinkhorn}, SK achieves total marginal
$\ell_1$ error $\varepsilon$ in
$O_{\gamma,\rho,\tau}(\log n+\log(1/\varepsilon))$ iterations.
For \Cref{item-gammalargeroneovertwo}, \Cref{lem:dense-normalized-spectral-gap}
gives $1-\sigma^2\ge(2\gamma-1)\rho^2$. Hence
\[
 1-s_2(B)^2
 \ge\frac{d^5\rho^2(2\gamma-1)}{T^2}>0,
\]
a positive constant depending only on $\gamma,\rho$.
Another application of \Cref{thm:endpoint-gap-global-sinkhorn} gives
$O_{\gamma,\rho}(\log n+\log(1/\varepsilon))$ iterations.
Each full cycle consists of two normalization steps, so the theorem's
iteration convention changes these bounds only by a constant factor.
The theorem is proved.
\end{proof}

\subsection{Lower bound for Dense Matrices}\label{sec-dense-lowerbound}
In this subsection, we prove \Cref{thm-tight-dense}.

\begin{proof}[Proof of \Cref{thm-tight-dense}]
Let $n=2m$ be even, and let $J_m$ denote the $m\times m$ all-ones
matrix. Define
\[
 c_n=n^{-4},
 \qquad
 b_n=\frac{1-c_n}{1+3c_n},
 \qquad
 p_n=\sqrt{b_nc_n}=\Theta(n^{-2}),
\]
and set
\[
 D_n=
 \begin{pmatrix}
  J_m & b_nJ_m\\
  c_nJ_m & J_m
 \end{pmatrix}.
\]
Every entry of $D_n$ belongs to $(0,1]$, and each row and column
contains exactly $m=n/2$ entries equal to $1$.
Thus $D_n$ is $(1/2,1)$-dense.
Writing $U_m=J_m/m$, we have
\[
 E_n=
 \begin{pmatrix}
  \dfrac{1}{\sqrt{(1+b_n)(1+c_n)}}
  &
  \dfrac{b_n}{1+b_n}\\[6pt]
  \dfrac{c_n}{1+c_n}
  &
  \dfrac{1}{\sqrt{(1+b_n)(1+c_n)}}
 \end{pmatrix}
 \otimes U_m.
\]
The largest singular value of $E_n$ is $1$, and the product of its
two nonzero singular values equals the determinant of the displayed
$2\times2$ factor. Consequently,
\[
 \sigma_n=s_2(E_n)
 =\frac{1-b_nc_n}{(1+b_n)(1+c_n)}.
\]
The definition of $b_n$ gives
$b_n+c_n+3b_nc_n=1$,
and hence
$(1+b_n)(1+c_n)=2(1-b_nc_n)$.
Therefore, $\sigma_n=\frac12$.

Define 
\begin{align*}
 s_0&=\sqrt{\frac{b_n}{c_n}}, \qquad \quad\qquad\qquad\qquad \qquad s_{k+1}
 =
 \frac{(1+p_n^2)s_k+2p_n}
 {2p_ns_k+1+p_n^2},\qquad \qquad \forall k\geq 0.\\
P_0&=
 \begin{pmatrix}
  \dfrac1{1+b_n} & \dfrac{b_n}{1+b_n}\\[6pt]
  \dfrac{c_n}{1+c_n} & \dfrac1{1+c_n}
 \end{pmatrix},\qquad \qquad\qquad P_k=
 \begin{pmatrix}
  \dfrac1{1+p_ns_k} & \dfrac{p_ns_k}{1+p_ns_k}\\[6pt]
  \dfrac{p_n}{s_k+p_n} & \dfrac{s_k}{s_k+p_n}
 \end{pmatrix},\qquad \forall k\geq 0.
\end{align*}
By $A_n^{(0)}=R_n^{-1}D_n$ and the block structure of $D_n$, one can verify
$A_n^{(2k)}=P_k\otimes U_m$.
Define
\[
 \lambda_n=
 \left(\frac{1-p_n}{1+p_n}\right)^2,
 \qquad
 z_k=\frac{s_k-1}{s_k+1}.
\]
The preceding recurrence implies
\[
 z_{k+1}=\lambda_n z_k,
 \qquad
 z_k=z_0\lambda_n^k,
 \qquad
 z_0=
 \frac{\sqrt{b_n}-\sqrt{c_n}}
 {\sqrt{b_n}+\sqrt{c_n}}.
\]

Write the off-diagonal entries of $P_k$ as
\[
 u_k=\frac{p_ns_k}{1+p_ns_k},
 \qquad
 v_k=\frac{p_n}{s_k+p_n}.
\]
Since $s_0>1$ and $0<\lambda_n<1$, we have $s_k>1$ and $u_k>v_k$
for every finite $k$.
Moreover,
\[
 u_k-v_k
 =
 \frac{p_n(s_k^2-1)}{(1+p_ns_k)(s_k+p_n)}
 =
 \frac{(1-\lambda_n)z_k}{1-\lambda_n z_k^2}.
\]
Thus the column sums of $P_k$ are $1-(u_k-v_k)$ and
$1+(u_k-v_k)$.
Each of these column sums is repeated $m=n/2$ times in
$A_n^{(2k)}$, whose row sums all equal $1$.
Therefore,
\begin{equation}\label{eq:boundary-exact-sk-error}
 \sum_{i=1}^{n}\left|r_i(A_n^{(2k)})-1\right|
 +
 \sum_{j=1}^{n}\left|c_j(A_n^{(2k)})-1\right|
 =
 n\,\frac{(1-\lambda_n)z_k}{1-\lambda_n z_k^2}.
\end{equation}

Since $z_0\to1$ and $p_n\to0$, for all sufficiently large $n$ we have
\[
 z_0\ge\frac23,
 \qquad
 2p_n\le1-\lambda_n
 =\frac{4p_n}{(1+p_n)^2}
 \le4p_n.
\]
For every integer $0\le k\le1/(16p_n)$, Bernoulli's inequality gives
\[
 \lambda_n^k
 \ge1-k(1-\lambda_n)
 \ge1-4kp_n
 \ge\frac34.
\]
Consequently, $z_k=z_0\lambda_n^k\ge1/2$.
Since $0<1-\lambda_n z_k^2\le1$, 
\eqref{eq:boundary-exact-sk-error} yields
\begin{equation}\label{eq:boundary-even-sk-error}
 \sum_{i=1}^{n}\left|r_i(A_n^{(2k)})-1\right|
 +
 \sum_{j=1}^{n}\left|c_j(A_n^{(2k)})-1\right|
 \ge np_n > \frac{1}{400n},\qquad \forall\, 0\le k\le1/(16p_n).
\end{equation}

The intervening odd iterations also have a comparable error.
To see this, set $\delta_k=u_k-v_k$.
Column normalization of $P_k$ produces row sums
\[
 1+\frac{\delta_k\det P_k}{1-\delta_k^2},
 \qquad
 1-\frac{\delta_k\det P_k}{1-\delta_k^2}.
\]
Furthermore,
\[
 \det P_0
 =\frac{1-b_nc_n}{(1+b_n)(1+c_n)}
 =\frac12.
\]
The determinant of the $2\times2$ factor cannot decrease under
either normalization: the two margins being normalized have sum
$2$ and hence product at most $1$, and normalization divides the
determinant by their product.
It follows that $\det P_k\ge1/2$.
Therefore, for each $0\le k\le1/(16p_n)$,
\begin{align*}
 &\sum_{i=1}^{n}\left|r_i(A_n^{(2k+1)})-1\right|
 +
 \sum_{j=1}^{n}\left|c_j(A_n^{(2k+1)})-1\right|=
 n\,\frac{\delta_k\det P_k}{1-\delta_k^2}
 \ge\frac{n\delta_k}{2}\\
 &\qquad=
 \frac12\left(
 \sum_{i=1}^{n}\left|r_i(A_n^{(2k)})-1\right|
 +
 \sum_{j=1}^{n}\left|c_j(A_n^{(2k)})-1\right|
 \right)
 \ge\frac{np_n}{2} >\frac{1}{400n}.
\end{align*}
Combined with \eqref{eq:boundary-even-sk-error} and $1/(16p_n) = \Omega(n^2)$,
the theorem is immediate.
\end{proof}

\newpage

%% file: MissingProof.tex
\section{Buffered robust vertex expansion implies strict Hall}
\label{sec-bve-stricthall}
In this section, we prove \Cref{prop:strict-hall}.

\begin{proof}[Proof of \Cref{prop:strict-hall}]
Fix a nonempty proper row set $I\subsetneq[n]$.  We distinguish the two
exhaustive cases $|I|\le qn$ and $|I|>qn$.

Suppose first that $|I|\le qn$.  Assume for contradiction that
$|N_A(I)|\le(1+\eta)|I|$, and set $J\triangleq N_A(I)$.  The pair $(I,J)$
has $|J|\le(1+\eta)|I|$.  Moreover, the definition of $N_A(I)$ gives
$A(I,[n]\setminus J)=0$.  Since $I\neq\varnothing$ and $\phi>0$, we have
$\phi|I|>0$.  On the other hand, \eqref{eq:buffered-forward} gives
$A(I,[n]\setminus J)\ge\phi|I|$.  This contradicts
$A(I,[n]\setminus J)=0$.  Thus
$|N_A(I)|>(1+\eta)|I|>|I|$.

Suppose now that $|I|>qn$, and set $K\triangleq[n]\setminus I$.  Since $I$
is proper, $K\neq\varnothing$.  The inequalities $|I|>qn$ and $q>1/2$
give
\[
 0<|K|=n-|I|<(1-q)n<qn.
\]
Assume for contradiction that $|N_A(I)|\le|I|$.  Taking complements in
$[n]$ gives
\[
 |[n]\setminus N_A(I)|=n-|N_A(I)|\geq n-|I|=|K|.
\]
Choose $J\subseteq[n]\setminus N_A(I)$ with $|J|=|K|$.  Then
$J\neq\varnothing$, $|J|<qn$, and $|K|=|J|\le(1+\eta)|J|$.  Thus the
column set $J$ and the row buffer $K$ lie in the admissible range of
\eqref{eq:buffered-reverse}, which gives
\[
 A([n]\setminus K,J)\geq\phi|J|>0.
\]
The final strict inequality holds because $J\neq\varnothing$ and $\phi>0$.
On the other hand, $[n]\setminus K=I$ and
$J\subseteq[n]\setminus N_A(I)$.  Since $A$ is nonnegative, every entry
$A_{ij}$ with $i\in I$ and $j\in J$ equals $0$.  Therefore
$A([n]\setminus K,J)=A(I,J)=0$, a contradiction.  This proves
\eqref{eq:strict-hall}.

Furthermore, \eqref{eq:strict-hall} implies that $A$ has total support.
Indeed, fix any positive entry $A_{ij}$. After deleting row $i$
and column $j$, every nonempty set $S$ of remaining rows has
at least
\[
 |N_A(S)\setminus\{j\}|
 \ge |N_A(S)|-1
 \ge |S|
\]
neighbors. 
By Hall's marriage theorem~\cite{hall1935representatives},
the remaining rows and columns admit a perfect matching.
Together with $(i,j)$, this yields a perfect matching
containing the edge corresponding to $A_{ij}$.
Thus every positive entry belongs to a positive diagonal,
so $A$ has total support. By the Sinkhorn--Knopp theorem
\cite{sinkhorn1967concerning}, there exist positive diagonal
matrices $X$ and $Y$ such that $XAY$ is doubly stochastic.
The proposition is proved.
\end{proof}

\section{Spectral-gap comparison under bounded entrywise rescaling}\label{sec-spectralgapcomparison}
In this section, we prove \Cref{lem-spectragap-relation}.
The following lemma is used in the proof of \Cref{lem-spectragap-relation}.

\begin{lemma}
\label{lem-strict_spectral_gap}
Let $D \in \mathbb{R}_{\ge 0}^{n \times n}$ be a symmetric doubly stochastic matrix. Let $G \in \mathbb{R}_{\ge 0}^{n \times n}$ be a symmetric matrix with its largest eigenvalue $\lambda_1(G) = 1$, and let $\=v \in \mathbb{R}_{>0}^n$ be its corresponding strictly positive eigenvector. 
Suppose there exists a constant $\alpha \in (0, 1]$ such that $D_{i,j} \ge \alpha G_{i,j}$ for all $i,j \in [n]$.  
Then all eigenvalues of $D$ and $G$ are real.
Moreover,
$$ 1 - \lambda_2(D) \ge \alpha \left( \frac{\min_i v_i}{\max_i v_i} \right)^2\left( 1 - \lambda_2(G) \right).$$
\end{lemma}

\begin{proof}
Since $D$ and $G$ are real symmetric matrices, their eigenvalues are all real.
Define 
\begin{align}\label{eq-def-pdiagvinversegdiagv}
\qquad P = (\diag(\=v))^{-1}\, G \,\diag(\=v).
\end{align}
Since $P$ is similar to $G$, we have $ \lambda_2(P) = \lambda_2(G)$.
Thus, to prove this lemma, it is sufficient to prove
\begin{align}\label{eq-relation-spectragap-dvsp}
1 - \lambda_2(D) \ge \alpha \left( \frac{\min_i v_i}{\max_i v_i} \right)^2\left( 1 - \lambda_2(P) \right).
\end{align}
In the following, we prove \eqref{eq-relation-spectragap-dvsp}.

By \eqref{eq-def-pdiagvinversegdiagv}, we obtain
\begin{align}\label{eq-pij-vjvigij}
P_{i,j} = \frac{v_j}{v_i} G_{i,j}.
\end{align}
Given that $G \=v = \=v$, it follows that
$$ r_i(P) = \frac{1}{v_i} \sum_j G_{i,j} v_j = \frac{v_i}{v_i} = 1.$$
Thus, $P$ is a row-stochastic matrix. 
Furthermore, define $\=\pi = (\pi_1,\cdots,\pi_n)$
where $\pi_i = \frac{v_i^2}{\|v\|^2}$. 
It is straightforward to verify that $\sum_{i \in [n]} \pi_i = 1$; thus, $\={\pi}$ represents a valid probability distribution.
Moreover,
by \eqref{eq-pij-vjvigij} and that $G$ is symmetric, we obtain
$$ \pi_i P_{i,j} = \frac{v_i^2}{\|v\|^2} \frac{v_j}{v_i} G_{i,j} = \frac{v_j^2}{\|v\|^2} \frac{v_i}{v_j}G_{i,j} = \frac{v_j^2}{\|v\|^2} \frac{v_i}{v_j}G_{j,i} = \pi_j P_{j,i} ,\qquad \forall\  i,j\in [n].$$
Hence, 
$P$ is reversible with respect to $\=\pi$.
Let $\=\mu$ denote the discrete uniform distribution $\left(\frac{1}{n},\cdots,\frac{1}{n}\right)$.
By $D$ is symmetric, we have
$$ \mu_i D_{i,j} = \mu_j D_{j,i} ,\qquad \forall\  i,j\in [n].$$
Hence, 
$D$ is reversible with respect to $\=\mu$.

Recall the definitions of $\mathcal{E}$ and $\text{Var}$ in \eqref{eq-def-epxx-varpix}.
By that $D_{i,j} \ge \alpha G_{i,j}$ for each $i,j\in [n]$, we obtain
\begin{align*}
\mathcal{E}_D(x,x) &= \frac{1}{2} \sum_{i,j} \mu_i D_{i,j} (x_i - x_j)^2 = \frac{1}{2n} \sum_{i,j} D_{i,j} (x_i - x_j)^2 \ge \frac{\alpha}{2n} \sum_{i,j} G_{i,j} (x_i - x_j)^2.
\end{align*}
Meanwhile,
\begin{align*}
\mathcal{E}_P(x,x) &= \frac{1}{2} \sum_{i,j} \pi_i P_{i,j} (x_i - x_j)^2 = \frac{1}{2\|v\|^2} \sum_{i,j} G_{i,j} v_i v_j (x_i - x_j)^2 \le \frac{\max_k v_k^2}{2\|v\|^2} \sum_{i,j} G_{i,j} (x_i - x_j)^2.
\end{align*}
Hence,
\begin{equation}\label{eq-dirichlet_comp}
\mathcal{E}_D(x,x) \ge   \frac{\alpha\|v\|^2}{n \max_k v_k^2}\, \mathcal{E}_P(x,x).
\end{equation}
In addition, we have
$$ \mu_i = \frac{1}{n} = \left( \frac{\|v\|^2}{n v_i^2} \right) \pi_i \le \left( \frac{\|v\|^2}{n \min_k v_k^2} \right) \pi_i \qquad \forall\  i\in [n].$$
Thus, 
\begin{equation}\label{eq-var_comp}
\text{Var}_{\mu}(x) = \min_{c} \sum_i \mu_i(x_i - c)^2 \leq \frac{\|v\|^2}{n \min_k v_k^2} \min_{c} \sum_i \pi_i(x_i - c)^2 
= \frac{\|v\|^2}{n \min_k v_k^2}  \text{Var}_{\pi}(x).
\end{equation}
By \Cref{lem-variational_characterization}, \eqref{eq-dirichlet_comp} and \eqref{eq-var_comp}, we obtain 
\begin{align*} 
\quad 1 - \lambda_2(D)  &= \inf_{\substack{x \in \mathbb{R}^n \\ \mathrm{Var}_{\mu}(x) > 0}} \frac{\mathcal{E}_D(x,x)}{\mathrm{Var}_{\mu}(x)}  
\ge \inf_{\substack{x \in \mathbb{R}^n \\ \mathrm{Var}_{\mu}(x) > 0}}\alpha \left( \frac{\min_k v_k}{\max_k v_k} \right)^2\frac{\mathcal{E}_P(x,x)}{\mathrm{Var}_{\pi}(x)}  \\
&= 
\inf_{\substack{x \in \mathbb{R}^n \\ \mathrm{Var}_{\pi}(x) > 0}}\alpha \left( \frac{\min_k v_k}{\max_k v_k} \right)^2\frac{\mathcal{E}_P(x,x)}{\mathrm{Var}_{\pi}(x)} 
=\alpha \left( \frac{\min_k v_k}{\max_k v_k} \right)^2\inf_{\substack{x \in \mathbb{R}^n \\ \mathrm{Var}_{\pi}(x) > 0}}\frac{\mathcal{E}_P(x,x)}{\mathrm{Var}_{\pi}(x)} \\
& = \alpha \left( \frac{\min_k v_k}{\max_k v_k} \right)^2 \left( 1 - \lambda_2(P)\right). 
\end{align*}
Hence, \eqref{eq-relation-spectragap-dvsp} is proved.
The lemma is immediate.
\end{proof}

Now we can prove \Cref{lem-spectragap-relation}.

\begin{proof}[Proof of \Cref{lem-spectragap-relation}]
Write $d=\gamma\rho\in(0,1]$, and define $G=(A')^\top A'$.
Then $\lambda_2(G)=s_2(A')^2$, so it suffices to prove
\begin{equation}\label{eq-relation-spectragap-dg}
 1-\lambda_2(D)\ge\frac{d^5}{t^2}\bigl(1-\lambda_2(G)\bigr).
\end{equation}
Density and $A\in[0,1]^{n\times n}$ give
\begin{equation}\label{eq-range-row-column-sum}
 dn\le r_i(A),c_j(A)\le n\le n/d
 \qquad\forall i,j\in[n].
\end{equation}
In particular, all these marginal sums are positive, and
\[
 A'_{ij}=\frac{A_{ij}}{\sqrt{r_i(A)c_j(A)}}
 \le\frac{A_{ij}}{dn}.
\]
Let $B'=Q^{-1/2}B$. By \eqref{eq-rib-bound-cib-eqone}
and \eqref{eq-relation-aijbij},
\[
 B'_{ij}\ge\sqrt d\,B_{ij}
 \ge\frac{\sqrt d}{tn}A_{ij}
 \ge\frac{d^{3/2}}{t}A'_{ij}.
\]
All entries are nonnegative, so
\begin{equation}\label{eq-sumkinnbprimekibprimekj-bound}
 \sum_k B'_{ki}B'_{kj}
 \ge\frac{d^3}{t^2}\sum_k A'_{ki}A'_{kj}
 \qquad\forall i,j\in[n].
\end{equation}
Since $D=(B')^\top B'$ and $G=(A')^\top A'$, this yields
\begin{equation}\label{eq-range-elements-dg}
 D_{ij}=\sum_k B'_{ki}B'_{kj}
 \ge\frac{d^3}{t^2}G_{ij}
 \qquad\forall i,j\in[n].
\end{equation}
Moreover,
\[
 D\one=B^\top Q^{-1}B\one=B^\top\one=\one.
\]
Thus $D$ is symmetric and doubly stochastic.

To identify the top eigenvalue of $G$, set
$P=Q_2^{-1}A^\top Q_1^{-1}A$. Then
\begin{equation}\label{eq-relation-g-rowstochastic}
 G=Q_2^{-1/2}A^\top Q_1^{-1}AQ_2^{-1/2}
  =Q_2^{1/2}P Q_2^{-1/2},
\end{equation}
and
\begin{equation}\label{eq-eigenvalue-qaqa}
 P\one=Q_2^{-1}A^\top Q_1^{-1}A\one
        =Q_2^{-1}A^\top\one=\one.
\end{equation}
Hence $P$ is row-stochastic, so its spectral radius is $1$.
By similarity, $\lambda_1(G)=1$, with strictly positive eigenvector
$v=Q_2^{1/2}\one$.

The comparison parameter $d^3/t^2$ belongs to $(0,1]$:
indeed, column-stochasticity of $B$ and its entrywise lower bound imply
\[
 1=\sum_i B_{ij}\ge\frac{c_j(A)}{tn}\ge\frac dt,
\]
so $t\ge d$ and $d^3/t^2\le d\le1$.
We may therefore apply \Cref{lem-strict_spectral_gap} to obtain
\[
 1-\lambda_2(D)
 \ge\frac{d^3}{t^2}
       \frac{\min_j c_j(A)}{\max_j c_j(A)}
       \bigl(1-\lambda_2(G)\bigr).
\]
Finally, \eqref{eq-range-row-column-sum} gives
\[
 \frac{\min_j c_j(A)}{\max_j c_j(A)}\ge d\ge d^2.
\]
Since $1-\lambda_2(G)\ge0$, this proves
\eqref{eq-relation-spectragap-dg} and the lemma.
\end{proof}

\vspace{5cm}